\documentclass[acmlarge, imwut]{acmart}

\setcopyright{cc}
\setcctype{by}
\acmJournal{IMWUT}
\acmYear{2026} \acmVolume{10} \acmNumber{4} \acmArticle{252}
\acmMonth{12} \acmDOI{10.1145/3857956}

\usepackage[shortlabels]{enumitem}
\usepackage{tabularx}
\usepackage{multirow}
\usepackage[most]{tcolorbox}
\usepackage{fvextra}
\usepackage{subcaption}
\usepackage{seqsplit}
\usepackage{color, colortbl}
\setlist{nosep}

\definecolor{revisionblue}{RGB}{0,90,230}
\newcommand{\projectname}[1]{TRACE}

\newcommand{\rw}[1]{{#1}}

\begin{document}
\title{\projectname{}: Temporal Reasoning over Context and Evidence for Activity Recognition in Smart Homes}

\author{Yingtian Shi}
\orcid{0000-0001-8733-7041}
\email{yshi457@gatech.edu}
\affiliation{%
  \institution{School of Interactive Computing, Georgia Institute of Technology}
  \city{Atlanta, Georgia}
  \country{USA}}

\author{Abivishaq Balasubramanian}
\orcid{0000-0003-1984-7765}
\email{b.abivshaq@gmail.com}
\affiliation{%
  \institution{Georgia Institute of Technology}
  \city{Atlanta}
  \state{GA}
  \country{USA}
}

\author{Jessica Herring}
\orcid{0000-0001-5094-266X}
\email{jherring40@gatech.edu}
\affiliation{%
  \institution{Berry College}
  \city{Rome, Georgia}
  \country{USA}
}

\author{Jiachen Li}
\orcid{0000-0002-6084-5131}
\email{li.jiachen4@northeastern.edu}
\affiliation{%
  \institution{Northeastern University}
  \city{Boston}
  \state{MA}
  \country{USA}
}

\author{Juan Macias Romero}
\orcid{0009-0008-1179-7568}
\email{juanrmaci@gmail.com}
\affiliation{%
  \institution{Universidad Carlos III de Madrid}
  \city{Madrid}
  \state{Madrid}
  \country{Spain}
}

\author{Rosemarie Santa Gonzalez}
\orcid{0000-0002-4759-5756}
\email{rosemarie.santa@gatech.edu}
\affiliation{%
  \institution{Georgia Institute of Technology}
  \city{Atlanta}
  \state{GA}
  \country{USA}
}

\author{Varun Mishra}
\orcid{0000-0003-3891-5460}
\email{v.mishra@northeastern.edu}
\affiliation{%
  \institution{Northeastern University}
  \city{Boston}
  \state{MA}
  \country{USA}
}

\author{Agata Rozga}
\orcid{0000-0002-5558-9786}
\email{agata@gatech.edu}
\affiliation{
  \institution{School of Interactive Computing, Georgia Institute of Technology}
  \city{Atlanta, Georgia}
  \country{USA}}

\author{Xiang Zhi Tan}
\orcid{0000-0002-6455-4972}
\email{zhi.tan@northeastern.edu}
\affiliation{%
  \institution{Northeastern University}
  \city{Boston}
  \state{MA}
  \country{USA}
}

\author{Thomas Plötz}
\email{thomas.ploetz@gatech.edu}
\orcid{0000-0002-1243-7563}
\affiliation{%
  \institution{School of Interactive Computing, Georgia Institute of Technology}
  \city{Atlanta, Georgia}
  \country{USA}}

\renewcommand{\shortauthors}{Shi et al.}

\begin{abstract}
Human activity recognition (HAR) in smart homes remains challenging because many daily activities exhibit similar local sensor patterns, while minimally intrusive sensing provides sparse and ambiguous observations. As a result, methods based on short temporal or event windows often fail to capture the broader temporal and behavioral context needed for reliable activity understanding. We present \projectname{} (Temporal Reasoning over Context and Evidence), a contextual activity recognition framework for smart homes that integrates multi-source sensor evidence with user-specific contextual priors to improve activity interpretation.
Rather than treating recognition as a \rw{temporally isolated} classification problem, \projectname{} leverages contextual reasoning to resolve ambiguities and reduce fragmented predictions.
We evaluate ~\projectname{} on public benchmarks and in \rw{a practical case study} conducted in our own smart-home environment.
Results show that \projectname{} improves recognition accuracy for semantically complex activities, produces more temporally coherent predictions that better align with user-specific routines, and maintains robust performance under cross-domain transfer and missing-modality conditions.
These findings demonstrate the value of contextual reasoning for advancing smart-home HAR.
\end{abstract}





\maketitle

\section{Introduction}

Human activity recognition (HAR) in smart homes has great practical potential for a range of application domains and thus has been at the forefront of ubiquitous computing for years. 
HAR facilitates capabilities like fall detection~\cite{al2024human}, sleep analysis~\cite{sathyanarayana2016robust}, and adaptive automation~\cite{gladence2017home, liu2023understanding, shi2024bridging}, and as such HAR systems provide the technical foundation for health monitoring~\cite{chen2011knowledge, jalal2012recognition} and personalized interaction~\cite{smisek20113d} for users.
In real-world deployments, these systems are often built on minimally intrusive sensors to preserve usability and privacy.

However, this design choice comes at the cost of greater ambiguity in activity recognition. In many smart-home HAR systems~\cite{yang2015deep,yao2017deepsense,comparative_quigley_2018}, recognition is still performed over short temporal or event windows, with predictions made mainly from local sensor patterns. 
As a result, \rw{such fixed, short 
window-based inference supports efficient and responsive recognition, but
provides limited access to broader behavioral context when local observations
alone are insufficient.}

With minimally intrusive sensing, different activities can produce highly similar observations. For example, a system may detect that a user is sitting at the dining table and incorrectly classify the activity as eating, while the user may actually be working or reading at the table. 
Without contextual information such as time of day, preceding activities, or the user’s routines, the model lacks the broader situational cues needed to distinguish between plausible but confounding interpretations, which can result in unreliable predictions and fragmented timelines with frequent jumps across similar activities.

\rw{Prior work has incorporated temporal context through probabilistic graphical models~\cite{patterson2005fine} and hierarchical deep models~\cite{10.1145/3410531.3414306, radu2018multimodal, zhang2021harmi}. Although these approaches demonstrate the value of contextual information, graphical models often depend on manually specified structures, while deep models require substantial labeled data and primarily learn statistical regularities. Both remain limited in flexibly applying common-sense knowledge and user-specific routines to reconsider predictions that are implausible in context.}

\rw{To complement short-window recognition with broader behavioral information,} we introduce routine priors and context into the inference process
and explore Large Language Models (LLMs) as a reasoning layer for contextual activity refinement.
Rather than replacing sensor-based activity recognizers, LLMs are used to reinterpret recognition outputs in light of general knowledge about everyday behavior and user-specific context. 
They can help evaluate whether a predicted activity is reasonable given the time of day, recent activity history, and the user’s typical routines. Because such contextual information can be injected directly through prompting, the framework can adapt its inference to individual users without requiring additional labeled training data for each home. 
By combining low-level recognition results with higher-level contextual constraints, the framework can better resolve ambiguous predictions and produce activity labels that are more consistent with the user’s daily behavior.

We introduce \projectname{} (Temporal Reasoning over Context and Evidence), 
\rw{a framework that augments local activity predictions with progressively broader contextual
information.}
\projectname{} first produces minute-level interpretations by aligning and combining multiple sensing and inference sources, such as ambient events and wearable-derived activity estimates, and then incrementally refines them over longer temporal horizons. By incorporating common-sense knowledge and user-specific routines as contextual priors, the framework can correct noisy local predictions, merge fragmented segments, and suppress context-inconsistent false positives.


To validate the effectiveness of ~\projectname{}, we conducted a two-phase evaluation.
First, we assess system performance on public benchmark datasets, including CASAS and MARBLE, under both in-domain and cross-domain settings. 
Second, we examine the contribution of different information sources and reasoning modules in a \rw{practical case study} conducted in our smart-home environment.
Across these evaluations, \projectname{} demonstrates stronger performance in temporally complex and cross-domain settings, produces more coherent activity segments.

The contributions of this work are as follows:
\begin{itemize}
    
    \item We introduce \projectname{}, a smart-home HAR framework that integrates multi-source sensor evidence, user-specific contextual priors, and progressive temporal refinement into the activity inference process.

    \item \rw{ We conduct an extensive experimental evaluation of \projectname{} on multiple public benchmarks and in a smart-home case study. The results demonstrate that \projectname{} is particularly effective in challenging settings where activities are difficult to distinguish from short local sensor windows alone. By reasoning jointly over local sensor evidence, spatial context, user routines, and recent activity history, \projectname{} can resolve otherwise ambiguous observations and produce more temporally coherent predictions than baseline methods.}
 
    \item Our analysis offers guidance for the design of contextual HAR systems by showing how different modalities, contextual priors, refinement stages, and LLM backbones contribute to recognition performance and activity interpretation.

\end{itemize}

\section{Related Work}
Although sensor-based Human Activity Recognition (HAR) in smart homes has made significant progress, existing approaches still face limitations in understanding long-term behaviors and capturing high-level context. 
To provide an overview of this field and highlight remaining gaps, we organize the related work into three main directions. First, we review traditional and deep learning-based HAR methods. Second, we summarize approaches that incorporate contextual information and sequential modeling to capture dependencies between activities. Third, we discuss recent advances in knowledge-driven and large language model (LLM)-based methods, which aim to enhance semantic understanding and reasoning. 

\subsection{Sensor-Based Human Activity Recognition in Smart Homes}
Human Activity Recognition (HAR) in smart homes aims to infer residents’ daily activities from ambient sensor data~\cite{krishnan2014activity, Wang_2019,survey_bouchabou_2021,sensorbased_dang_2020, leng2026agentsense} such as motion sensors, contact sensors, and wearable sensors~\cite{lara2012survey,zhang2022wearable,deep_hammerla_2016,leng2023generating}. With the increasing deployment of Internet of Things (IoT) devices, sensor-based HAR has become a key enabling technology for applications such as health monitoring~\cite{chen2011knowledge, jalal2012recognition,health_sahu_2022}, fall detection~\cite{al2024human}, and assisted living~\cite{mpdmodel_gong_2012}.
Early research in this area mainly relied on probabilistic and statistical models like Hidden Markov Models (HMM) and Conditional Random Fields (CRF) to capture temporal dependencies between sensor observations and human activities~\cite{van2008accurate,review_kulsoom_2022,learning_cook_2012,activity_fahad_2014,evaluating_alshammari_2018}. 
With the increasing availability of smart home datasets and the rise of deep learning, more recent studies have adopted neural network architectures such as convolutional neural networks (CNNs)~\cite{jiang2015human,sharma2025human,multitask_duan_2023}, recurrent neural networks (RNNs)~\cite{murad2017deep,deep_foumani_2024}, and Long Short-Term Memory (LSTM)~\cite{deepConvLSTM, thukral2025layout,deep_chen_2020} networks to model sequential sensor events and improve recognition performance. These models typically transform sensor streams into fixed-length segments and perform activity classification on each segment independently.

A common design in many HAR systems is the use of sliding temporal or event windows, where sensor events within a short time are aggregated to extract features and predict the activity label~\cite{yang2015deep,yao2017deepsense,comparative_quigley_2018}. While this approach enables real-time inference, it primarily focuses on short-term sensor patterns, often ignoring broader behavioral context.
However, human activities in daily life are rarely isolated~\cite{human_chen_2019}. Instead, they are embedded in longer user routines and contextual patterns, such as daily schedules and habitual activity sequences. For example, the activity of preparing breakfast is not only characterized by sensor activations in the kitchen but is also associated with time-of-day and preceding activities in the morning routine. Short-window models that rely only on local sensor observations may therefore struggle to distinguish activities with similar sensor patterns but different contextual meanings~\cite{chen2011knowledge}.
Although some studies incorporate longer temporal dependencies through sequential or hierarchical modeling~\cite{krishnan2014activity,joint_hamad_2020}, most systems still center on local classification, limiting their ability to understand long-term routines and achieve robust performance in realistic deployment.

\subsection{Context-Aware and Sequential Activity Modeling}
To address the limitations of short-window activity recognition, several studies have explored incorporating richer temporal and contextual information into activity modeling~\cite{efficient_hamad_2020,human_chen_2019,ye2012situation}. 
Early approaches often relied on knowledge-driven frameworks, such as ontologies~\cite{chen2009ontology,cosar_riboni_2011}, rule-based systems~\cite{chen2011knowledge,van2008accurate,activity_tapia_2004}, or probabilistic graphical models~\cite{zhu2014context,patterson2005fine,activity_fahad_2014}, to explicitly represent relationships between sensors, activities, locations, and context.
These methods facilitate reasoning over activity sequences using predefined structures and domain knowledge. 
However, as the number of activities and contextual variables increases, these methods face state-space explosion, making inference computationally expensive and difficult to scale.
Moreover, constructing and maintaining such knowledge bases often requires extensive manual effort and expert knowledge.

More recent studies adopt data-driven approaches to model temporal dependencies between activities using deep learning techniques. 
These methods attempt to capture relationships between activities by learning activity transition patterns or sequential dependencies from sensor data~\cite{recognizing_singla_2010, rashidi2010discovering,feature_pltz_2011,mohd2022deep,omolaja2022context}. 
While these approaches capture temporal correlations between activities, they require large amounts of labeled data and primarily reflect statistical patterns~\cite{krishnan2014activity}, lacking deeper semantic understanding of human routines.
As a result, although these methods can capture frequent activity transitions, they often struggle to interpret higher-level routine structures or contextual relationships~\cite{activity_cook_2015}. 
This limitation restricts their ability to reason about contextual activity relationships or uncommon behavioral patterns.
Despite progress in sequential modeling, these approaches remain limited in understanding long-term daily routines and common-sense knowledge. This motivates the use of large language models (LLMs) for enhancing HAR. By leveraging knowledge of daily routines and contextual information, LLMs can guide or refine activity recognition~\cite{wei2022chain}.

\subsection{Knowledge-Driven and LLM-Based Activity Reasoning}

Recent advances in large language models (LLMs) provide a new avenue for enhancing human activity recognition. A growing body of work has explored how LLMs and knowledge‑based techniques can enhance HAR~\cite{yan2025largelanguagemodelguidedsemantic, arrotta2024contextgpt,cumin2026knowledge}. Unlike deep learning approaches that require large amounts of labeled data and focus primarily on local sensor patterns, LLM‑based methods leverage the rich semantic and contextual knowledge encoded in pretrained language models to reason about activities at a higher level and improve generalization.

One direction of research converts sensor data into representations that LLMs can interpret~\cite{llmguided_ronando_2025, llms_demirel_2025}. For example, SensorLLM~\cite{sensorllm_2025_2025} introduces a two‑stage framework that aligns multivariate sensor time‑series with human‑intuitive trend descriptions, enabling LLMs to capture numerical variations and channel‑specific features and perform HAR classification competitively with state‑of‑the‑art models. Similarly, studies~\cite{hargpt_ji_2024} have demonstrated that with carefully designed prompts, powerful LLMs like GPT‑4 can perform zero‑shot activity recognition from raw inertial measurement unit (IMU) data, recognizing activities such as walking or climbing stairs without task‑specific training.

Other research investigates the use of LLMs as agents or multimodal reasoners~\cite{llm4har_hong_2025, multidimensional_hasan_2024, llasa_ye_2025, le2025multi}. For instance, on‑device multimodal LLM agents~\cite{ondevice_siam_2025} have been developed to not only classify activities from heterogeneous sensors but also provide interpretable reasoning and Q\&A capabilities, bridging the gap between classification and higher‑level semantic understanding. Retrieval‑augmented generation (RAG) frameworks take this further by combining lightweight statistical descriptors with LLM inference, enabling training‑free activity recognition that performs robustly across diverse benchmarks without fine‑tuning~\cite{raghar_sivaroopan_2025}.

However, existing LLM-HAR methods mainly focus on short-term alignment or single-modality sensors, lacking background knowledge of long-term routines. 
Building on these advances, our work leverages LLMs to incorporate routine and common-sense knowledge into HAR, extending the temporal horizon beyond short sensor windows. By reasoning over typical activity sequences and temporal dependencies, the system can disambiguate activities with similar sensor patterns but different contextual meanings, effectively bridging the gap between low-level activity recognition and high-level behavioral understanding.

\section{Methodology}
Many existing smart-home HAR systems infer activities from short sensor windows and thus have limited access to broader behavioral context.
As a result, recognition errors caused by local ambiguity or noisy sensor readings are often difficult to correct after the fact.
To address this limitation, we introduce ~\projectname{} (Temporal Reasoning over Context and Evidence), which extends inference beyond local sensor patterns by explicitly organizing behavioral representation at three levels: 
\textit{i)} sensor-observation level evidence; 
\textit{ii)} activity level predictions; and 
\textit{iii)} routine level context.

\begin{figure}[t]
    \centering
    \includegraphics[width=0.9\linewidth]{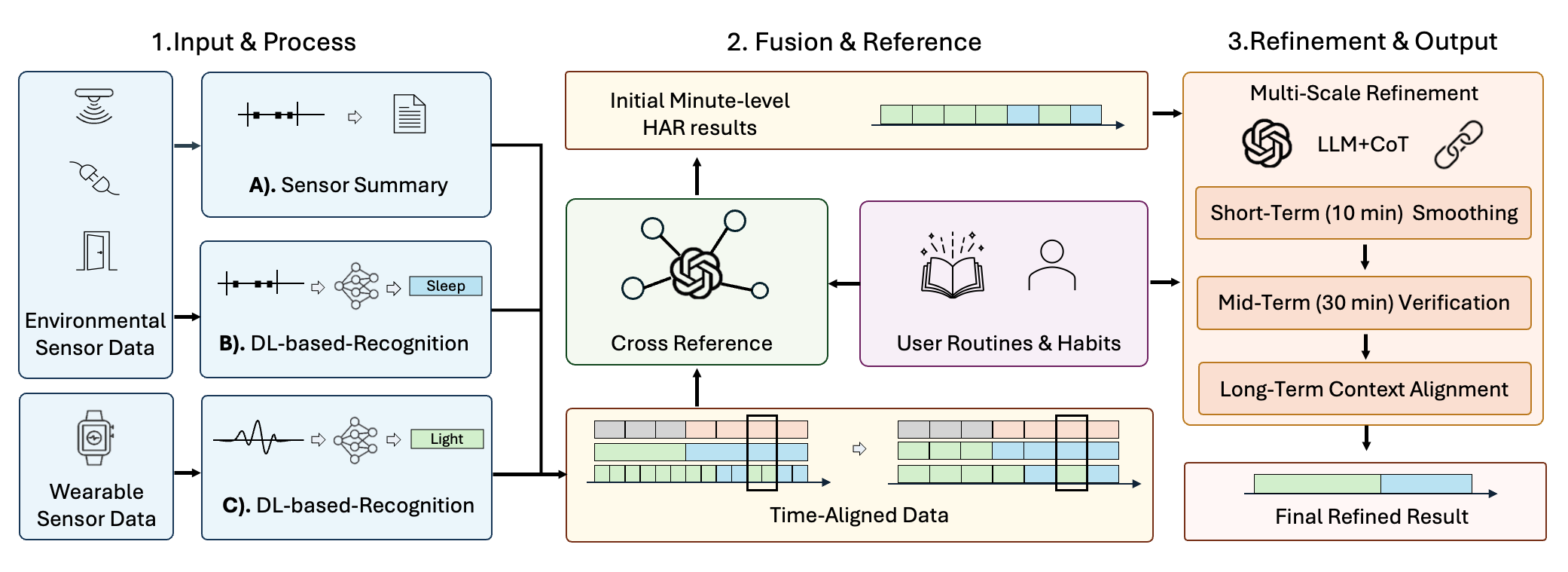}
    \caption{Overview of the \projectname{} framework. The system processes multimodal data through three stages: (1) Data Input and Feature-specific Processing, (2) Time-aligned Fusion and Cross-reference, and (3) Multi-scale Refinement via an LLM with Chain-of-Thought (CoT) reasoning, incorporating prior knowledge and user-specific routines. }
    \label{fig:outline}
\end{figure}

Based on this representation, ~\projectname{} consists of three stages (Fig.~\ref{fig:outline}). Raw sensor streams are first processed by modality-specific pipelines to generate sensor summaries and activity predictions. The outputs are aligned to a shared timeline and cross-referenced by an LLM to produce an initial interpretation for each minute. 
Then, the fused results are progressively refined over longer temporal horizons using routine-level priors.

The goal of this design is not to replace sensor-based recognizers, but to augment them with a contextual reasoning layer. After multi-scale inference, the final output is consistent with the behavioral context, stays temporally coherent with neighboring activity segments, and matches the user’s longer-term routines. 
This multi-scale reasoning process improves robustness to noisy activity predictions while preserving logical consistency. In addition to sensor signals, \projectname{} incorporates user-specific contextual information, such as daily routines and activity preferences, when available.

\subsection{Behavior Representation Levels}
For clarity, we distinguish three levels of behavioral representation in ~\projectname{}. This distinction is introduced to clarify what kind of information is preserved and how different forms of context contribute to the final activity interpretation.

\textbf{Sensor-observation level.} 
\rw{This level captures fine-grained, short-term evidence close to the original sensor stream rather than semantic activities. It includes sensor triggers and state or value changes within a short duration. Environmental examples include a door contact opening, a kitchen motion sensor firing, or temperature and humidity changes. For wearable sensing, it includes signal variations over a few seconds, such as acceleration bursts or transient physiological fluctuations.}

\textbf{Activity level.} 
\rw{This level represents goal-oriented sequences of observations with clear semantic meaning, such as “cooking” or “taking a shower”. It abstracts away from individual sensor events to describe what the user is doing.}

\textbf{Routine level.} 
\rw{This level captures recurring temporal and spatial organizations of activities, such as “a weekday morning routine consisting of waking up, bathroom use, and breakfast preparation around 7:00 AM”. It describes when and where activities typically occur and how they are commonly ordered.}

This three-level representation is inspired by prior hierarchical accounts of behavior (e.g., movement/ event/ action/ history distinctions in 
\cite{nagel1988image,bobick1997movement}), but is adapted here for smart-home activity inference.
~\projectname{} uses these levels jointly so that activity inference is guided not only by local observations, but also by broader behavioral context.

\subsection{Data Input and Feature-Specific Processing}
\label{process}
\projectname{} supports heterogeneous sensing streams but does not require multimodal input. At minimum, it requires one smart-home environmental stream; when additional streams are available, such as wearable-derived activity estimates, the framework treats them as complementary evidence sources.

We organize these inputs into two complementary levels of behavioral evidence: 
\textit{i)} sensor-observation level summaries; and 
\textit{ii)} activity level prediction results. The sensor-observation level preserves short-term contextual evidence. Specifically, the summary is a structured description of three aspects within each local window: the user’s location, the user’s interactions in the environment, and environmental conditions such as temperature and humidity. 
In contrast, the activity level infers candidate activity labels from the available sensing streams to describe what the user is most likely doing at a given time.
In the following, we first explain how summaries are constructed from raw sensor observations and then describe how activity labels are obtained from the available sensing streams.

\subsubsection{Sensor-observation Level Context Extraction}

Following prior definitions of context~\cite{dey2000context}, ~\projectname{} summarizes each local window into three components: 
\textit{i)} user location;
\textit{ii)} user-object interaction; and 
\textit{iii)} environmental condition.
Let $U_k^{sum}$ denote the $k$-th fixed-duration window used for summary construction. 
For each $U_k^{sum}$, the system generates a structured summary defined in Eq.~\ref{eq:summary} (Fig \ref{fig:outline}A):
\begin{equation}
S_k = \{L_k, I_k, E_k\},
\label{eq:summary}
\end{equation}
where $L_k$ denotes the user’s location summary, $I_k$ is the interaction summary, and $E_k$ is the environmental summary.

For the location summary $L_k$, the system examines all activations from location-associated sensors (e.g., a motion sensor) within $U_k^{sum}$ and records the sequence of triggered locations in temporal order. 
Formally, if $M_k$ denotes the ordered set of location-associated sensor events within $U_k^{sum}$, then $L_k$ is obtained by mapping $M_k$ to its corresponding ordered location sequence. 
If the same location appears multiple times within \(U_k^{sum}\), we retain only its last occurrence in the ordered sequence. This simplification is based on the assumption that the selected time window is typically short, so the user is unlikely to exhibit highly complex location transitions within a single window.
If no motion sensor is triggered within \(U_k^{sum}\), the system retains the most recent location from the previous window, i.e., $L_k \leftarrow L_{k-1}$. As a special case, the system infers an ``out of home'' state using a simple rule: if the last triggered event comes from a sensor annotated as an entrance sensor (e.g., the front-door contact sensor or entrance motion sensor) and no other sensor is activated during the following three minutes, then the user location is set to ``out of home''.

For the interaction summary $I_k$, the system converts sensor activations associated with household objects or appliances into interaction records using a predefined sensor-to-object mapping. This mapping is derived from sensor metadata collected during installation, including the object/device and room information. For example, contact sensors could be mapped to an entry such as ``open the cabinet door''. If multiple interaction events occur within \(U_k^{sum}\), they are recorded in temporal order. Thus, $I_k$ represents the ordered set of object- and device-related interactions.

For the environment summary \(E_k\), the system summarizes continuous environmental measurements, including temperature and humidity, within $U_k^{sum}$. Specifically, for each environmental variable $v$, the system computes three statistics: minimum, maximum, and mean, as shown in Eq.~\ref{eq:environment_summary}:
\begin{equation}
   E_k(v) = \bigl(\min_{x \in U_k^{sum}} v(x),\; \max_{x \in U_k^{sum}} v(x),\; \mathrm{mean}_{x \in U_k^{sum}} v(x)\bigr). 
\label{eq:environment_summary}
\end{equation}

Here, $x \in U_k^{sum}$ denotes an environmental sensor reading observed within the current window.
If no new environmental reading is observed within $U_k^{sum}$, the system carries forward the most recent environmental summary from the previous window, i.e., $E_k \leftarrow E_{k-1}$.

\begin{figure}[t]
    \centering
    \includegraphics[width=0.8\linewidth]{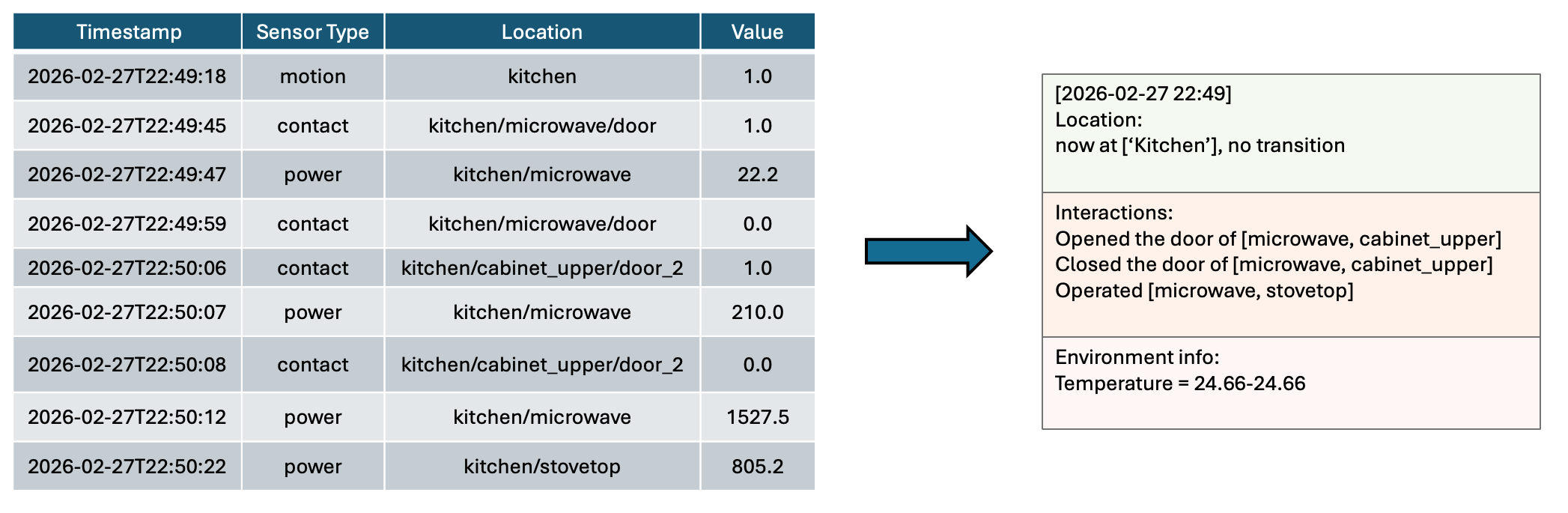}
    \vspace*{-0.5em}
    \caption{Example of the sensor-observation level summary constructed from environmental sensor events within a fixed-duration window. The summary contains three components: location summary, interaction summary, and environmental summary.}
    \label{fig:sigsummary}
        \vspace*{-0.5em}
\end{figure}

The summary $S_k $ is a structured JSON object that captures the contextual evidence available in the current window: where the user appears to be, which objects they interacted with, and what environmental conditions were observed. Figure \ref{fig:sigsummary} shows an example of a sensor-observation level summary.

\subsubsection{Activity Level Prediction}
At the activity level, \projectname{} generates candidate activity labels from the available modalities using modality-specific pipelines. 
These pipelines operate on their native input segmentations rather than on the final aligned timeline. In particular, the environmental data is processed using event-based windows, where each window contains a fixed number of consecutive sensor activations, whereas the wearable data is processed using fixed-duration time windows.

For the environmental data, let $U_j^{env}$ denote the $j$-th event window and let $a_j^{env}$ denote the activity-level prediction. For the wearable data, let $U_k^{wear}$ denote the $k$-th fixed-duration window and let $a_k^{wear}$ denote the corresponding prediction. The modality-specific activity inference can be written as Eq.~\ref{eq:modality_prediction}:
\begin{equation}
   a_j^{env} = f_{env}(U_j^{env}), \qquad
a_k^{wear} = f_{wear}(U_k^{wear}), 
\label{eq:modality_prediction}
\end{equation}

where $f_{env}$ and $f_{wear}$ denote the environmental and wearable recognition pipelines.

For the environmental sensor, the data usually consists of discrete event sequences that are sparse but exhibit strong temporal dependencies. 
To adapt across different household layouts, we used the event-to-text encoding strategy proposed by~\citet{thukral2025layout}, which converts each sensor event into a textual trigger description and then encodes it using a pretrained language model to obtain semantic event embeddings. These embeddings are further processed by a biLSTM-based recognizer to model temporal dependencies across events and produce activity-level predictions. 
This environmental prediction serves as the reference for activities that are strongly associated with object use and spatial context, such as ``cooking'', ``sleeping'', or ``relaxing''.

For the wearable sensor, the input consists of continuous multimodal time-series data collected at high frequency. We use the open-source pipeline developed by the Oxford Wearables Group~\cite{walmsley2022reallocation,doherty2018gwas,willetts2018statistical,doherty2017large} to extract predictions from the accelerometer and heart rate data. These wearable-derived outputs provide sleep-related and sedentary-state predictions. Compared with the environmental stream, the wearable stream provides stronger evidence for activities primarily reflected in body-motion patterns.

\subsection{Time-aligned Data Fusion \& Cross Reference}
\label{align}
In \projectname{}, the environmental recognizer operates on event-based windows $U_j^{env}$, whereas the wearable recognizer and sensor summary construction operate on fixed-duration windows $U_k^{wear}$ and $U_k^{sum}$. To enable joint reasoning across these heterogeneous sources, \projectname{} projects all outputs onto a shared minute-level timeline and then performs cross-reference through late fusion.

\subsubsection{Data Alignment}
Let $T_t$ denote the $t$-th aligned one-minute interval on the shared timeline. 
For each $T_t$, the system constructs the synchronized evidence bundle in Eq.~\ref{eq:aligned_bundle}, consisting of the sensor summary and the activity-level predictions available for that minute:
\begin{equation}
    X_t = \{ S_t, A_t \}, \qquad A_t = \{A_t^{env}, A_t^{wear}\}
\label{eq:aligned_bundle}
\end{equation}

where $S_t$ denotes the sensor summary within $T_t$, and $A_t^{env}$ and $A_t^{wear}$ denote the environmental and wearable activity predictions after temporal alignment. 

\begin{figure}[t]
    \centering
        \vspace*{-0.5em}
    \includegraphics[width=0.9\linewidth]{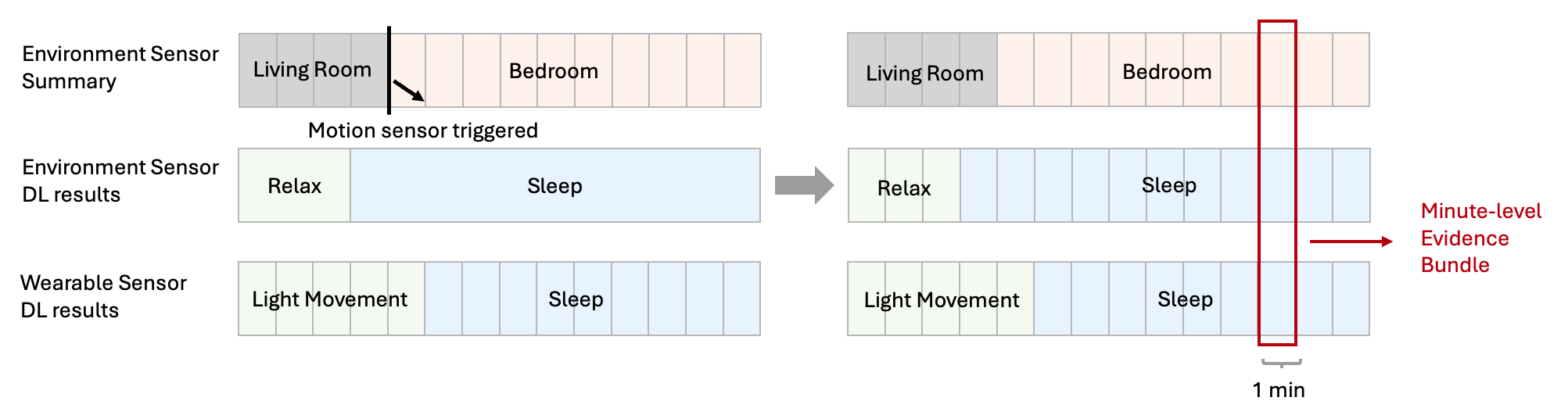}
        \vspace*{-1em}
    \caption{Time-aligned fusion inputs in \projectname{}. From top to bottom, the rows correspond to sensor summaries, environmental activity predictions, and wearable activity predictions after alignment to the shared one-minute timeline.}
    \label{fig:alignment}
\end{figure}

We use one minute as the shared alignment unit because it is consistent with prior work on coarse-grained event segmentation and with standard epoch settings in wearable-based activity and sleep analysis ~\cite{zacks2001event,ancoli2003role}.
The alignment procedure differs by source type (Fig.\ \ref{fig:alignment}). For the environmental recognizer, each raw prediction $a_j^{env}$ is generated on an event window $U_j^{env}$ and then assigned to the one-minute interval by majority overlap with that interval. For the wearable stream and the observation summary, the window is already fixed-duration; therefore, $S_t$ and $A_t^{wear}$ 
are directly associated with their corresponding interval. 
Finally, all sources are represented on the same timeline and can be jointly referenced minute by minute. 
All results are synchronized using a unified sensor timestamp.

\subsubsection{Constructing the User-Specific Contextual Prior}

In addition to the minute-specific evidence in $X_t$, \projectname{} integrates the user-specific contextual prior $C$, which includes user routine descriptions, habitual activity patterns, and other preferences. Unlike $S_t$ and $A_t$, which vary across aligned minutes, $C$ is independent of $t$ and remains fixed during the inference process. In our framework, 
$C$ captures user-level contextual information that is not directly observable from the current sensor window but is useful for constraining activity interpretation at a higher level.

The contextual prior $C$ consists of four components:
\textit{i)} a layout description of the home and its activity-relevant objects; 
\textit{ii)} typical times of common activities;  
\textit{iii)} typical locations of common activities; and
\textit{iv)} user-specific habits such as common transitions, preferences, and exceptions. Together, these four components provide user-specific prior knowledge about which activities are possible in the home, when they usually occur, where they are likely to take place, and how they are typically ordered over time.
In our implementation, these components are represented in semi-structured natural language templates so that they can be injected directly into the LLM prompt. They can be obtained either from user interviews or from historical data, depending on the deployment setting.

\subsubsection{Cross-Reference}
After temporal alignment, \projectname{} performs a cross-reference step to produce a single minute-level activity interpretation from the heterogeneous evidence sources.
We use a 10-minute segment as the basic unit, and the corresponding 10 aligned evidence bundles $\{X_t\}$ are populated into the prompt in JSON format together with their timestamps. Each $X_t$ contains the observation summary and the activity predictions from different sensing streams, while the prompt also provides the LLM with source descriptions, allowed labels, and user-specific contextual priors $C$. 
Compared with fixed fusion rules, an LLM offers a more flexible way to combine heterogeneous information by using explicit instructions and contextual priors.

The prompt guides the model through a structured decision process.
It first uses contextual information to narrow down the candidate labels, then considers evidence of interactions with activity-related objects or devices when available, and finally makes the decision by referring to the predictions from different sources.
To resolve the conflicts among different results, several priorities are explicitly enforced in the prompt. For example, the label ``Sleeping'' is only assigned when the wearable stream provides sleep-related evidence. 
When the available evidence is not sufficient, the model outputs a fallback label ``Other'' rather than forcing an over-specific decision. 
For each minute, the fusion module returns the final label, the strongest alternative label, and a short evidence-based rationale, which are then passed to the later temporal refinement stages. 
Formally, the cross-reference step maps the aligned evidence bundles and contextual prior to the minute-level fused output sequence in Eq.~\ref{eq:cross_reference}:
\begin{equation}
  \hat{Y}_{w} = g\bigl(\{X_t\}_{t \in w},\, C\bigr),  
\label{eq:cross_reference}
\end{equation}
where $g(\cdot)$ is the LLM-based late-fusion function over the current inference window.
Per minute the fusion output is then represented as Eq.~\ref{eq:minute_output}:
\begin{equation}
    \hat{y}_t = \bigl(l_t,\; \tilde{l}_t,\; r_t\bigr), \qquad t \in w,
\label{eq:minute_output}
\end{equation}

where \(l_t\) is the final label, \(\tilde{l}_t\) is the strongest alternative label, and \(r_t\) is a short evidence-based rationale.
The left part in Figure~\ref{fig:1_crossref} shows an example of the prompt structure, and the full prompt is provided in the Appendix.

\begin{figure}[t]
    \centering
    \includegraphics[width=0.8\linewidth]{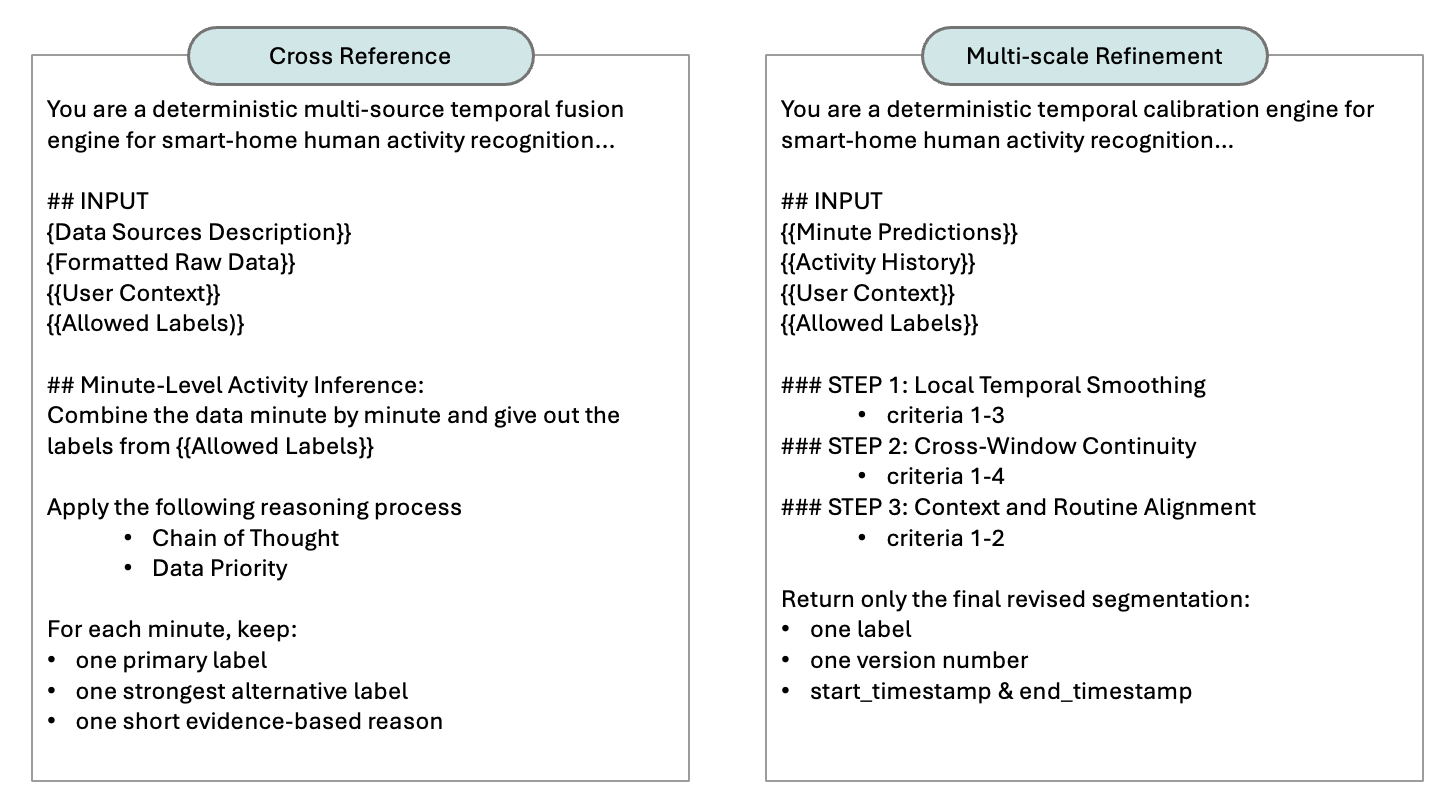}
    \caption{Prompt structures used in \projectname{}. The left panel shows the cross-reference prompt, which takes aligned multi-source evidence, user context, and the allowed label set as input, and produces minute-level fused predictions including a primary label, the strongest alternative label, and a short evidence-based rationale. The right panel shows the multi-scale refinement prompt, which takes the fused minute-level predictions, recent activity history, and user context as input, and applies an ordered three-step reasoning procedure: short-term temporal smoothing, mid-term cross-window verification, and long-term context alignment.}
        \vspace*{-0.5em}
    \label{fig:1_crossref}
\end{figure}

    \vspace*{-0.5em}
\subsection{Multi-Scale Refinement \& Context Alignment}
The cross-reference stage produces minute-level predictions, but these may still contain local fluctuations, fragmented segments, and context-inconsistent labels. To address this, ~\projectname{} applies a multi-scale refinement procedure that incrementally revises the current activity inference result over longer temporal horizons. Rather than treating the current window as a hard segmentation boundary, the refinement module outputs versioned activity segments.

For each refinement step, the module takes the minute-level fused prediction sequence, the refined history from previous steps, and the user-specific contextual prior $C$ as input. 
The cross-reference output for the current window $w$ follows the format in Eq.~\ref{eq:refinement_input}: 
\begin{equation}
    \hat{Y}_w = \{\hat{y}_t\}_{t\in w},
\qquad
\hat{y}_t = (l_t,\tilde{l}_t,r_t),
\label{eq:refinement_input}
\end{equation}

During refinement, the model does not rely only on the primary label $l_t$, but also uses the alternative label $\tilde{l}_t$ and the reason $r_t$ as supplementary information when re-evaluating ambiguous or weakly supported predictions.

The refinement module outputs the updated set of versioned activity segments defined in Eq.~\ref{eq:versioned_segments}:
\begin{equation}
\bar{Y} = \{z_i\}_{i=1}^{N},
\qquad
z_i = (t_i^{\mathrm{start}}, t_i^{\mathrm{end}}, l_i, v_i),
\label{eq:versioned_segments}
\end{equation}
where $t_i^{\mathrm{start}}$ and $t_i^{\mathrm{end}}$ are the start and end times of the segment, $l_i$ is the refined activity label, and $v_i$ is the segment version. Importantly, the segment boundaries are not constrained to lie within the current window. 
Instead, newly generated segments are treated as the latest result of the corresponding time range and are used to overwrite older fragmented results on overlapping intervals.

\subsubsection{Short-term Smoothing}
The first refinement stage corrects local noise within a short temporal horizon. Behavior Setting theory suggests that location strongly constrains possible activities, so the window must be long enough to determine whether the user remains in a relatively stable setting~\cite{hall1969ecological}. Prior work on event segmentation and experience sampling also indicates that human activities and states are often stable over several minutes, while still changing across larger goal-directed episodes~\cite{zacks2001event,kurby2008segmentation,csikszentmihalyi1987validity,myin2018experience}.
Guided by these findings, we use a 10-minute window for smoothing. This window is long enough to reduce minute-level noise and misclassifications, while still short enough to preserve meaningful transitions between successive activity segments and behavior settings.

Given the fused minute-level predictions in the current range, the model performs short-term refinement based on three criteria: 
\textit{i) temporal stabilization}: the refined result should prefer continuous activity segments rather than frequent minute-to-minute changes; 
\textit{ii) noise awareness}: very short isolated fragments are treated as likely noise unless they are clearly supported by surrounding evidence; and
\textit{iii) contextual consistency}: when local evidence is unclear, interpretations that better agree with nearby predictions and the immediate context are preferred.

The short-term refinement step is formalized in Eq.~\ref{eq:short_term_refinement}:
\begin{equation}
\bar{Y}^{\text{short}} = h_{\mathrm{short}}(\hat{Y}_w, C),
\label{eq:short_term_refinement}
\end{equation}
where $\hat{Y}_w$ denotes the current activity interpretation before this update, and $h_{\mathrm{short}}(\cdot)$ denotes the LLM-based short-term refinement operator. At this stage, the model mainly suppresses local fluctuations and merges short, inconsistent fragments into temporally smoother activity segments.

\vspace*{-0.5em}
\subsubsection{Mid-term Verification}

After short-term refinement, \projectname{} performs a second update over a longer temporal horizon to correct fragmented boundaries and implausible short-range transitions. Prior studies suggest that many daily activities persist for tens of minutes, and that aggregating sensor observations over such intervals improves the recognition of long-duration activities~\cite{dumazedier1975use,agre1988dynamic,van2010activity}. Guided by this, we use a 30-minute horizon for this stage. Rather than reprocessing a fully observed 30-minute window, the model revises the current interpretation using recent recognized activity history together with the newly refined results, which helps maintain continuity across updates and corrects fragmentation introduced by earlier local decisions.

Let $H_w$ denote the recognized activity history immediately preceding the current window.
The mid-term verification stage revises the interpretation based on four criteria: 
\textit{i) temporal continuity}: neighboring regions with the same activity label should be merged when the evidence suggests that they belong to one continuous activity;
\textit{ii) transition regularity}: activity transitions should follow plausible daily patterns or the user’s routine, and unlikely transitions require stronger supporting evidence; 
\textit{iii) duration plausibility}: activity labels should not be assigned to highly implausible durations unless strong evidence supports them; and
\textit{iv) boundary correction}: when a single activity is split into multiple segments by brief interruptions, these fragments will be recombined unless there is strong evidence that a real transition occurred. This step is formalized in Eq.~\ref{eq:mid_term_refinement}:
\begin{equation}
\bar{Y}^{\text{mid}} = h_{\mathrm{mid}}(\bar{Y}^{\text{short}}, H_w, C),
\label{eq:mid_term_refinement}
\end{equation}

where $h_{\mathrm{mid}}(\cdot)$ is the LLM-based mid-term verification operator. Compared with the first stage, this step focuses less on local minute-level noise but on whether previously separated segments should be revised into longer coherent units.


\subsubsection{Long-term Context Alignment}
Prior work~\cite{schank2013scripts} suggests that individuals follow structured routines in their daily lives. So the final refinement stage aligns the current interpretation with longer-term user routines and contextual priors. Unlike the previous two stages, which focus primarily on temporal consistency, this stage uses the user context $C$ as a soft behavioral prior to resolve ambiguity and improve context specificity.

More specifically, the model re-evaluates the current activity interpretation based on two criteria: 
\textit{i) context consistency}: the inferred activities should agree with the surrounding context, including time, location, recent activity flow, and the user’s routine patterns. Predictions that do not fit this context are discouraged unless they are strongly supported by the available evidence; and
\textit{ii) context-conditioned specification}: when both the local evidence and the broader context support a more specific interpretation, the model refines a generic activity label into the most suitable specific label within the allowed label set. 
This long-term alignment step is defined in Eq.~\ref{eq:long_term_refinement}:
\begin{equation}
\bar{Y}^{\text{long}} = h_{\mathrm{long}}(\bar{Y}^{\text{mid}}, C),
\label{eq:long_term_refinement}
\end{equation}
where $h_{\mathrm{long}}(\cdot)$ denotes the LLM-based long-term alignment operator. This stage is particularly useful for refining generic labels into more context-specific activities, such as interpreting a cooking-related segment as preparing breakfast when it occurs in the morning and agrees with the user’s routine.

The output of the final stage, $\bar{Y}^{\text{long}}$, is treated as the latest activity interpretation. 
When newly generated segments overlap with previously stored results, the system keeps the newer version and overwrites the older, fragmented interpretation in the overlapping interval. 
In this way, the activity timeline is updated incrementally over time, while preserving the boundaries of non-overlapping segments in the latest output.

In practice, \projectname{} uses a single prompt for temporal refinement. The three conceptual refinement stages are summarized by Eqs.~\ref{eq:short_term_refinement}--\ref{eq:long_term_refinement}, but they are implemented within one unified prompt. The intermediate results ($\bar{Y}^{\text{short}}$ and $\bar{Y}^{\text{mid}}$) are kept only within the internal reasoning process. 
We adopt this unified design because these three stages are closely connected, and decisions made in earlier stages may need to be adjusted when later temporal or contextual information becomes available. 
A single prompt allows the model to revise the interpretation in one continuous reasoning process, rather than passing fixed intermediate outputs from one stage to the next. 
In our experiments, separating the refinement into multiple prompts led to a clear drop in performance, suggesting that stage-wise decomposition weakens information sharing across refinement steps and makes the overall process less effective. 
An example of the prompt structure is shown in the right part of Figure~\ref{fig:1_crossref}, and the full prompt is provided in the Appendix.

To provide a systematic overview of the design rationale behind \projectname{}, we summarize the key decisions made during the development of our framework, along with their justifications and supporting references (Table \ref{tab:design-heuristics}).

\begin{table}[t]
  \caption{Summary of \projectname{} Design Heuristics and Justifications}
      \vspace*{-0.5em}
  \label{tab:design-heuristics}
  \footnotesize
  \begin{tabularx}{\textwidth}{|p{2cm}|X|X|l|}
    \hline
    \textbf{Decision} & \textbf{Implementation} & \textbf{Justification} & \textbf{References} \\ \hline
    \textbf{Sensor-observation Level \newline Extraction} & Rule-based and structured event summarization & Captures information from causally-driven state transitions and provides fine-grained evidence to resolve activity ambiguity & ~\cite{doi:10.34133/research.0467,dey2000context} \\ \hline
    \textbf{LLM-based \newline Late Fusion} & Chain-of-Thought reasoning incorporating source reliability & Dynamically resolves sensor conflicts and enables semantic alignment across heterogeneous data & ~\cite{wei2022chain} \\ \hline
    
    \textbf{Time Window \newline Alignment} & 1-minute unified synchronization window across all data streams & Established standard for activity intensity scoring. Located in goal-directed coarse-grained events & ~\cite{zacks2001event,ancoli2003role} \\ \hline
    \textbf{Short-Term Smoothing} & 10-minute sliding window with LLM-based CoT reasoning & Based on Behavior Setting theory, aligns with human grains of parsing continuous activities into meaningful events & ~\cite{hall1969ecological,zacks2001event,kurby2008segmentation,csikszentmihalyi1987validity,myin2018experience}\\ \hline
    \textbf{Mid-Term \newline Verification} & 30-minute sliding window with LLM-based CoT reasoning  & Based on typical duration of daily activities; balances capturing long-duration patterns while avoiding fragmentation & ~\cite{dumazedier1975use,agre1988dynamic,van2010activity} \\ \hline
    \textbf{Long-Term \newline Context \newline Alignment} & Integrating long-term user routines and habits as behavioral priors & Leverages structured historical patterns to refine general labels into context-specific descriptions & ~\cite{schank2013scripts,suchman1987plans} \\ \hline
  \end{tabularx}
\end{table}

    \vspace*{-0.5em}

\section{Experimental Evaluation}
\label{eval}
To demonstrate the effectiveness of \projectname{}, we perform a two-stage experimental evaluation: 
\textit{i)} A technical evaluation and comparison to state-of-the-art methods based on publicly available, widely used benchmark datasets and methods; and
\textit{ii)} \rw{An additional evaluation through a practical case study in our own smart home environment.} 

In what follows we concentrate on the benchmark-based evaluation, whereas Section \ref{sec:deployment} focuses on our case study.
Specifically, in this section our benchmark experiments focus on three research questions:
\begin{description}
\item[RQ1:] Whether contextual reasoning improves recognition under in-domain and cross-domain transfer?
\item[RQ2:] Whether multi-scale refinement improves temporal coherence and segment-level quality?
\item[RQ3:] Whether \projectname{} remains robust across different or missing modality settings?
\end{description}
Together, these questions assess the overall performance and robustness of \projectname{} across a range of relevant real-world conditions.

    \vspace*{-0.5em}
\subsection{Metrics}

We evaluate recognition performance primarily using a time-based evaluation protocol, which measures the proportion of the total observation time during which the predicted activity label matches the ground-truth label.
\rw{This protocol is also referred to as frame-based or time-slice evaluation in continuous activity recognition ~\cite{activity_tapia_2004, wang2011recognizing, ward2011performance}. This protocol has previously been used in smart-home HAR. }
Specifically, we convert both predictions and ground truth into consecutive fine-grained temporal intervals and compare their labels interval by interval. The final score is computed as the percentage of correctly labeled intervals over the entire observation period. This metric is particularly appropriate for continuous smart-home HAR because the task aims to recover the activity timeline over time, rather than only classify isolated windows or pre-segmented episodes.

\begin{figure}[t]
    \centering
    \vspace*{-0.5em}
    \includegraphics[width=0.5\linewidth]{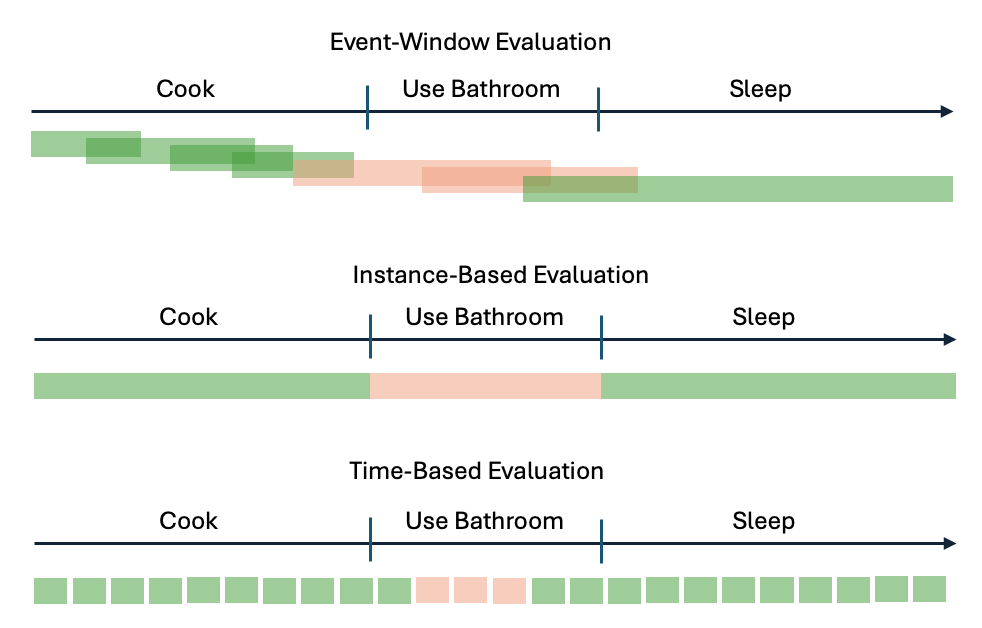}
    \caption{Illustration of three evaluation protocols in smart-home HAR. In event-window-based evaluation, activities with dense sensor interactions, such as cooking, may contribute more overlapping evaluation windows, while long but sparsely observed activities, such as sleep, may contribute fewer windows. Instance-based evaluation treats each activity segment as one unit regardless of duration. In contrast, time-based evaluation measures prediction correctness directly along the temporal axis, making errors contribute proportionally to the amount of time affected.}
    \vspace*{-1em}
    \label{fig:metric_comparison}
\end{figure}

\rw{Window-based and instance-based protocols remain useful for evaluating classification performance, but they do not fully capture the temporal consequences of errors in continuous monitoring (Fig.\ \ref{fig:metric_comparison}). }
Environmental sensors in smart homes often produce event-driven readings, such as motion detections or contact state changes, rather than uniformly sampled observations. This means that sensor-event density is not temporally uniform across activities. 
Activities like ``cooking'' may trigger many events, while behaviorally stable activities such as ``sleeping'' may generate few events despite lasting for hours.

A metric that counts ``event windows'' or ``instances'' would over-represent active, short-duration tasks and under-represent behaviorally stable but long-duration activities. 
And some instance-based metrics assume clear boundaries that rarely exist in the real world, where transitions between activities are often ambiguous or sensor-dependent. 
Our goal is not only to measure whether each activity is classified correctly, but also to evaluate whether a model preserves the temporal coverage and duration of activities along the continuous timeline. 
This reflects the requirement of real-world deployments, where an activity recognition system must assign an activity label to each timestep rather than only classify isolated windows or instances.

To complement this time-based view, we additionally report Ward’s segment-level metrics~\cite{ward2011performance} and Earth Mover’s Distance (EMD)~\cite{rubner2000earth}. Fragmentation and merge measure whether activity segments are incorrectly split or combined, while overfill and underfill quantify boundary mismatches. EMD compares the predicted and ground-truth segment-length distributions, with lower values indicating more similar temporal structure. Together, these metrics provide a complementary assessment of temporal coherence and segmentation quality beyond label correctness over time.

\subsection{Datasets}
\rw{For benchmark evaluation, we use several publicly available smart-home activity datasets: three CASAS~\cite{cook2012casas} datasets, Orange4Home~\cite{cumin2017dataset}, UCI-ADL~\cite{ordonez2013activity,ordonez2013uciadl}, and MARBLE~\cite{arrotta2021marble}. 
These datasets are widely used single-resident benchmarks in smart-home HAR and collectively cover diverse sensor modalities, home layouts, sensor configurations, and activity categories.}
In CASAS, we use the Aruba, Milan and Kyoto7 homes as cross-home benchmark targets.
Since cross-domain is designed to operate without assuming access to target-home training data, we map the original annotations from all datasets into a shared label space and retain only the overlapping activity categories, following prior work ~\cite{thukral2025layout, liciotti2020sequential}.
Compared with other ambient datasets, MARBLE includes both environmental and wearable/mobile sensing, allowing us to evaluate the framework under richer multimodal conditions. Since the recordings are short and do not provide sufficient longitudinal behavioral patterns, we do not extract user routine priors for MARBLE. Table~\ref{tab:dataset_info_sources} summarizes the information sources and contextual inputs used for each dataset. Additional details on the datasets are provided in Appendix~\ref{app:benchmarks}.

\rw{Multi-resident datasets such as ARAS~\cite{alemdar2013aras} and
MuRAL~\cite{chen2026mural} involve simultaneous sensor activations from
different residents, which introduces a distinct data-association problem.
We therefore focus on single-resident settings in the main evaluation and
report a preliminary multi-resident test in Appendix~\ref{app:aras_multistudy}.}

\begin{table*}[t]
\centering
\caption{Available information sources for each dataset.}
\label{tab:dataset_info_sources}
\vspace*{-0.5em}
\renewcommand{\arraystretch}{1.15}
\setlength{\tabcolsep}{8pt}
\footnotesize
\resizebox{\textwidth}{!}{
\begin{tabular}{lcccc}
\toprule
\textbf{Information type} & \textbf{CASAS (Aruba \& Milan \& Kyoto7)} & \rw{\textbf{Orange4Home}} & \rw{\textbf{UCI-ADL (Home A \& B)}} & \textbf{MARBLE} \\
\midrule
Env. prediction ($A_t^{env}$) & \checkmark & \rw{\checkmark} & \rw{\checkmark} & \checkmark \\
Wear. prediction ($A_t^{wear}$) & -- & \rw{--} & \rw{--} & \checkmark \\
Summary-location ($L_t \in S_t$) & \checkmark & \rw{\checkmark} & \rw{\checkmark} & \checkmark \\
Summary-interaction ($I_t \in S_t$) & \checkmark & \rw{\checkmark} & \rw{\checkmark} & \checkmark \\
Summary-env ($E_t \in S_t$) & \checkmark & \rw{\checkmark} & \rw{\checkmark} & -- \\
Context ($C$) & \checkmark & \rw{\checkmark} & \rw{\checkmark} & -- \\
\bottomrule
\end{tabular}
}
\vspace*{-0.5em}
\end{table*}

\subsection{Dataset Split \& Experimental Setup }
For the CASAS datasets, we use a time-based blocked evaluation protocol~\cite{hammerla2015let, plotz2023know}. 
Instead of randomly splitting individual windows, we hold out continuous time blocks for testing, since adjacent sensor windows are highly correlated and random window-level splits can leak short-term temporal patterns across train and test sets~\cite{plotz2021applying}. 
Specifically, we select three non-overlapping continuous 20-day segments from different parts of the timeline as test folds. This duration provides a sufficiently long test period to cover both frequent and long-duration activities, while preserving enough remaining data for training and validation. 
In each fold, the 20-day segment is held out as test, while the remaining data are used for training and validation. 
This design preserves temporal continuity in the test data and evaluates generalization to unseen periods of daily life rather than to isolated windows from the same local context.  
To construct user priors, we use only the training split and derive statistics-based routine descriptions from it, preventing leakage from the test. 

\rw{For the Orange4Home dataset, we also construct three temporally separated evaluation folds. In each
fold, one continuous period is held out for testing, while the remaining data
are used for training and validation. User routine priors are
derived exclusively from the training portion of each fold.}

\rw{The UCI-ADL dataset contains recordings from two independent
single-resident smart homes, denoted Home A and Home B. 
For each home, we construct three temporally separated folds for evaluation.
 User routine priors are constructed independently for each home using
only its corresponding training split. Results are reported separately for
Home A and Home B.}

\rw{For CASAS, we follow the sliding-window protocol from prior work~\cite{fiori2026improving}. Specifically, we apply a window of 30 events with a step size of 10 events over the environmental sensor stream. 
Because sensor-event densities differ across datasets, we adjust the number
of events per window while maintaining a broadly comparable temporal coverage: 40 events with a step of 20 for
Orange4Home, and 20 events with a step of 10 for UCI-ADL.}

We compare \projectname{} against three state-of-the-art baselines: 
\begin{itemize}
    \item[(i)] E-FCN~\cite{bouchabou2021fully}, a fully convolutional network designed for sliding-window-based smart-home activity recognition;
    \item[(ii)] DeepCASAS-BiLSTM~\cite{liciotti2020sequential}, a bidirectional LSTM that captures dependencies in sequential sensor events;
    
    \item[(iii)] TDOST~\cite{thukral2025layout}, a stronger semantic transfer that incorporates contextualized event representations. 
\end{itemize}

For the MARBLE dataset, we follow a cross-session evaluation protocol, where data from one script is used for testing, and data from the remaining scripts are used for training and validation. We follow the benchmark setting used by prior multimodal methods and adopt 6-second windows with 80\% overlap. 
Because MARBLE is a scripted dataset with much shorter activity segments than real-world daily routines, we use 6-second intervals as the alignment and evaluation unit. Accordingly, we reduce the short-term refinement window to 1 minute, while the mid-term stage again uses the two most recent inference results as historical input. In addition, because environmental events in MARBLE are too sparse, TDOST-style event-window prediction is less suitable. We therefore use DeepConvLSTM as the environmental recognizer and compare \projectname{} against two baselines:
\begin{itemize}
    \item[(i)] DualLSTM~\cite{arrotta2021marble}, a two-stream model that jointly uses environmental and wearable signals;
    \item[(ii)] DeepConvLSTM~\cite{deepConvLSTM}, which is trained only on the environmental sensor stream;
\end{itemize}

\rw{Detailed model versions, inference parameters, and determinism assumptions are provided in Appendix~\ref{app:llm-configurations}.}

\subsection{Results}

\subsubsection{Robustness in In-Domain and Cross-Domain Scenarios}
\label{sec:results_benchmark_robustness}

\begin{table*}[htbp]
\small
\centering
\caption{\rw{In-domain and cross-domain performance comparison across the evaluated smart-home activity-recognition datasets. In-domain evaluation includes CASAS Aruba, Milan, and Kyoto7, Orange4Home, and UCI-ADL Homes A and B, while cross-domain evaluation considers transfer among CASAS Aruba, Milan, and Kyoto7. Results are reported as mean $\pm$ standard deviation (\%) over multiple runs. Higher values are better for accuracy and F1 scores, whereas lower values are better for fragmentation rate (FR), merge rate (MR), overfill (OF), underfill (UF), and Earth Mover's Distance (EMD). Bold values indicate the best result within each setting.}}
\label{tab:cross_domain_aruba_milan_kyoto}
\resizebox{\textwidth}{!}{
\begin{tabular}{l l c c c c c c c c}
\toprule
\textbf{Setting} & \textbf{Model} & \textbf{Accuracy} & \textbf{Weighted F1} & \textbf{Macro F1} & \textbf{FR $\downarrow$} & \textbf{MR $\downarrow$} & \textbf{OF $\downarrow$} & \textbf{UF $\downarrow$} & \textbf{EMD $\downarrow$} \\
\midrule
\multicolumn{10}{c}{\textbf{\rw{In-domain evaluation}}} \\
\midrule
\multirow{4}{*}{CASAS Aruba}
& E-FCN            & 84.64 $\pm$ 1.43 & 84.51 $\pm$ 1.57 & 54.33 $\pm$ 2.31 & 3.28 $\pm$ 0.28 & 0.11 $\pm$ 0.09 & 0.56 $\pm$ 0.18 & 9.20 $\pm$ 1.65 & 1.40 $\pm$ 0.56 \\
& DeepCASAS        & \textbf{88.62 $\pm$ 0.53} & \textbf{88.55 $\pm$ 0.52} & \textbf{59.42 $\pm$ 0.76} & 2.44 $\pm$ 1.40 & \textbf{0.06 $\pm$ 0.04} & \textbf{0.36 $\pm$ 0.04} & \textbf{7.47 $\pm$ 3.53} & 1.22 $\pm$ 0.36 \\
& TDOST            & 86.89 $\pm$ 1.19 & 86.26 $\pm$ 1.66 & 51.74 $\pm$ 1.29 & 2.05 $\pm$ 0.29 & 0.11 $\pm$ 0.08 & 0.93 $\pm$ 0.26 & 10.44 $\pm$ 3.62 & 1.41 $\pm$ 0.37 \\
& \projectname{}   & 85.93 $\pm$ 1.73 & 85.88 $\pm$ 1.66 & 52.02 $\pm$ 3.51 & \textbf{1.97 $\pm$ 1.29} & 0.29 $\pm$ 0.16 & 1.04 $\pm$ 0.10 & 11.07 $\pm$ 4.01 & \textbf{0.89 $\pm$ 0.42} \\

\midrule
\multirow{4}{*}{CASAS Milan}
& E-FCN            & 16.11 $\pm$ 6.41 & 15.81 $\pm$ 4.11 & 8.29 $\pm$ 1.29 & 4.90 $\pm$ 1.81 & 1.23 $\pm$ 0.94 & 3.24 $\pm$ 0.97 & 11.65 $\pm$ 3.92 & 21.83 $\pm$ 1.17 \\
& DeepCASAS        & 15.69 $\pm$ 6.92 & 15.75 $\pm$ 5.02 & 7.69 $\pm$ 2.16 & \textbf{3.80 $\pm$ 0.47} & 0.97 $\pm$ 0.99 & 3.32 $\pm$ 1.99 & \textbf{10.07 $\pm$ 1.03} & 21.68 $\pm$ 1.13 \\
& TDOST            & 14.70 $\pm$ 6.00 & 14.83 $\pm$ 4.26 & 7.30 $\pm$ 2.03 & 4.05 $\pm$ 0.86 & 1.62 $\pm$ 0.49 & 2.14 $\pm$ 0.77 & 13.97 $\pm$ 8.35 & 19.96 $\pm$ 0.23 \\
& \projectname{}   & \textbf{53.97 $\pm$ 2.61} & \textbf{55.09 $\pm$ 2.61} & \textbf{29.12 $\pm$ 1.53} & 5.32 $\pm$ 1.83 & \textbf{0.46 $\pm$ 0.18} & \textbf{1.99 $\pm$ 0.21} & 12.81 $\pm$ 3.63 & \textbf{6.73 $\pm$ 0.64} \\

\midrule
\multirow{4}{*}{CASAS Kyoto7}
& E-FCN & 31.24 $\pm$ 9.71 & 28.36 $\pm$ 10.35 & 15.57 $\pm$ 2.72 & 3.43 $\pm$ 0.49 & 0.14 $\pm$ 0.20 & 4.03 $\pm$ 0.60 & \textbf{8.47 $\pm$ 0.73} & 21.83 $\pm$ 1.55 \\
& DeepCASAS & 19.32 $\pm$ 4.87 & 17.00 $\pm$ 8.75 & 13.30 $\pm$ 2.68 & 3.65 $\pm$ 1.75 & 0.13 $\pm$ 0.10 & 5.05 $\pm$ 0.79 & 15.43 $\pm$ 3.87 & 7.98 $\pm$ 2.93 \\
& TDOST & 19.61 $\pm$ 10.05 & 16.03 $\pm$ 7.76 & 15.10 $\pm$ 7.16 & 3.81 $\pm$ 0.24 & \textbf{0.10 $\pm$ 0.07} & 3.81 $\pm$ 1.15 & 18.96 $\pm$ 0.78 & \textbf{7.45 $\pm$ 2.74} \\
& \projectname{} & \textbf{42.77 $\pm$ 1.82} & \textbf{51.35 $\pm$ 1.17} & \textbf{35.77 $\pm$ 4.08} & \textbf{0.52 $\pm$ 0.40} & 4.35 $\pm$ 0.24 & \textbf{2.53 $\pm$ 0.08} & 11.06 $\pm$ 0.55 & 29.88 $\pm$ 1.73 \\
\midrule

\multirow{4}{*}{\rw{Orange4Home}} & \rw{E-FCN} & \rw{$94.31 \pm 4.41$} & \rw{$95.00 \pm 3.14$} & \rw{$59.92 \pm 4.82$} & \rw{$6.32 \pm 1.03$} & \rw{$\mathbf{0.00 \pm 0.00}$} & \rw{$\mathbf{0.00 \pm 0.00}$} & \rw{$4.28 \pm 1.39$} & \rw{$18.64 \pm 1.06$} \\
& \rw{DeepCASAS} & \rw{$94.55 \pm 4.14$} & \rw{$95.82 \pm 2.21$} & \rw{$65.46 \pm 0.98$} & \rw{$7.00 \pm 0.71$} & \rw{$\mathbf{0.00 \pm 0.00}$} & \rw{$0.01 \pm 0.00$} & \rw{$3.64 \pm 0.56$} & \rw{$17.35 \pm 0.39$} \\
& \rw{TDOST} & \rw{$76.59 \pm 9.50$} & \rw{$84.26 \pm 6.79$} & \rw{$63.68 \pm 3.02$} & \rw{$4.39 \pm 0.35$} & \rw{$\mathbf{0.00 \pm 0.00}$} & \rw{$0.01 \pm 0.00$} & \rw{$\mathbf{3.35 \pm 1.52}$} & \rw{$14.05 \pm 1.22$} \\
& \rw{\projectname{}} & \rw{$\mathbf{95.06 \pm 4.07}$} & \rw{$\mathbf{95.93 \pm 2.63}$} & \rw{$\mathbf{70.19 \pm 2.12}$} & \rw{$\mathbf{1.48 \pm 0.11}$} & \rw{$\mathbf{0.00 \pm 0.00}$} & \rw{$0.04 \pm 0.01$} & \rw{$10.63 \pm 3.42$} & \rw{$\mathbf{9.80 \pm 0.43}$} \\

\midrule

\multirow{4}{*}{\rw{UCI-ADL Home A}} & \rw{E-FCN} & \rw{$97.06 \pm 3.96$} & \rw{$98.09 \pm 2.46$} & \rw{$82.43 \pm 6.66$} & \rw{$0.00 \pm 0.00$} & \rw{$0.01 \pm 0.00$} & \rw{$0.21 \pm 0.28$} & \rw{$0.68 \pm 0.96$} & \rw{$\mathbf{5.11 \pm 1.58}$} \\
& \rw{DeepCASAS} & \rw{$94.60 \pm 6.20$} & \rw{$96.27 \pm 3.95$} & \rw{$68.30 \pm 13.27$} & \rw{$0.08 \pm 0.10$} & \rw{$\mathbf{0.00 \pm 0.00}$} & \rw{$0.11 \pm 0.15$} & \rw{$0.20 \pm 0.12$} & \rw{$5.67 \pm 3.02$} \\
& \rw{TDOST} & \rw{$84.04 \pm 22.13$} & \rw{$78.86 \pm 29.45$} & \rw{$69.42 \pm 23.25$} & \rw{$\mathbf{0.00 \pm 0.00}$} & \rw{$0.00 \pm 0.00$} & \rw{$1.99 \pm 2.80$} & \rw{$\mathbf{0.11 \pm 0.08}$} & \rw{$5.76 \pm 3.08$} \\
& \rw{\projectname{}} & \rw{$\mathbf{98.13 \pm 2.42}$} & \rw{$\mathbf{98.76 \pm 1.56}$} & \rw{$\mathbf{87.39 \pm 8.72}$} & \rw{$0.65 \pm 0.56$} & \rw{$0.00 \pm 0.00$} & \rw{$\mathbf{0.00 \pm 0.00}$} & \rw{$0.87 \pm 0.86$} & \rw{$11.93 \pm 4.96$} \\

\midrule

\multirow{4}{*}{\rw{UCI-ADL Home B}} & \rw{E-FCN} & \rw{$96.18 \pm 1.79$} & \rw{$96.19 \pm 1.40$} & \rw{$72.49 \pm 2.67$} & \rw{$\mathbf{0.00 \pm 0.00}$} & \rw{$0.08 \pm 0.08$} & \rw{$0.15 \pm 0.10$} & \rw{$1.13 \pm 0.22$} & \rw{$4.41 \pm 1.51$} \\
& \rw{DeepCASAS} & \rw{$95.16 \pm 1.28$} & \rw{$95.38 \pm 1.12$} & \rw{$\mathbf{76.20 \pm 1.55}$} & \rw{$\mathbf{0.00 \pm 0.00}$} & \rw{$\mathbf{0.02 \pm 0.02}$} & \rw{$0.05 \pm 0.02$} & \rw{$0.95 \pm 0.77$} & \rw{$3.94 \pm 0.54$} \\
& \rw{TDOST} & \rw{$95.42 \pm 1.28$} & \rw{$95.38 \pm 1.32$} & \rw{$62.94 \pm 5.74$} & \rw{$\mathbf{0.00 \pm 0.00}$} & \rw{$0.04 \pm 0.03$} & \rw{$0.13 \pm 0.04$} & \rw{$0.49 \pm 0.33$} & \rw{$5.67 \pm 0.91$} \\
& \rw{\projectname{}} & \rw{$\mathbf{96.33 \pm 1.68}$} & \rw{$\mathbf{96.30 \pm 1.28}$} & \rw{$72.93 \pm 2.89$} & \rw{$0.54 \pm 0.40$} & \rw{$0.04 \pm 0.05$} & \rw{$\mathbf{0.03 \pm 0.04}$} & \rw{$\mathbf{0.31 \pm 0.27}$} & \rw{$\mathbf{2.11 \pm 0.87}$} \\

\midrule
\multicolumn{10}{c}{\textbf{\rw{Cross-domain evaluation}}} \\
\midrule

\multirow{4}{*}{\shortstack{Milan\\$\downarrow$\\Aruba}}
& E-FCN            & 30.38 $\pm$ 1.17 & 26.21 $\pm$ 5.43 & 10.47 $\pm$ 1.59 & \textbf{0.99 $\pm$ 0.29} & 5.17 $\pm$ 0.88 & 4.63 $\pm$ 1.04 & 3.46 $\pm$ 2.03 & 20.26 $\pm$ 0.29 \\
& DeepCASAS        & 24.93 $\pm$ 1.99 & 21.54 $\pm$ 4.12 & 8.21 $\pm$ 1.69 & 2.37 $\pm$ 0.14 & 5.97 $\pm$ 1.57 & 4.45 $\pm$ 1.83 & \textbf{0.96 $\pm$ 0.19} & 17.84 $\pm$ 0.33 \\
& TDOST            & 28.15 $\pm$ 4.25 & 18.46 $\pm$ 3.04 & 14.64 $\pm$ 1.92 & 3.17 $\pm$ 0.58 & 2.51 $\pm$ 1.04 & 6.63 $\pm$ 0.82 & 7.37 $\pm$ 2.84 & 17.90 $\pm$ 0.84 \\
& \projectname{}   & \textbf{55.77 $\pm$ 4.00} & \textbf{58.86 $\pm$ 3.51} & \textbf{47.26 $\pm$ 3.02} & 3.39 $\pm$ 1.00 & \textbf{0.53 $\pm$ 0.16} & \textbf{3.12 $\pm$ 0.51} & 12.36 $\pm$ 1.75 & \textbf{2.32 $\pm$ 0.22} \\

\midrule
\multirow{4}{*}{\shortstack{Aruba\\$\downarrow$\\Milan}}
& E-FCN            & 44.05 $\pm$ 13.92 & 34.28 $\pm$ 13.38 & 10.67 $\pm$ 1.93 & 3.68 $\pm$ 1.10 & 1.12 $\pm$ 0.84 & 4.05 $\pm$ 1.40 & 7.27 $\pm$ 1.24 & 22.65 $\pm$ 0.85 \\
& DeepCASAS        & 44.49 $\pm$ 13.97 & 33.71 $\pm$ 14.23 & 10.52 $\pm$ 2.43 & 3.02 $\pm$ 1.54 & 2.02 $\pm$ 0.58 & 4.46 $\pm$ 1.37 & 8.91 $\pm$ 2.18 & 27.60 $\pm$ 0.37 \\
& TDOST            & 44.24 $\pm$ 13.00 & 33.25 $\pm$ 13.48 & 10.06 $\pm$ 1.43 & \textbf{2.38 $\pm$ 0.88} & 3.56 $\pm$ 0.25 & 3.82 $\pm$ 1.08 & \textbf{6.09 $\pm$ 0.67} & 28.24 $\pm$ 0.63 \\
& \projectname{}   & \textbf{55.67 $\pm$ 2.15} & \textbf{57.60 $\pm$ 1.77} & \textbf{31.81 $\pm$ 1.90} & 3.10 $\pm$ 0.95 & \textbf{0.65 $\pm$ 0.17} & \textbf{2.96 $\pm$ 0.55} & 11.11 $\pm$ 1.63 & \textbf{7.04 $\pm$ 0.32} \\

\midrule
\multirow{4}{*}{\shortstack{Aruba\\$\downarrow$\\Kyoto7}}
& E-FCN & 28.56 $\pm$ 12.67 & 23.48 $\pm$ 8.23 & 13.30 $\pm$ 6.32 & 3.13 $\pm$ 0.87 & 1.03 $\pm$ 0.37 & 6.54 $\pm$ 1.94 & 11.65 $\pm$ 6.06 & 33.33 $\pm$ 4.99 \\
& DeepCASAS & 14.12 $\pm$ 3.18 & 14.36 $\pm$ 3.72 & 10.67 $\pm$ 2.56 & 8.31 $\pm$ 1.08 & \textbf{0.28 $\pm$ 0.20} & 4.77 $\pm$ 0.36 & 13.90 $\pm$ 5.69 & \textbf{20.62 $\pm$ 1.30} \\
& TDOST & \textbf{37.49 $\pm$ 10.43} & 21.36 $\pm$ 9.55 & 8.99 $\pm$ 1.89 & \textbf{0.10 $\pm$ 0.14} & 12.94 $\pm$ 1.54 & 3.70 $\pm$ 1.50 & \textbf{0.00 $\pm$ 0.00} & 44.66 $\pm$ 2.58 \\
& \projectname{} & 36.03 $\pm$ 1.41 & \textbf{45.99 $\pm$ 1.93} & \textbf{13.80 $\pm$ 1.11} & 0.28 $\pm$ 0.12 & 5.80 $\pm$ 1.95 & \textbf{3.30 $\pm$ 0.56} & 8.56 $\pm$ 2.15 & 25.87 $\pm$ 1.37 \\
\midrule
\multirow{4}{*}{\shortstack{Milan\\$\downarrow$\\Kyoto7}}
& E-FCN & 28.86 $\pm$ 13.12 & 21.20 $\pm$ 10.96 & 11.08 $\pm$ 3.37 & 6.04 $\pm$ 0.85 & 0.63 $\pm$ 0.51 & 5.59 $\pm$ 1.77 & 6.94 $\pm$ 1.24 & 43.68 $\pm$ 0.81 \\
& DeepCASAS & 37.23 $\pm$ 9.57 & 27.00 $\pm$ 10.40 & 13.89 $\pm$ 3.26 & 5.26 $\pm$ 1.23 & \textbf{0.31 $\pm$ 0.29} & 5.49 $\pm$ 1.70 & 12.73 $\pm$ 3.17 & 38.83 $\pm$ 3.83 \\
& TDOST & 16.50 $\pm$ 10.11 & 12.85 $\pm$ 8.68 & 7.19 $\pm$ 4.22 & 0.56 $\pm$ 0.24 & 0.74 $\pm$ 1.05 & 5.72 $\pm$ 3.10 & 4.51 $\pm$ 2.94 & 36.94 $\pm$ 3.47 \\
& \projectname{} & \textbf{38.65 $\pm$ 0.66} & \textbf{47.70 $\pm$ 1.30} & \textbf{18.74 $\pm$ 0.50} & \textbf{0.21 $\pm$ 0.11} & 3.45 $\pm$ 0.28 & \textbf{2.71 $\pm$ 0.97} & \textbf{2.81 $\pm$ 2.44} & \textbf{12.21 $\pm$ 0.41} \\

\midrule
\multirow{4}{*}{\shortstack{Kyoto7\\$\downarrow$\\Aruba}}
& E-FCN & 25.62 $\pm$ 5.10 & 25.82 $\pm$ 4.19 & 14.21 $\pm$ 2.44 & 7.02 $\pm$ 1.51 & 0.08 $\pm$ 0.06 & 1.57 $\pm$ 0.38 & 16.51 $\pm$ 1.39 & 17.35 $\pm$ 1.41 \\
& DeepCASAS & 12.80 $\pm$ 1.34 & 14.45 $\pm$ 0.40 & 9.31 $\pm$ 0.57 & 7.73 $\pm$ 0.98 & \textbf{0.02 $\pm$ 0.02} & \textbf{0.72 $\pm$ 0.22} & 20.77 $\pm$ 4.48 & 28.83 $\pm$ 2.06 \\
& TDOST & 7.12 $\pm$ 0.45 & 8.71 $\pm$ 1.32 & 5.80 $\pm$ 0.57 & 6.22 $\pm$ 1.09 & 0.04 $\pm$ 0.04 & 1.23 $\pm$ 0.19 & 17.19 $\pm$ 5.95 & 28.42 $\pm$ 1.14 \\
& \projectname{} & \textbf{49.76 $\pm$ 0.04} & \textbf{56.63 $\pm$ 0.44} & \textbf{24.49 $\pm$ 3.44} & \textbf{1.46 $\pm$ 0.02} & 4.54 $\pm$ 0.36 & 2.48 $\pm$ 0.68 & \textbf{7.49 $\pm$ 3.68} & \textbf{15.63 $\pm$ 1.01} \\

\midrule
\multirow{4}{*}{\shortstack{Kyoto7\\$\downarrow$\\Milan}}
& E-FCN & 39.90 $\pm$ 18.44 & 41.72 $\pm$ 18.50 & 17.83 $\pm$ 5.13 & 8.21 $\pm$ 1.03 & 0.14 $\pm$ 0.16 & 2.69 $\pm$ 0.54 & 18.70 $\pm$ 3.11 & \textbf{13.45 $\pm$ 1.07} \\
& DeepCASAS & 7.88 $\pm$ 1.91 & 7.25 $\pm$ 1.85 & 8.72 $\pm$ 0.76 & 10.60 $\pm$ 1.64 & \textbf{0.01 $\pm$ 0.01} & \textbf{1.47 $\pm$ 0.32} & 21.18 $\pm$ 2.45 & 22.56 $\pm$ 1.99 \\
& TDOST & 6.45 $\pm$ 2.66 & 6.01 $\pm$ 2.67 & 5.19 $\pm$ 2.00 & 8.95 $\pm$ 2.95 & 0.16 $\pm$ 0.13 & 2.66 $\pm$ 0.29 & 18.62 $\pm$ 2.11 & 21.15 $\pm$ 0.58 \\
& \projectname{} & \textbf{44.42 $\pm$ 2.98} & \textbf{52.80 $\pm$ 2.55} & \textbf{22.31 $\pm$ 2.08} & \textbf{1.82 $\pm$ 0.55} & 3.39 $\pm$ 0.42 & 3.55 $\pm$ 0.67 & \textbf{9.63 $\pm$ 3.11} & 19.23 $\pm$ 1.02 \\
\bottomrule
\end{tabular}
}
\end{table*}
\rw{Table~\ref{tab:cross_domain_aruba_milan_kyoto} reports the in-domain results across all evaluated datasets and the cross-domain results across Aruba, Milan, and Kyoto7. }
\rw{The results show clear differences across datasets of varying difficulty.
 In the Aruba, Orange4Home, and UCI-ADL in-domain settings, most methods achieve relatively strong performance, and \projectname{} performs comparably to the best-performing baselines while often providing better macro F1 or temporal coherence.}
However, in the more difficult Milan and Kyoto7 in-domain settings, \projectname{} substantially outperforms all baselines. This suggests that \projectname{} is 
especially effective in settings where activity patterns are more ambiguous, temporally extended, or difficult to infer from local sensor evidence alone -- which resembles typical, real-world application scenarios.

This trend is even clearer under cross-domain transfer. Across the six transfer settings, baseline methods generally show large performance drops, indicating substantial domain shift across homes with different layouts, sensor configurations, activity distributions, and user routines. In contrast, \projectname{} achieves the best results for all cross-domain settings. These results suggest that incorporating contextual priors and recent temporal history helps \projectname{} make more robust predictions when the target home differs from the training environment.

A closer look at the results shows that long-duration activities with weak sensor signals, especially sleep-related behaviors, are one of the main challenges in Milan. In time-based evaluation, mistakes on these activities matter more because they take up a large part of the full timeline. In such cases, methods that are solely based on local sensor patterns often fail to maintain stable predictions over time. In contrast, \projectname{} can better recognize these long-duration activities by using contextual information and temporal continuity. 
We present the main results in the confusion matrix shown in Fig.\ \ref{fig:confusion} (complete results can be found in Appendix \ref{app:detailed_results}). When trained on Milan and tested on Aruba, \projectname{} clearly improves the recognition of important long activities such as Sleep and Relax, and also better identifies meaningful categories like Work. The tradeoff is that \projectname{} is less likely to predict the general label Other, \rw{since contextual evidence often supports one of the more semantically informative activity categories in the benchmark label set.} 
Overall this leads to better semantic recognition and stronger cross-domain robustness.
\begin{figure}[t]
    \centering
    \includegraphics[width=0.7\linewidth]{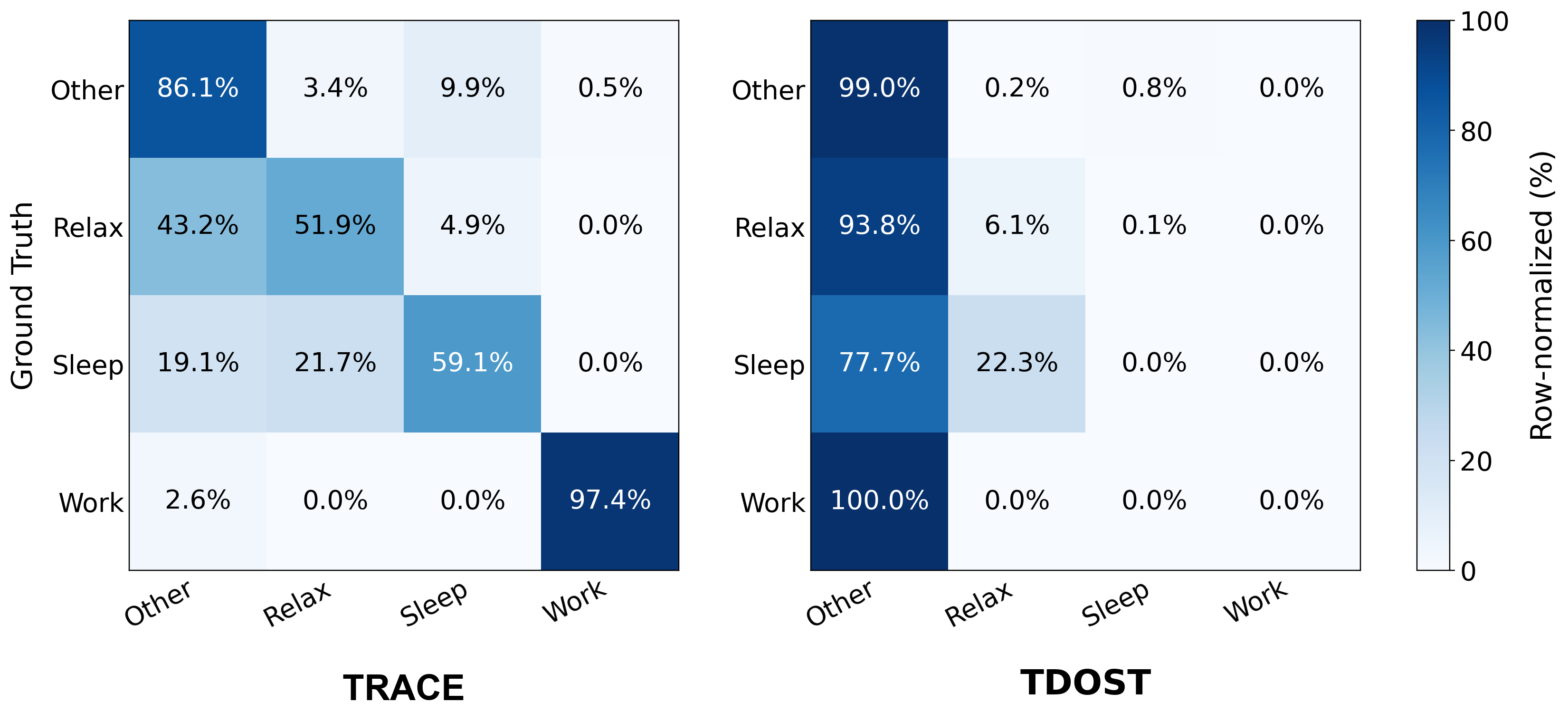}
    \caption{Row-normalized confusion matrices for selected classes under the Milan → Aruba cross-domain setting. The left panel corresponds to \projectname{} and the right panel to the baseline TDOST. The figure highlights differences in class-wise prediction behavior for selected activity categories. Full confusion matrices are provided in the Appendix~\ref{app:confusion_matrix}.}
    \label{fig:confusion}
\end{figure}

\subsubsection{Enhancing Temporal Continuity and Stability}
\label{sec:results_benchmark_temporal}
Beyond time-based classification performance, we further evaluate whether each method produces temporally coherent activity timelines. To this end, we analyze Ward's segment metrics together with the distribution of predicted segment lengths.

\rw{As shown in Table~\ref{tab:cross_domain_aruba_milan_kyoto}, \projectname{} often improves temporal coherence, although the gains vary across datasets. In the in-domain evaluation, it achieves the lowest EMD on Aruba, Milan, Orange4Home, and UCI-ADL Home B. It also substantially reduces fragmentation on Orange4Home and obtains the lowest overfill, underfill, and EMD on UCI-ADL Home B. Under cross-domain transfer, \projectname{} achieves the lowest EMD in four of the six settings and frequently reduces fragmentation, merge, or overfill errors. These improvements are particularly evident in Milan-related settings, suggesting that temporal refinement helps preserve long-duration activity structure when local sensor evidence is sparse but contextual and historical cues remain informative.}

However, improvements in semantic recognition do not always translate into more accurate segment boundaries. 
\rw{ For example, \projectname{} achieves the strongest classification performance on UCI-ADL Home A but has a higher EMD than the baselines. The Kyoto7-related evaluations also show mixed results across the segment metrics.}
One possible reason is that some activities are weakly observed by the sensor setup. For example, the workspace is not directly instrumented, making ``Work'' difficult to separate from nearby or contextually similar activities. In such cases, \projectname{} may rely more on contextual feasibility and recent history, improving semantic recognition but sometimes smoothing over short transitions or producing imperfect boundary alignment.

\begin{figure}[t]
    \centering
    \includegraphics[width=0.9\linewidth]{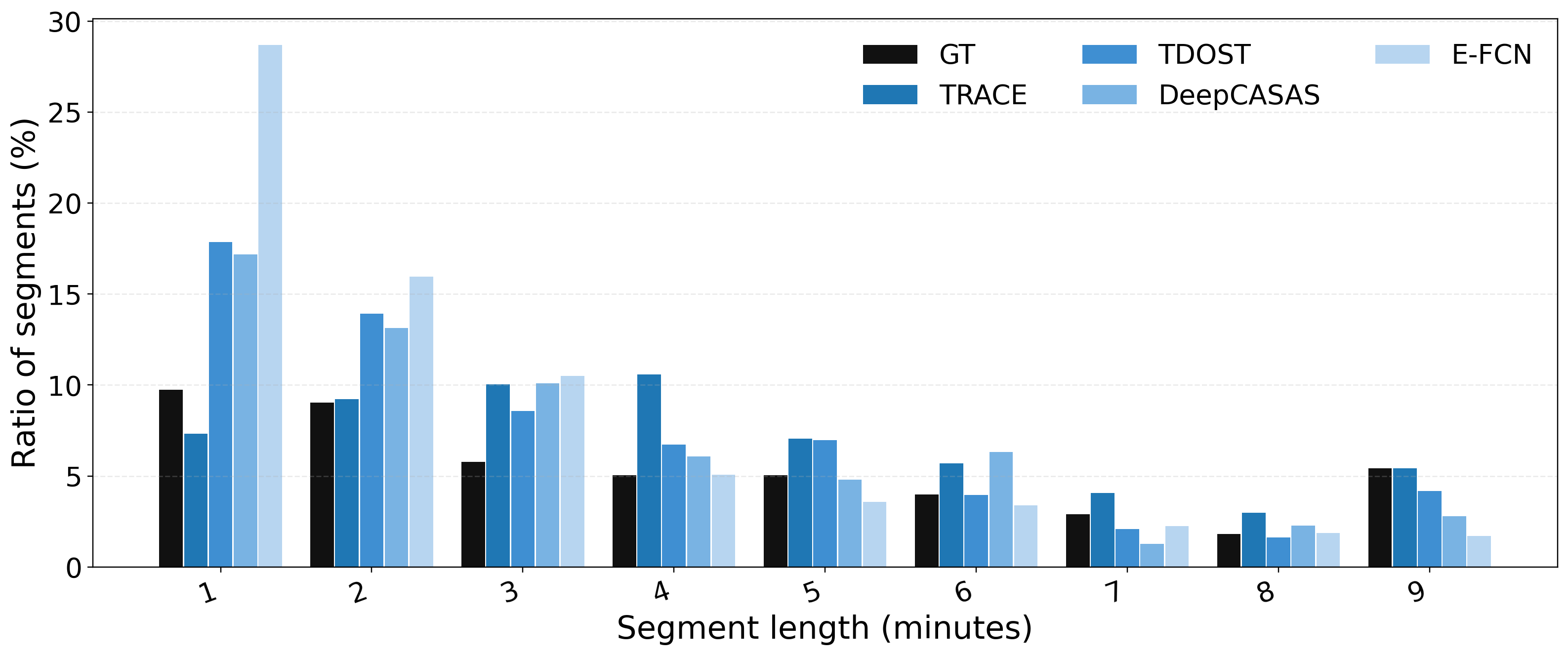}
    \caption{
Distribution of short activity segments in the Aruba$\rightarrow$Aruba in-domain setting. The figure reports the percentage of predicted segments with lengths from 1 to 9 minutes, compared against the ground truth (GT). 
Compared with the baseline models, \projectname{} produces a short-segment distribution that more closely follows the ground truth, especially for 1--2 minute segments, where the baselines substantially overproduce short fragments.
}
\label{fig:short_segment_distribution_aruba_aruba}
    \label{fig:distribution}
\end{figure}

The short-segment distribution (Figure ~\ref{fig:distribution}) shows that \projectname{} produces fewer 1--2 minute fragments than the baselines, indicating that the refinement module helps suppress unstable label switching and produces a smoother activity timeline. But this smoothing can also reduce sensitivity to brief or weakly observed transitions. Therefore, higher merge or underfill in some settings does not necessarily contradict the classification gains; rather, it shows that \projectname{} sometimes favors semantic continuity over fine-grained boundary precision when the available sensor evidence is insufficient.

\rw{Overall, the primary temporal benefit of \projectname{} lies in reducing unstable fragmentation and preserving longer-term semantic continuity, whereas precise boundary alignment remains dependent on the available sensing evidence.}

\subsubsection{Performance Across Diverse Modality Configurations}

\label{sec:results_benchmark_missing_modality}

MARBLE presents a substantially different modality setting than CASAS, providing a useful test of whether \projectname{} remains effective when the available sensor inputs change.
Because one test split does not contain ground-truth instances of “EATING,” “SETTING\_UP\_TABLE,” or “CLEARING\_TABLE,” we retain a fixed label space and assign these classes an F1 score of 0 for that split, providing a conservative aggregate estimate.
As shown in Table~\ref{tab:marble_class_results}, \projectname{} achieves the best overall performance on MARBLE in terms of weighted and macro F1 despite this mismatch in sensing configuration. 
\rw{ Importantly, TRACE without wearable sensing still achieves 71.45\% macro F1, outperforming both DualLSTM and DeepConvLSTM. This suggests that the framework is not tightly coupled to a fixed modality combination, but remains effective as long as useful contextual and sensor evidence are available.}

The class-level results further support this interpretation. Clear gains are observed for activities involving transitions or contextual cues, such as ``ENTERING\_HOME'' and ``LEAVING\_HOME''. For several semantically ambiguous activities (e.g., ``ANSWERING\_PHONE'' and ``MAKING\_PHONE\_CALL''), \projectname{} remains comparable to or slightly better than the baselines, indicating that contextual reasoning improves difficult cases without degrading easier ones.

However, \projectname{} does not outperform the baselines on every category. It shows weaker performance on some short or overlapping activities, suggesting that such cases may be more sensitive to smoothing or contextual bias. Overall, the results show that \projectname{} is particularly beneficial for activities that require contextual disambiguation, supporting the effectiveness of multi-source fusion and reasoning under varying modality conditions.

\begin{table}[t]
\centering
\small
\caption{Per-class F1 scores (\%) of \projectname{}, \rw{\projectname{} without wearable sensing}, DualLSTM, and DeepConvLSTM on MARBLE, together with overall weighted F1 and macro F1. One test split does not contain instances of ``EATING'', ``SETTING\_UP\_TABLE'', and ``CLEARING\_TABLE''; for consistency, their F1 scores are set to 0 in that split when computing aggregate metrics.}
\label{tab:marble_class_results}
\begin{tabular}{lcccc}
\toprule
\textbf{Class}
& \textbf{\projectname{}} & \rw{\textbf{\projectname{} (No Wearable)}} & \textbf{DualLSTM} & \textbf{DeepConvLSTM} \\
\midrule
ANSWERING\_PHONE & \textbf{99.65 $\pm$ 0.20} & \rw{99.01 $\pm$ 0.87} & 99.49 $\pm$ 0.51 & 97.00 $\pm$ 4.39 \\
CLEARING\_TABLE & 22.49 $\pm$ 23.38 & \rw{\textbf{23.41 $\pm$ 23.91}} & 19.41 $\pm$ 19.67 & 8.90 $\pm$ 15.42 \\
COOKING & \textbf{85.41 $\pm$ 4.99} &\rw{83.88 $\pm$ 5.54} & 84.72 $\pm$ 2.86 & 84.49 $\pm$ 6.08 \\
EATING & \textbf{70.96 $\pm$ 40.97} & \rw{69.97 $\pm$ 40.41} & 68.73 $\pm$ 39.68 & 68.67 $\pm$ 39.67 \\
ENTERING\_HOME & \textbf{81.52 $\pm$ 10.29} &\rw{ 79.99 $\pm$ 7.30} & 38.28 $\pm$ 20.45 & 54.98 $\pm$ 16.61 \\
LEAVING\_HOME & 80.63 $\pm$ 11.32 & \rw{\textbf{84.34 $\pm$ 10.23 }}& 62.81 $\pm$ 10.43 & 55.79 $\pm$ 12.37 \\
MAKING\_PHONE\_CALL& \textbf{99.18 $\pm$ 0.39}& \rw{98.54 $\pm$ 0.74}& 97.77 $\pm$ 1.61& 98.57 $\pm$ 1.04 \\
PREPARING\_COLD\_MEAL& 55.80 $\pm$ 10.09& \rw{55.84 $\pm$ 11.03}& \textbf{57.20 $\pm$ 17.00}& 56.37 $\pm$ 10.17 \\
SETTING\_UP\_TABLE& 25.04 $\pm$ 23.71&\rw{ 14.42 $\pm$ 24.97}& \textbf{36.00 $\pm$ 22.38}& 31.65 $\pm$ 26.99 \\
TAKING\_MEDICINES& \textbf{63.26 $\pm$ 14.24}&\rw{ 59.55 $\pm$ 19.55}& 58.90 $\pm$ 15.39& 62.10 $\pm$ 15.19 \\
USING\_PC& \textbf{97.27 $\pm$ 3.02}&\rw{ 95.11 $\pm$ 3.87}& 93.83 $\pm$ 3.53& 95.36 $\pm$ 3.42 \\
WASHING\_DISHES& 64.60 $\pm$ 8.24&\rw{ 65.16 $\pm$ 10.36}& \textbf{68.53 $\pm$ 7.00}& 65.50 $\pm$ 8.08 \\

WATCHING\_TV& \textbf{99.61 $\pm$ 0.22}&\rw{ 99.61 $\pm$ 0.27}& 99.53 $\pm$ 0.33& 99.46 $\pm$ 0.42 \\

\midrule
Weighted F1
& \textbf{82.65 $\pm$ 4.05}& \rw{81.42 $\pm$ 4.85}& 80.90 $\pm$ 5.85& 80.08 $\pm$ 4.22 \\

Macro F1
& \textbf{72.72 $\pm$ 4.19}&\rw{ 71.45 $\pm$ 3.61}& 68.09 $\pm$ 0.80& 67.60 $\pm$ 3.46 \\
\bottomrule
\end{tabular}
\end{table}

\section{\rw{Case Study}}
\label{sec:deployment}
Public benchmarks provide standardized settings for comparing HAR methods, but they do not fully capture the practical constraints involved in deploying a smart-home HAR system in a new environment. The motivation for studying this challenging deployment scenario is real-world applicability: for everyday users, it is difficult to justify long setup phases, extensive annotation, or frequent maintenance before a HAR system becomes useful \cite{hiremath2022bootstrapping,hiremath2023lifespan,hiremath2024maintenance}.

In practice, a deployed system may need to operate with limited environment-specific labeled data, incomplete or changing sensing modalities, user-specific routines, and computational constraints imposed by continuous inference. These challenges are difficult to isolate using public datasets alone. We therefore conduct \rw{a case study} in our smart-home environment to examine how \projectname{} behaves as a practical, deployed system, rather than only as a method evaluated on standardized benchmarks.
This case study focuses on two practical questions:
\begin{description}
\item[RQ4:] Can \projectname{} operate in a newly deployed environment without specific adaptation?
\item[RQ5:] Can \projectname{} maintain useful performance under practical reasoning-resource constraints?
\end{description}

\subsection{\rw{Smart-Home Environment and Study Protocol}}
We instrumented a studio-style smart-home environment with ambient and wearable sensors and evaluated \projectname{} on continuous sensor streams from this deployed setting. 
The environment includes a separate bathroom, while the remaining open-plan space is functionally partitioned into several zones, including the bedroom, living area, kitchen, closet, and dining area. A total of 24 ambient sensors were installed, including motion sensors, contact sensors, and smart plugs. The motion sensors also provide temperature and humidity readings. During the study, participants lived in the environment while wearing a wristband that continuously recorded heart rate and accelerometer signals. 
We use a coarse-grained and a fine-grained activity set. 
The former consists of eight activities, and the latter further refines several coarse categories into more specific activities, resulting in 13 classes overall. 
\rw{The main case-study evaluation uses the coarse-grained labels, while the context-supported fine-grained interpretation results are reported separately in Appendix~\ref{app:results_deployment_fine_grained}.}
This study is intended as a case study and was approved by our university's Institutional Review Board (IRB).\footnote{The case study was approved by the IRB at the Georgia Institute of Technology (protocol number: IRB2024-31).}
We collected data from two participants affiliated with the research team, each of whom lived in the environment for ca.\ one day following their usual daily habits. In total, the study produced about 50 hours of multimodal recordings. User routine priors were collected separately through questionnaires. Ground-truth activity labels were derived from participant activity diaries and post-session reviews with the experimenter.
More detailed descriptions of the recorded datasets, sensor modalities, and activity labels are provided in  Appendix~\ref{app:deployment_dataset}.

\emph{Consistent with a realistic scenario, we use no data for model training, validation, prompt tuning, or model adaptation.} The activity annotations collected during the study are used only for post-hoc evaluation. 
The environmental recognizer is based on TDOST trained on CASAS Milan, whose activity categories largely overlap with those in our dataset. However, because our deployed environment differs in sensor types, layout, and triggering frequency, the event density is substantially different from the original CASAS setting. We therefore use non-overlapping 20-event windows for environmental prediction, as this more closely matches the recognizer's effective temporal span during training.
We also include ADL-LLM~\cite{civitarese2025large} as a zero-shot LLM-based baseline. We implement this baseline following its main design, using GPT-5-mini as the reasoning LLM. This baseline directly infers labels from the available sensor observations. We repeat the evaluation three times and report the mean and standard deviation.

A central challenge in a real deployment is that a system may not have access to all information assumed by a fully configured HAR pipeline. Labeled data may be unavailable for environment-specific training, users may not always wear a device, sensor streams may be incomplete or temporarily unavailable and contextual priors may be missing or partial. We therefore evaluate \projectname{} under limited-information conditions, including zero-shot conditions and controlled removal of sensing and contextual inputs.

Table~\ref{tab:ablation_results} shows the ablation study results in the practical setting. Compared with the original TDOST baseline and the zero-shot ADL-LLM baseline, the full \projectname{} achieves the best overall performance. These results indicate that \projectname{} can operate without specific training while benefiting from the integration of heterogeneous sensor evidence, user context, and temporal refinement. 

\begin{table*}[t]
\centering
\scriptsize
\caption{Ablation study evaluating the contribution of different components in \projectname{} on our smart-home \rw{case study}. Per-class F1, weighted F1, and macro F1 are reported as mean$\pm$std in percentage points (\%). TDOST is deterministic in this setting and is therefore reported without standard deviation. Bold values indicate the best result in each row.}
\label{tab:ablation_results}
\resizebox{\textwidth}{!}{%
\begin{tabular}{lcc|ccccc}
\toprule
Class 
& ADL-LLM
& TDOST
& Env-Only 
& No Wearable 
& No Summary 
& No Context 
& \projectname{} \\
\midrule
Other          & $7.10 \pm 5.10$   & $0.00$  & $0.00 \pm 0.00$  & $\mathbf{9.82 \pm 2.67}$  & $0.26 \pm 0.45$  & $2.61 \pm 4.53$  & $7.16 \pm 1.47$  \\
Relax          & $41.24 \pm 2.47$  & $26.41$ & $27.47 \pm 0.28$ & $31.42 \pm 5.10$ & $72.81 \pm 5.39$ & $90.74 \pm 0.57$ & $\mathbf{93.12 \pm 0.18}$ \\
Cook           & $\mathbf{52.35 \pm 8.18}$  & $16.18$ & $15.87 \pm 0.07$ & $27.04 \pm 6.49$ & $25.51 \pm 0.84$ & $20.22 \pm 0.46$ & $36.19 \pm 4.45$ \\
Leave\_Home    & $22.89 \pm 4.11$  & $6.65$  & $6.06 \pm 0.00$  & $\mathbf{80.51 \pm 7.50}$ & $6.06 \pm 0.00$  & $67.52 \pm 3.71$ & $75.83 \pm 6.61$ \\
Sleep          & $41.91 \pm 2.81$  & $67.20$ & $67.45 \pm 0.07$ & $67.59 \pm 0.69$ & $\mathbf{98.55 \pm 0.56}$ & $97.99 \pm 0.58$ & $98.22 \pm 0.96$ \\
Eat            & $\mathbf{39.59 \pm 0.74}$  & $0.00$  & $0.00 \pm 0.00$  & $28.17 \pm 7.21$ & $7.13 \pm 2.51$  & $15.64 \pm 12.08$ & $38.59 \pm 2.84$ \\
Work           & $0.00 \pm 0.00$   & $0.00$  & $0.00 \pm 0.00$  & $43.82 \pm 7.65$ & $\mathbf{51.78 \pm 6.51}$ & $0.00 \pm 0.00$  & $44.12 \pm 2.93$ \\
Use\_Bathroom  & $59.92 \pm 7.36$  & $52.63$ & $52.52 \pm 1.70$ & $72.31 \pm 1.53$ & $54.40 \pm 1.84$ & $68.85 \pm 2.24$ & $\mathbf{78.02 \pm 2.19}$ \\
\midrule
Weighted F1
& $30.72 \pm 0.95$ 
& $34.15$ 
& $34.54 \pm 0.13$ 
& $51.26 \pm 2.31$ 
& $65.22 \pm 2.17$ 
& $71.58 \pm 0.96$ 
& $\mathbf{80.22 \pm 0.44}$ \\
Macro F1   
& $33.13 \pm 0.31$ 
& $21.13$ 
& $21.17 \pm 0.25$ 
& $45.09 \pm 2.40$ 
& $39.56 \pm 1.25$ 
& $45.45 \pm 1.96$ 
& $\mathbf{58.91 \pm 1.18}$ \\
\bottomrule
\end{tabular}%
}
\end{table*}

\begin{figure*}[t]
    \centering
    \begin{subfigure}[t]{\textwidth}
        \centering
        \includegraphics[width=\textwidth]{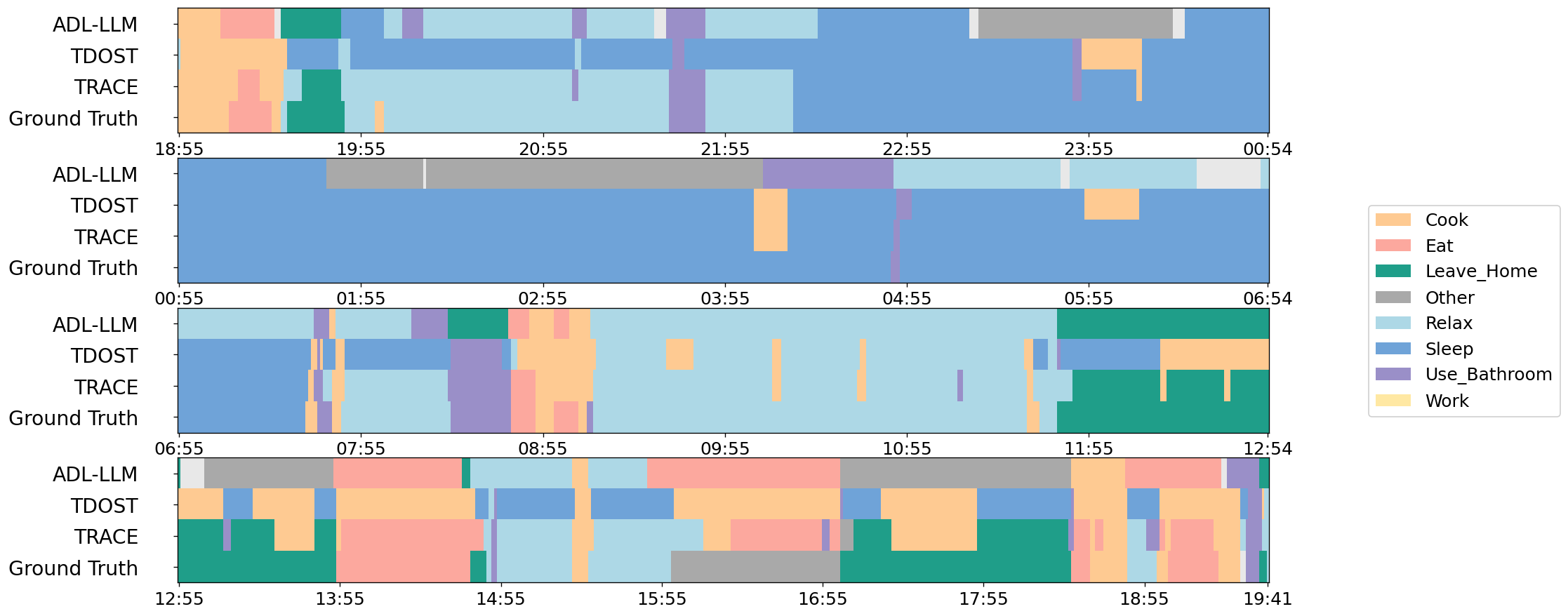}
        \caption{Participant 1}
        \label{fig:deployment_timeline_p1}
    \end{subfigure}
    \vspace{0.5em}
    \begin{subfigure}[t]{\textwidth}
        \centering
        \includegraphics[width=\textwidth]{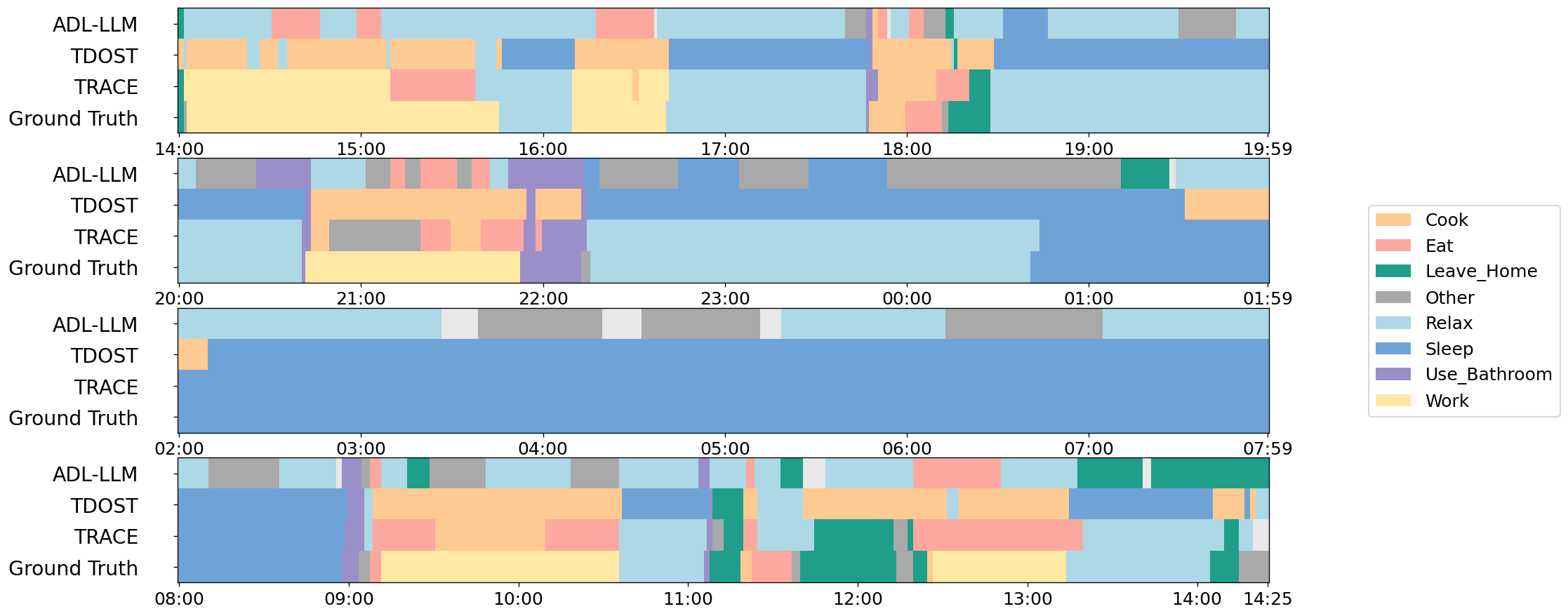}
        \caption{Participant 2}
        \label{fig:deployment_timeline_p2}
    \end{subfigure}
    \caption{Qualitative comparison of activity predictions on the smart-home study timelines for two participants. Each timeline compares ADL-LLM, TDOST, and \projectname{} against the ground-truth annotation. Compared with the baselines, \projectname{} generally produces more temporally coherent predictions and better preserves long-duration and context-dependent activities in continuous daily sequences.}
    \label{fig:deployment_timeline}
\end{figure*}

Figure~\ref{fig:deployment_timeline} provides qualitative examples from the two participants. Each subfigure shows a continuous activity timeline for one participant. 
Compared with ADL-LLM and TDOST, \projectname{} produces more temporally coherent predictions that better align with the ground truth, especially for long-duration states such as ``Sleep'' and ``Relax'' and for context-dependent activities such as ``Eat'' and ``Work''. The baselines often produce extended misclassifications or fragmented transitions, while \projectname{} better preserves activity continuity and reduces isolated label changes.

The ablation results further reveal that the information sources contribute differently across activities.
Even when wearable input is removed, \projectname{} still substantially outperforms the baselines, suggesting that environmental sensing and the proposed reasoning pipeline already provide useful information in the deployed environment. However, the wearable channel remains important for low-motion and long-duration states such as ``Sleep'' and ``Relax'', where environmental sensors alone may provide limited evidence.
The sensor-observation summary is especially important for ``Leave\_Home''. When the summary is removed, the F1 score for this class drops from $75.83\%$ in the full model to $6.06\%$. This indicates that sensor-observation level summaries provide critical cues for coarse transitions such as leaving home, where entrance-related motion, contact events, and location changes are more informative than activity-level predictions alone.
User context is particularly important for ``Eat'' and ``Work''. In our deployed environment, only the dining area contains a table, making table-related behavior ambiguous without additional contextual information. Without contextual priors and user routines, it is difficult to determine whether the user is eating, working, or simply staying in the dining area. 
Overall, these results suggest that the different components in \projectname{} are complementary, with each one contributing most strongly to a different subset of activities.

\subsection{Practical Deployment Feasibility}

\rw{
The practical deployment of \projectname{} also requires consideration of the computational overhead associated with LLM inference.
\projectname{} processes the sensor stream in 10-minute blocks. Each block generates two LLM queries for cross-reference and temporal refinement, resulting in 12 queries per hour. Based on our execution logs, each query contained approximately 2,700 prompt tokens and 600 completion tokens and took approximately 15 seconds. The two-stage inference therefore adds approximately 30 seconds of processing per block, corresponding to 5\% of the 10-minute observation window. This supports near-real-time, although not instantaneous, operation for smart-home monitoring applications that do not require sub-second responses.}

Large cloud-hosted models may provide stronger reasoning, but relying on them continuously can introduce latency, cost, and privacy concerns. 
\rw{Using the GPT-5-mini pricing applied in our estimate (\$0.25 per million input tokens and \$2.00 per million output tokens), the average cost is approximately \$0.0019 per query, corresponding to about \$0.55 per day of continuous operation. Although \projectname{} does not transmit raw sensor streams, its prompts may contain structured summaries of user locations, object interactions, recent activity history, and routine descriptions, which can reveal sensitive behavioral patterns. Related work has similarly identified privacy exposure, dependence on external services, and instability across model versions as limitations of prompt-based LLM approaches to smart-home HAR, and has explored a prompt-free alternative based on language embeddings of sensor observations and activity descriptions~\cite{dhekane2025thou}. We therefore evaluate both cloud-hosted and locally deployed LLM backbones to examine the trade-offs among recognition performance, computational requirements, cost, and privacy.
}

Table~\ref{tab:llm_refinement_results} reports performance before and after temporal refinement under different LLM backbones. Across all backbones, the full pipeline improves over cross-reference alone, suggesting that temporal refinement contributes beyond simply combining evidence from different sources. The improvement is more pronounced for stronger reasoning models: weighted F1 increases by $2.85$ points for GPT-5-mini and by $2.51$ points for Qwen3-32B-Reason. 
\rw{ Importantly, the locally deployed Qwen3-32B-Reason model achieves a competitive F1 score compared with GPT-5-mini, indicating that cloud inference is not functionally required when sufficient local computational resources are available.
}

These results show that the main deployment trade-off is among reasoning capability, hardware requirements, and privacy.  These overheads remain modest relative to the substantial performance gains observed in the challenging settings for which \projectname{} is designed.
\rw{
Cloud-based GPT-5-mini provides the strongest performance with modest latency and monetary cost, but requires the transmission of sensitive behavioral context. Local Qwen3-32B-Reason achieves comparable performance while keeping data within the deployment environment, although it requires dedicated GPU resources. In contrast, smaller local models are easier to deploy but currently provide substantially weaker temporal and contextual reasoning. Hybrid configurations that retain sensitive context locally and invoke cloud models only for uncertain cases may offer a useful compromise.}

\begin{table}[t]
\centering
\small
\caption{Effect of multi-stage temporal refinement under different LLM backbones. \rw{``Cross-Ref Only''} denotes performance after multi-source cross-reference only, and \rw{``Full TRACE''} denotes performance after applying the full multi-stage refinement pipeline. All values are reported in percentage points (\%) from one run.}
\label{tab:llm_refinement_results}
\begin{tabular}{lcccccc}
\toprule
\multirow{2}{*}{LLM Backbone}
& \multicolumn{3}{c}{Weighted F1}
& \multicolumn{3}{c}{Macro F1} \\
\cmidrule(lr){2-4} \cmidrule(lr){5-7}
& \rw{Cross-Ref Only} & \rw{Full TRACE} & Gain 
&\rw{ Cross-Ref Only} &\rw{ Full TRACE} & Gain \\
\midrule
GPT-5-mini         & 77.79 & 80.64 & +2.85 & 56.96 & 59.39 & +2.43 \\
Qwen3-32B-Reason   & 75.63 & 78.14 & +2.51 & 57.17 & 58.85 & +1.68 \\
Qwen3-32B          & 36.49 & 38.34 & +1.85 & 26.17 & 28.12 & +1.95 \\
Llama-3.1-8B       & 19.87 & 21.60 & +1.73 & 34.28 & 34.55 & +0.27 \\

\bottomrule
\end{tabular}
\end{table}

Overall, the \rw{practical case study} complements the benchmark evaluation by examining \projectname{} under realistic deployment constraints. The results show that \projectname{} can be applied without environment-specific training, but its effectiveness depends on available sensing evidence, contextual information, and the reasoning capacity of the LLM backbone.

\section{Discussion}
The contribution of \projectname{} lies not only in improved recognition accuracy, but also in an alternative, more practical formulation of smart-home HAR: rather than treating recognition as isolated short-window classification, the system performs activity interpretation under temporal and contextual constraints. In what follows, we discuss four implications of this design.

\subsection{Context as a Constraint for Activity Interpretation}

An important finding is that contextual information is useful not simply because it provides additional input, but because it constrains how ambiguous sensor evidence is interpreted. In smart-home settings, similar local sensor patterns may correspond to different activities, particularly under sparse and minimally intrusive sensing. TRACE uses temporal continuity, environmental conditions, object interactions, and user routines as soft constraints on which activities are plausible at a particular moment. The goal is therefore not merely to combine more modalities or employ a stronger classifier, but to interpret local observations within a broader behavioral context. The ablation results further show that different contextual components support different activity classes, indicating that the improvements arise from resolving specific sources of ambiguity rather than uniformly altering all predictions. 
\rw{Although the benefit is smaller in the relatively easy Aruba in-domain setting, the gains are substantial in the more challenging Milan, cross-domain, and \rw{in-situ settings}, where local sensor evidence is sparse, ambiguous, or affected by domain shift. In these challenging settings, the additional reasoning layer addresses failure modes that lightweight classifiers do not reliably resolve, making the added system complexity worthwhile. Overall, \projectname{} remains competitive when local sensor evidence is sufficiently discriminative while providing greater benefits when that evidence is sparse, ambiguous, or affected by domain shift.}

\subsection{Temporal Coherence and Boundary Tradeoffs}

Arguably, HAR in practical application scenarios is not only about predicting the correct label at each time step, but also about recovering a coherent activity timeline. Compared with the baselines, \projectname{} generally produces a more coherent activity timeline, with fewer very short activity pieces and segment-length distributions that better match the ground truth. This indicates that temporal refinement does more than correct isolated labels: it yields a more stable interpretation of ongoing behavior. This is important in practice because many smart-home applications depend on temporally consistent activity segments rather than minute-level predictions.

At the same time, our results reveal a tradeoff. The same refinement process that suppresses short-term label switches may still introduce boundary errors, such as splitting long activities into incomplete segments or shifting transition points. In other words, better continuity does not always imply better boundary preservation. This suggests that future refinement strategies should more explicitly account for uncertainty, for example, by performing secondary checks on activity boundaries or by introducing additional boundary optimization mechanisms beyond smoothing.

\subsection{Adaptive Context and Bidirectional Routine Modeling}

\rw{In \projectname{}, contextual priors are derived from home layouts, object-function mappings, historical observations, and user surveys, and remain fixed throughout the evaluated sequence. However, user routines may vary across days or gradually change because of aging, health conditions, or changes in the home environment. Because routines are treated as soft guidance, TRACE can accommodate isolated deviations when sensor evidence is sufficiently strong. Nevertheless, the current framework cannot determine whether repeated mismatches reflect sustained routine drift, temporary variation, or errors from upstream recognizers. Long-term adaptation therefore remains an important limitation.}

This limitation also points to a bidirectional relationship between activity recognition and routine modeling. In the current design, routines constrain activity inference by making some interpretations more plausible than others. Over longer periods, however, reliable activity predictions could also be used to maintain and personalize the routine model itself. 
\rw{In other words, contextual HAR could move from a one-way process, in which a statically defined routine prior informs recognition, toward a closed-loop process in which recognition results provide evidence for updating that prior. Such a formulation may improve adaptation to long-term behavioral drift, but it requires confidence-aware updating to prevent recognition errors from being reinforced as routine knowledge.}

\rw{A possible extension is a closed-loop routine model in which high-confidence recognition results are periodically used to update contextual priors. Recent activity episodes could be aggregated over weekly or monthly windows into a routine profile $\hat{R}_t$, capturing typical timing, duration, location, and transition patterns. If discrepancies between $\hat{R}_t$ and the existing routine prior ${R}_t$ persist across multiple periods, the prior could be updated using Eq. \ref{eq:routine}.}
\rw{
\begin{equation}
R_{t+1} = (1-\alpha)R_t + \alpha \hat{R}_t,
\label{eq:routine}
\end{equation}}

\rw{where $\alpha$ balances long-term stability against responsiveness to sustained change.
Confidence-aware filtering would reduce the risk of reinforcing recognition errors, while a human-in-the-loop mechanism could further validate uncertain updates: repeated or consequential deviations could trigger selective confirmation from the resident or caregiver. This approach is related to prior work on maintaining bootstrapped smart-home HAR systems~\cite{hiremath2022bootstrapping,hiremath2023lifespan,hiremath2024maintenance}, but its implementation and evaluation require longer-term longitudinal deployments and remain future work.}

\subsection{Zero-Shot Potential and Practical Limits of LLM-based HAR}

In \projectname{}, the LLM is used not as a raw sensor classifier, but as a reasoning layer that integrates heterogeneous evidence under temporal and contextual constraints. The results suggest that this is a promising role for LLMs, particularly in smart-home environments where sensing pipelines are heterogeneous, partially missing, and difficult to fuse in a single end-to-end model.

More broadly, \projectname{} also exhibits a limited form of zero-shot potential. Because the system reasons over high-level contextual descriptions and structured summaries, it is not limited to memorizing previously seen low-level sensor patterns and can sometimes infer plausible activities in unfamiliar settings. This property is particularly relevant in cross-domain scenarios. However, this capability remains partial: it still depends on meaningful summaries, sufficiently accurate upstream predictions, and informative contextual descriptions.

The effectiveness of this design also depends strongly on the reasoning capability of the underlying model. Stronger LLMs improve both cross-reference and refinement, but they also increase inference cost, latency, and deployment complexity. In addition, the framework remains dependent on upstream recognizers, and its contextual priors may be incomplete or drift over time. Together, these limitations point to several important next steps for deployment-oriented HAR, including adaptive routine updating, stronger uncertainty handling, privacy-preserving inference, and lighter-weight reasoning backbones that can support continuous use in real homes.

\section{Conclusion}

We presented \projectname{}, a smart-home activity recognition framework that moves beyond short-window sensor classification by incorporating multi-source sensor evidence and user-specific contextual priors into an LLM-based reasoning pipeline. Through time-aligned cross-reference and multi-scale temporal refinement, \projectname{} improves both the semantic plausibility and temporal coherence of activity predictions.
Results on public benchmarks and a practical case study show that \projectname{} is particularly effective in challenging settings where local observations are ambiguous, incomplete, or difficult to generalize across homes. The framework improves robustness in cross-domain settings, better supports semantically complex activities, and produces more coherent activity timelines than baseline methods.
This work shows that contextual reasoning is a promising direction for smart-home HAR. By combining sensor evidence with temporal structure and user context, \projectname{} takes a step toward more robust and meaningful activity understanding in real-world smart-home environments.

\begin{acks}
This work was partially supported by the National Science Foundation, Grant number IIS-2112633. 
\end{acks}

\bibliographystyle{ACM-Reference-Format}
\bibliography{base}

\newpage
\appendix

\section{Datasets}
\subsection{Benchmarks }
\label{app:benchmarks}
For CASAS, we select the Aruba, Milan, and Kyoto7 datasets as benchmark targets. All datasets were collected from smart homes occupied by a single resident. Aruba contains 219 days of recordings, Milan contains 83 days, and Kyoto7 contains 62 days. Both provide environmental sensing streams, including motion, door, and temperature sensors. Since our system is designed to operate without assuming access to target-home training data, we map the original activity annotations in the datasets into a shared label space and consider only their overlapping activity categories following the mapping method (Table~\ref{tab:casas_label_mapping}) in prior work.

The MARBLE dataset contains data collected from 12 participants, including both single-occupancy and multi-occupancy living scenarios. In our experiments, we consider only the single-resident cases. MARBLE includes multimodal data from both environmental sensors and wearable/mobile devices. Specifically, the environmental sensing setup includes magnetic sensors for cabinet-door openings, pressure sensors for seating detection, and smart plugs for appliance usage monitoring. In addition, the smartphone records phone interaction events, such as call initiation and termination, while the smartwatch provides Bluetooth beacon-based location signals and inertial sensor data. The dataset covers a diverse set of daily activities, including answering a phone call, clearing the table, preparing a hot meal, preparing a cold meal, eating, entering home, leaving home, making a phone call, setting the table, taking medication, using a computer, washing dishes, and watching TV.

\rw{The Orange4Home dataset was collected from a single resident in a fully instrumented two-story apartment. It contains approximately 180 hours of recordings collected over 20 working days across four consecutive weeks. The apartment is equipped with 236 heterogeneous sensors, including binary sensors, water-consumption meters, power monitors, and environmental sensors measuring factors such as temperature, humidity, luminosity, CO$_2$, and noise. The dataset contains 493 activity instances annotated in situ and covers 20 classes of daily activities, including entering and leaving home, meal preparation, cooking, washing dishes, eating, computing, watching TV, dressing, reading, napping, showering, and toileting. Compared with the CASAS datasets, Orange4Home provides denser and more heterogeneous environmental sensing.}

\rw{The UCI-ADL dataset contains fully labeled daily-life recordings from two single-resident homes, referred to as Home A and Home B, each occupied by a different resident. The dataset comprises 35 days of recordings in total, including 14 days from Home A and 21 days from Home B. Both homes are equipped with binary environmental sensors, including PIR sensors for detecting movement around doorways and activity-relevant areas, pressure sensors on the bed and couch, magnetic sensors on cupboards and cabinets, and plug sensors for monitoring appliance usage. The sensing configurations differ slightly between the two homes, with Home B having a more limited set of object and appliance sensors but additional PIR sensors around several room entrances. The annotated activities include preparing breakfast, lunch, and dinner, snacking, personal care, showering, toileting, leaving home, relaxing or watching TV, and sleeping. Since the two homes differ in both their residents and sensing infrastructures, we evaluate Home A and Home B independently.}

\begin{table*}[t]
\centering
\small
\caption{Label mapping for the CASAS Aruba, Milan, and Kyoto7 datasets.}
\label{tab:casas_label_mapping}
\resizebox{\textwidth}{!}{
\begin{tabular}{ll|ll|ll}
\toprule
\multicolumn{2}{c|}{\textbf{Aruba}} & 
\multicolumn{2}{c|}{\textbf{Milan}} & 
\multicolumn{2}{c}{\textbf{Kyoto7}} \\
\textbf{Original Label} & \textbf{Mapped Label} & 
\textbf{Original Label} & \textbf{Mapped Label} & 
\textbf{Original Label} & \textbf{Mapped Label} \\
\midrule
Relax               & Relax         & Kitchen Activity         & Cook          & Meal Preparation      & Cook \\
Meal Preparation    & Cook          & Guest Bathroom           & Use Bathroom  & R1 Work               & Work \\
Enter Home          & Enter Home    & Read                     & Relax         & R1 Personal Hygiene   & Personal Hygiene \\
Leave Home          & Leave Home    & Master Bathroom          & Use Bathroom  & R2 Work               & Work \\
Sleeping            & Sleep         & Leave Home               & Leave Home    & R2 Bed to Toilet      & Bed to Toilet \\
Eating              & Eat           & Master Bedroom Activity  & Other         & R2 Personal Hygiene   & Personal Hygiene \\
Work                & Work          & Watch TV                 & Relax         & R1 Sleep              & Sleep \\
Bed to Toilet       & Bed to Toilet & Sleep                    & Sleep         & R2 Sleep              & Sleep \\
Wash Dishes         & Work          & Bed to Toilet            & Bed to Toilet & R1 Bed to Toilet      & Bed to Toilet \\
Housekeeping        & Work          & Desk Activity            & Work          & Watch TV              & Relax \\
Resperate           & Other         & Morning Meds             & Take Medicine & Study                 & Other \\
Other               & Other         & Chores                   & Work          & Clean                 & Work \\
                     &               & Dining Room Activity     & Eat           & Wash Bathtub          & Other \\
                     &               & Evening Meds             & Take Medicine & Other                 & Other \\
                     &               & Meditate                 & Other         &                       & \\
                     &               & Other                    & Other         &                       & \\
\bottomrule
\end{tabular}
}
\end{table*}

\subsection{Smart-Home Case Study}
\label{app:deployment_dataset}
\begin{figure}[htbp]
    \centering
    \includegraphics[width=0.5\linewidth]{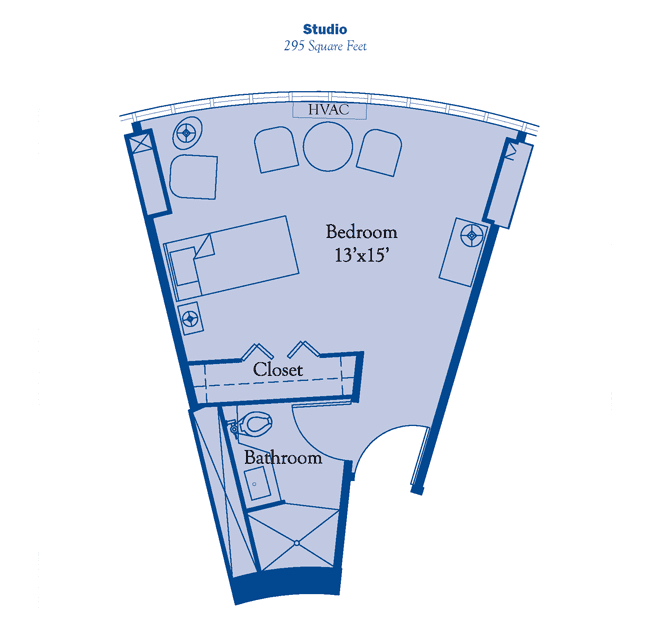}
    \caption{Floor plan of the studio-style smart-home environment used for data collection.}
    \label{fig:floorplan}
\end{figure}

To evaluate \projectname{} under realistic conditions, we evaluated the system in a studio apartment configured as a smart-home living environment. Unlike public benchmark datasets, this setting allowed us to examine how \projectname{} operates in a new physical space with its own layout, sensor placement, appliance configuration, and user routines. The apartment included a separate bathroom, while the remaining space formed an open-plan studio organized into several functional zones: foyer, kitchen, dining area, living area, bedroom area, and closet area. The floor plan is shown in Figure~\ref{fig:floorplan}.

\begin{table}[htbp]
\centering
\small
\caption{Dataset statistics by fine-grained activity label, with corresponding coarse-grained labels, instance count, total duration, and duration percentage.}
\label{tab:dataset_stats}
\begin{tabular}{llrrr}
\toprule
\textbf{Coarse-grained Label} & \textbf{Fine-grained Label} & \textbf{Count} & \textbf{Duration (min)} & \textbf{Duration (\%)} \\
\midrule
\multirow{3}{*}{Cook} 
    & Cook Breakfast & 4  & 16.0   & 0.54 \\
    & Cook Dinner    & 6  & 58.0   & 1.97 \\
    & Cook Lunch     & 3  & 10.0   & 0.34 \\
\midrule
\multirow{3}{*}{Eat}
    & Eat Breakfast  & 2  & 16.0   & 0.54 \\
    & Eat Dinner     & 4  & 52.0   & 1.76 \\
    & Eat Lunch      & 2  & 64.0   & 2.17 \\
\midrule
\multirow{2}{*}{Use Bathroom}
    & Showering      & 3  & 52.0   & 1.76 \\
    & Bathroom visit   & 9  & 27.0   & 0.91 \\
\midrule
Leave Home 
    & Leave Home     & 11 & 319.0  & 10.81 \\
Other
    & Other          & 8  & 93.0   & 3.15 \\
Relax
    & Relax          & 13 & 614.0  & 20.81 \\
Sleep
    & Sleep          & 11 & 1294.0 & 43.85 \\
Work
    & Work           & 5  & 336.0  & 11.39 \\
\bottomrule
\end{tabular}
\end{table}

The environment used 24 ambient sensors spanning three sensing modalities: 8 motion sensors, 12 door/contact sensors, and 4 smart-plug power sensors. These sensors were installed to capture activity-relevant interactions while remaining minimally intrusive. Sensor placement focused on semantically meaningful locations and objects that are strongly associated with everyday activities, such as food preparation, eating, relaxing, sleeping, and leaving home. In addition to motion events, the motion sensors also reported temperature and humidity measurements.

Motion sensors covered the dining area, foyer, living area, kitchen, bathroom, bed area, and closet area. Door/contact sensors were attached to the apartment entry door, bathroom door, upper and lower bathroom cabinets, upper and lower kitchen cabinets, two kitchen drawers, the microwave door, and the refrigerator and freezer doors. Smart-plug power sensors monitored four appliances: the stovetop, microwave, coffee maker, and bedside lamp. To provide complementary wearable evidence, participants also wore a Garmin watch continuously during the recording period, which provided heart rate and accelerometer measurements.

We collected data from two participants affiliated with the research team to evaluate the system. Each participant lived alone in the apartment for one continuous recording session lasting approximately 24-25 hours. Rather than following a predefined activity script, participants were asked to behave as naturally as possible and follow their usual daily routines. This design was intended to reflect a practical scenario in which the system must interpret naturally occurring activities from sparse, heterogeneous, and user-specific sensor evidence. Across both sessions, the study produced 50 hours of continuous multimodal recordings. Table~\ref{tab:dataset_stats} summarizes the resulting activity distribution. As expected in a naturalistic environment, the data are temporally imbalanced, with long-duration activities such as sleep, relaxation, work, and leaving home occupying a larger proportion of the timeline than short interaction-driven activities such as cooking, eating, and bathroom use.

Ground-truth labels were constructed after study through a two-stage annotation process. During the recording period, participants kept a detailed activity diary at approximately minute-level resolution, recording their activities in their own words. After each session, the participant and the experimenter jointly reviewed the diary and mapped the recorded activities to the study label set. Separately, user routine priors were collected through questionnaires and used as contextual information for \projectname{}.

\section{Additional Experiments}

\subsection{Context-Supported Fine-Grained Activity Interpretation}
\label{app:results_deployment_fine_grained}
Beyond recognizing coarse categories, a deployed smart-home HAR system should provide labels that are meaningful for downstream applications and users. For example, distinguishing ``Cook Breakfast'' from ``Cook Dinner'', or distinguishing a brief bathroom visit from showering, can be more useful than assigning only generic labels such as ``Cook'' or ``Use\_Bathroom''. We therefore evaluate whether \projectname{} can use context to refine coarse activity categories into more specific fine-grained labels.

Table~\ref{tab:coarse_to_fine} reports fine-grained recognition performance. \projectname{} performs well on several fine-grained activities, suggesting that deployment-specific context can support more detailed semantic interpretation when informative cues are available. 
For example, the model can successfully distinguish ``Eat Breakfast'' and ``Eat Lunch'' under the broader ``Eating'' category, likely because these labels are closely tied to routine and temporal context.
The model also distinguishes ``Bathroom visit'' from ``Showering'', which involve different interaction patterns and activity durations. However, the relatively lower performance on ``Cook Lunch'' suggests that some fine-grained interpretation remains difficult for labels with limited examples or weakly distinctive contextual evidence. Overall, these results show that context can enable meaningful fine-grained activity interpretation, but its benefit depends on whether fine-grained labels are associated with stable temporal, spatial, or interaction patterns.

\begin{table}[htbp]
\centering
\small
\caption{Per-class F1 scores for fine-grained activity recognition derived from coarse activity categories.}
\label{tab:coarse_to_fine}
\begin{tabular}{l l c c c l}
\toprule
\textbf{Coarse Label} & \textbf{Fine Label} & \textbf{\projectname{}}  \\
\midrule
Cook & Cook Breakfast & 53.57 \\
Cook & Cook Lunch     & 0.00 \\
Cook & Cook Dinner    & 58.33 \\
\addlinespace[0.15em]
Eat & Eat Breakfast            & 66.67 \\
Eat & Eat Lunch                & 84.21 \\
Eat & Eat Dinner               & 45.45 \\
\addlinespace[0.15em]
Use\_Bathroom & Bathroom visit & 55.81 \\
Use\_Bathroom & Showering      & 61.54 \\
\bottomrule
\end{tabular}
\end{table}

\rw{\subsection{Analysis of Context and Temporal Refinement}}
\noindent
\rw{Table~\ref{tab:temporal_context_ablation} consolidates the comparisons that
were previously reported across different parts of the evaluation. The Short
Window Baseline (TDOST) achieves a weighted F1 of 34.15\% and a macro F1 of
21.13\%, whereas the complete TRACE pipeline achieves 80.22\% and 58.91\%,
respectively. However, this overall difference should not be attributed to
temporal refinement alone. Without Temporal Refinement, TRACE already achieves
78.46\% weighted F1 and 57.76\% macro F1, showing that multi-source
cross-reference and contextual information account for a substantial part of
the improvement over the original short-window prediction.}

\rw{User-specific routine knowledge also provides an important complementary
contribution. Removing the user prior reduces weighted F1 from 80.22\% to
71.58\% and macro F1 from 58.91\% to 45.45\%. The largest class-level
reductions occur for Cook, Eat, and Work, which are difficult to distinguish
from local sensor evidence alone and benefit from information about typical
times, locations, and activity transitions.}

\rw{Temporal refinement provides a further, although smaller, improvement over
current-window cross-reference, increasing weighted F1 from 78.46\% to
80.22\% and macro F1 from 57.76\% to 58.91\%. The largest improvements are
observed for Sleep and Use\_Bathroom, suggesting that recent activity history
and longer-horizon continuity are particularly useful for stabilizing
temporally extended or weakly observed activities. Performance decreases
slightly for Other, Leave\_Home, and Work, indicating that refinement may
occasionally smooth over short transitions or revise locally correct
predictions. Overall, these results show that TRACE benefits jointly from
multi-source evidence integration, user-specific routine knowledge, and
longer-horizon temporal refinement, rather than from temporal reasoning alone.}

\begin{table*}[t]
\centering
\small
\caption{\rw{Controlled comparison of short-window recognition, user-specific context, and temporal refinement in the practical case study. \textit{Without User Prior} removes the user-specific routine prior while retaining the remaining TRACE pipeline. \textit{Without Temporal Refinement} uses the same sensing inputs and contextual information as Full TRACE but excludes the longer-horizon temporal refinement stage. Per-class F1, weighted F1, and macro F1 are reported as mean$\pm$std in percentage points (\%). The Short Window Baseline (TDOST) is deterministic and is therefore reported without standard deviation. Bold values indicate the best result in each row.}}
\label{tab:temporal_context_ablation}
\resizebox{0.8\textwidth}{!}{%
\begin{tabular}{l||c|ccc}
\toprule
\multirow{2}{*}{\rw{Class}} & \multirow{2}{*}{\rw{\shortstack[c]{Short Window\\Baseline (TDOST)}}} & \multicolumn{3}{|c}{\rw{\projectname{}}} \\
\cmidrule(lr){3-5}
& & \rw{\shortstack{Without\\User Prior}} & \rw{\shortstack{Without Temporal\\Refinement}} & \rw{\shortstack{Full\\\projectname{}}} \\
\midrule
\rw{Other} & \rw{$0.00$} & \rw{$2.61 \pm 4.53$} & \rw{$\mathbf{9.83 \pm 0.55}$} & \rw{$7.16 \pm 1.47$} \\
\rw{Relax} & \rw{$26.41$} & \rw{$90.74 \pm 0.57$} & \rw{$91.27 \pm 0.28$} & \rw{$\mathbf{93.12 \pm 0.18}$} \\
\rw{Cook} & \rw{$16.18$} & \rw{$20.22 \pm 0.46$} & \rw{$33.69 \pm 3.97$} & \rw{$\mathbf{36.19 \pm 4.45}$} \\
\rw{Leave\_Home} & \rw{$6.65$} & \rw{$67.52 \pm 3.71$} & \rw{$\mathbf{79.39 \pm 0.70}$} & \rw{$75.83 \pm 6.61$} \\
\rw{Sleep} & \rw{$67.20$} & \rw{$97.99 \pm 0.58$} & \rw{$93.90 \pm 0.22$} & \rw{$\mathbf{98.22 \pm 0.96}$} \\
\rw{Eat} & \rw{$0.00$} & \rw{$15.64 \pm 12.08$} & \rw{$38.25 \pm 1.60$} & \rw{$\mathbf{38.59 \pm 2.84}$} \\
\rw{Work} & \rw{$0.00$} & \rw{$0.00 \pm 0.00$} & \rw{$\mathbf{45.83 \pm 3.03}$} & \rw{$44.12 \pm 2.93$} \\
\rw{Use\_Bathroom} & \rw{$52.63$} & \rw{$68.85 \pm 2.24$} & \rw{$69.95 \pm 1.21$} & \rw{$\mathbf{78.02 \pm 2.19}$} \\
\midrule
\rw{Weighted F1} & \rw{$34.15$} & \rw{$71.58 \pm 0.96$} & \rw{$78.46 \pm 0.50$} & \rw{$\mathbf{80.22 \pm 0.44}$} \\
\rw{Macro F1} & \rw{$21.13$} & \rw{$45.45 \pm 1.96$} & \rw{$57.76 \pm 0.66$} & \rw{$\mathbf{58.91 \pm 1.18}$} \\
\bottomrule
\end{tabular}%
}
\end{table*}

\rw{\subsection{Benefits and Risks of Contextual Priors in LLM Refinement}}
\noindent
\rw{This experiment jointly examines when LLM refinement helps or harms activity recognition and how its behavior changes when contextual priors are absent, accurate, or inaccurate. We conduct the analysis on Orange4Home because the upstream sensor-based baseline already achieves relatively strong recognition performance on this dataset. This setting provides a meaningful test of whether LLM refinement can correct the remaining errors without unnecessarily overriding predictions that are already correct.}

\rw{To distinguish the contribution of general common-sense reasoning from that of user-specific routine knowledge, we compare three conditions: (1) \textit{Without Prior}, in which the LLM reasons only over the multi-source recognition outputs and temporal evidence; (2) \textit{Inaccurate prior}, in which it receives a deliberately distorted routine; and (3) \textit{Accurate prior}, in which it receives the correct user routine. The upstream predictions, LLM configuration, and refinement procedure are kept identical across conditions, with only the contextual prior being varied. For each activity class, we measure how often refinement corrects an initially incorrect prediction (\textit{wrong-to-correct}) and how often it changes an initially correct prediction into an incorrect one (\textit{correct-to-wrong}).}

\rw{The inaccurate prior is designed to remain behaviorally plausible while being temporally inconsistent with the actual routine, rather than containing arbitrary or obviously impossible descriptions. Specifically, we move \textit{Napping} from the afternoon to the morning and \textit{Reading} from midday to the late afternoon; shift the expected times of \textit{Showering}, sink use, and toilet use away from their original morning and midday distributions; and postpone the entire lunch sequence---\textit{Preparing}, \textit{Cooking}, \textit{Eating}, and \textit{Washing the Dishes}---to the afternoon while preserving its internal order. We additionally exchange the expected time periods of \textit{Computing} and \textit{Watching TV}, and partially swap the temporal priors associated with the \textit{Leaving}/\textit{Entering} and \textit{Going Up}/\textit{Going Down} transitions. These perturbations allow us to test whether the LLM follows conflicting routine information or appropriately prioritizes the available sensor evidence.}

\begin{table*}[t]
\centering
\small
\caption{\rw{Per-class effects of LLM refinement under different contextual-prior conditions on Orange4Home. W$\rightarrow$C denotes the percentage of initially incorrect predictions corrected by refinement, while C$\rightarrow$W denotes the percentage of initially correct predictions changed to incorrect labels. All results are reported as mean$\pm$std in percentage points (\%) over three folds. Bold values indicate the best result in each row, with higher W$\rightarrow$C and lower C$\rightarrow$W being better.}}
\label{tab:context_prior_correction}
\begin{tabular}{lcc|cc|cc}
\toprule
& \multicolumn{2}{c|}{\rw{Without Prior}}
& \multicolumn{2}{c|}{\rw{Inaccurate Prior}}
& \multicolumn{2}{c}{\rw{Accurate Prior}} \\
\rw{Class}
& \rw{W$\rightarrow$C} & \rw{C$\rightarrow$W}
& \rw{W$\rightarrow$C} & \rw{C$\rightarrow$W}
& \rw{W$\rightarrow$C} & \rw{C$\rightarrow$W} \\
\midrule
\rw{Other} & \rw{$3.51 \pm 4.96$} & \rw{$\mathbf{0.00 \pm 0.00}$} & \rw{$4.39 \pm 6.20$} & \rw{$\mathbf{0.00 \pm 0.00}$} & \rw{$\mathbf{7.72 \pm 5.61}$} & \rw{$\mathbf{0.00 \pm 0.00}$} \\
\rw{Entering} & \rw{$\mathbf{19.70 \pm 17.05}$} & \rw{$4.17 \pm 2.95$} & \rw{$\mathbf{19.70 \pm 17.05}$} & \rw{$4.17 \pm 2.95$} & \rw{$\mathbf{19.70 \pm 17.05}$} & \rw{$\mathbf{2.08 \pm 2.95}$} \\
\rw{Going\_up} & \rw{$\mathbf{69.44 \pm 14.16}$} & \rw{$\mathbf{0.00 \pm 0.00}$} & \rw{$\mathbf{69.44 \pm 14.16}$} & \rw{$\mathbf{0.00 \pm 0.00}$} & \rw{$\mathbf{69.44 \pm 14.16}$} & \rw{$\mathbf{0.00 \pm 0.00}$} \\
\rw{Showering} & \rw{$7.46 \pm 8.09$} & \rw{$6.71 \pm 3.33$} & \rw{$\mathbf{8.00 \pm 8.84}$} & \rw{$\mathbf{5.61 \pm 1.88}$} & \rw{$7.42 \pm 8.62$} & \rw{$6.53 \pm 2.90$} \\
\rw{Using\_the\_sink} & \rw{$\mathbf{14.34 \pm 6.72}$} & \rw{$\mathbf{21.75 \pm 14.70}$} & \rw{$13.50 \pm 7.88$} & \rw{$22.61 \pm 12.33$} & \rw{$13.50 \pm 6.42$} & \rw{$22.61 \pm 10.70$} \\
\rw{Going\_down} & \rw{$\mathbf{8.33 \pm 11.79}$} & \rw{$\mathbf{16.67 \pm 23.57}$} & \rw{$\mathbf{8.33 \pm 11.79}$} & \rw{$\mathbf{16.67 \pm 23.57}$} & \rw{$\mathbf{8.33 \pm 11.79}$} & \rw{$\mathbf{16.67 \pm 23.57}$} \\
\rw{Watching\_TV} & \rw{$\mathbf{0.68 \pm 0.75}$} & \rw{$1.55 \pm 1.34$} & \rw{$0.57 \pm 0.81$} & \rw{$\mathbf{1.09 \pm 0.59}$} & \rw{$0.57 \pm 0.81$} & \rw{$1.92 \pm 1.47$} \\
\rw{Using\_the\_toilet} & \rw{$32.37 \pm 16.08$} & \rw{$8.02 \pm 8.06$} & \rw{$33.96 \pm 14.88$} & \rw{$\mathbf{6.43 \pm 5.92}$} & \rw{$\mathbf{34.86 \pm 11.08}$} & \rw{$\mathbf{6.43 \pm 5.92}$} \\
\rw{Computing} & \rw{$\mathbf{0.27 \pm 0.12}$} & \rw{$0.50 \pm 0.10$} & \rw{$0.26 \pm 0.12$} & \rw{$0.44 \pm 0.08$} & \rw{$\mathbf{0.27 \pm 0.12}$} & \rw{$\mathbf{0.43 \pm 0.10}$} \\
\rw{Preparing} & \rw{$\mathbf{26.67 \pm 20.55}$} & \rw{$4.76 \pm 6.73$} & \rw{$24.29 \pm 17.96$} & \rw{$\mathbf{0.00 \pm 0.00}$} & \rw{$24.29 \pm 17.96$} & \rw{$4.76 \pm 6.73$} \\
\rw{Cooking} & \rw{$2.12 \pm 1.98$} & \rw{$2.12 \pm 1.98$} & \rw{$2.12 \pm 1.98$} & \rw{$1.84 \pm 0.38$} & \rw{$\mathbf{3.18 \pm 1.32}$} & \rw{$\mathbf{1.33 \pm 0.99}$} \\
\rw{Eating} & \rw{$\mathbf{8.50 \pm 4.64}$} & \rw{$\mathbf{3.53 \pm 2.16}$} & \rw{$\mathbf{8.50 \pm 4.64}$} & \rw{$4.12 \pm 1.97$} & \rw{$\mathbf{8.50 \pm 4.64}$} & \rw{$7.76 \pm 3.44$} \\
\rw{Washing\_the\_dishes} & \rw{$7.14 \pm 1.02$} & \rw{$34.85 \pm 2.18$} & \rw{$\mathbf{7.84 \pm 0.99}$} & \rw{$22.50 \pm 7.02$} & \rw{$\mathbf{7.84 \pm 0.99}$} & \rw{$\mathbf{22.12 \pm 2.66}$} \\
\rw{Cleaning} & \rw{$8.45 \pm 1.48$} & \rw{$\mathbf{3.04 \pm 2.22}$} & \rw{$8.93 \pm 5.04$} & \rw{$5.42 \pm 1.35$} & \rw{$\mathbf{9.26 \pm 4.40}$} & \rw{$\mathbf{3.04 \pm 2.22}$} \\
\rw{Dressing} & \rw{$\mathbf{3.03 \pm 4.29}$} & \rw{$6.06 \pm 4.29$} & \rw{$\mathbf{3.03 \pm 4.29}$} & \rw{$\mathbf{3.03 \pm 4.29}$} & \rw{$\mathbf{3.03 \pm 4.29}$} & \rw{$\mathbf{3.03 \pm 4.29}$} \\
\rw{Reading} & \rw{$5.40 \pm 1.18$} & \rw{$28.74 \pm 15.65$} & \rw{$7.46 \pm 0.79$} & \rw{$17.79 \pm 19.27$} & \rw{$\mathbf{12.55 \pm 4.37}$} & \rw{$\mathbf{10.47 \pm 7.11}$} \\
\rw{Napping} & \rw{$34.89 \pm 18.96$} & \rw{$\mathbf{13.19 \pm 9.07}$} & \rw{$\mathbf{36.46 \pm 16.08}$} & \rw{$13.30 \pm 7.96$} & \rw{$29.54 \pm 14.70$} & \rw{$15.15 \pm 8.50$} \\
\rw{Leaving} & \rw{$24.39 \pm 18.75$} & \rw{$\mathbf{0.00 \pm 0.00}$} & \rw{$23.11 \pm 18.62$} & \rw{$\mathbf{0.00 \pm 0.00}$} & \rw{$\mathbf{26.19 \pm 19.22}$} & \rw{$\mathbf{0.00 \pm 0.01}$} \\
\midrule
\rw{Overall} & \rw{$17.93 \pm 13.15$} & \rw{$0.94 \pm 0.12$} & \rw{$17.09 \pm 13.06$} & \rw{$\mathbf{0.72 \pm 0.12}$} & \rw{$\mathbf{19.20 \pm 13.43}$} & \rw{$0.74 \pm 0.16$} \\
\bottomrule
\end{tabular}
\end{table*}

\rw{As shown in Table~\ref{tab:context_prior_correction}, the benefits of LLM refinement outweigh the potential harm across all contextual-prior conditions. Overall, the LLM corrects a meaningful portion of the errors made by the upstream recognizer, while changing fewer than one percent of initially correct predictions. 
In particular, the high correction rate for \textit{Going\_up} can be attributed to its distinctive sensor patterns, which provide stronger evidence for refinement. Moreover, this class contains relatively few instances and is poorly recognized by TDOST, leaving a larger proportion of errors that the LLM can recover. 
In contrast, harmful modifications occur mainly for sensor-ambiguous classes, such as \textit{Using\_the\_sink}, \textit{Washing\_the\_dishes}, and \textit{Reading}, where contextual expectations and local observations may support multiple plausible activities. Overall, the LLM makes substantial corrections when the combined evidence strongly supports an alternative label while remaining conservative at the aggregate level.}

\rw{Accurate routine priors increase beneficial corrections while maintaining a similarly low rate of harmful changes; compared with using no prior, they also reduce unnecessary changes to initially correct predictions. Inaccurate priors influence the refinement behavior, but their overall impact remains limited and does not cause systematic degradation. This is because the routine prior is represented as soft contextual evidence rather than a hard decision rule. The LLM jointly considers the prior, multi-source sensor evidence, temporal relationships, and upstream recognition outputs; when these sources conflict, it can prioritize the evidence that is more strongly supported by the observations. These findings also indicate that user priors are most useful as guidance for resolving ambiguous cases, while the underlying sensor evidence remains essential for preventing overreliance on typical routines and accommodating deviations in actual behavior.}

\rw{\subsection{Multi-Resident Feasibility Evaluation on ARAS}}
\label{app:aras_multistudy}
\noindent
\rw{ARAS is a multi-resident smart-home dataset collected from two homes, each occupied by two residents. Unlike the single-resident datasets used in our main evaluation, sensor events in ARAS are generated within a shared environment and are not explicitly associated with individual residents. Consequently, concurrent or overlapping activities may produce interleaved sensor evidence, introducing ambiguity regarding which resident triggered each event. We conduct two complementary experiments to examine how \projectname{} behaves under this setting.}

\subsubsection{\rw{Controlled Single-Resident Approximation}}

\rw{We first construct a controlled approximation of a single-resident setting by selecting periods from Home A in which one resident was outside the home while the other remained inside. During these periods, the ambient sensor stream can be attributed primarily to the resident who remained at home, substantially reducing interference between residents. We do not conduct the same experiment on Home B because it contains too few periods satisfying this condition to support a meaningful evaluation. In addition, the version of the dataset shared with us contains annotations for only 20 activity classes, which limits the scope of the evaluation to these available classes.}

\rw{The selected samples do not form one continuous observation period. Instead, they consist of multiple disjoint intervals distributed across the original recording. To avoid introducing artificial temporal transitions, we treat each contiguous interval as an independent sequence, reset the historical state of \projectname{} at the beginning of each interval, and prevent input windows from crossing interval boundaries. This setting therefore provides only a controlled approximation of single-resident recognition and limits the amount of continuous historical context available to temporal refinement.}

\rw{We evaluate DeepConvLSTM, E-FCN, TDOST, and \projectname{} on this extracted subset to examine its effectiveness under approximately single-resident conditions. As shown in Table~\ref{tab:aras_single_resident}, \projectname{} achieves the highest macro F1 of 34.57\%, indicating more balanced recognition across activity classes. It performs particularly well on brushing teeth, changing clothes, and meal preparation, while DeepConvLSTM is stronger on sensor-distinct activities such as showering and sleeping. However, the disjoint intervals prevent \projectname{} from maintaining a continuous activity history, limiting longer-term contextual refinement and increasing its reliance on sensor observations and upstream predictions. This is challenging because activities such as studying, reading, listening to music, and using the Internet produce weak or similar ambient-sensor patterns. Overall, \projectname{} remains competitive, although limited temporal context and sensor ambiguity reduce the benefits of contextual refinement.}

\begin{table*}[t]
\centering
\small
\caption{\rw{Per-class and overall recognition performance on the controlled single-resident subset extracted from ARAS Home A. Results are reported as mean$\pm$std in percentage points (\%) over three runs. Bold values indicate the best result in each row.}}
\label{tab:aras_single_resident}
\resizebox{0.8\textwidth}{!}{%
\begin{tabular}{lcccc}
\toprule
\rw{\textbf{Class}} & \rw{\textbf{E-FCN}} & \rw{\textbf{DeepConvLSTM}} & \rw{\textbf{TDOST}} & \rw{\textbf{\projectname{}}} \\
\midrule
\rw{Brushing Teeth} & \rw{$9.10 \pm 9.30$} & \rw{$17.20 \pm 23.10$} & \rw{$11.30 \pm 3.70$} & \rw{$\mathbf{36.10 \pm 13.20}$} \\
\rw{Changing Clothes} & \rw{$15.90 \pm 6.80$} & \rw{$40.00 \pm 35.10$} & \rw{$11.50 \pm 6.30$} & \rw{$\mathbf{50.40 \pm 15.60}$} \\
\rw{Having Breakfast} & \rw{$31.50 \pm 13.40$} & \rw{$35.80 \pm 6.50$} & \rw{$\mathbf{45.70 \pm 14.40}$} & \rw{$35.00 \pm 4.60$} \\
\rw{Having Dinner} & \rw{$0.00 \pm 0.00$} & \rw{$0.00 \pm 0.00$} & \rw{$\mathbf{10.00 \pm 17.30}$} & \rw{$0.00 \pm 0.00$} \\
\rw{Having Lunch} & \rw{$5.00 \pm 6.20$} & \rw{$0.00 \pm 0.00$} & \rw{$\mathbf{28.20 \pm 10.40}$} & \rw{$7.50 \pm 7.10$} \\
\rw{Having Shower} & \rw{$35.50 \pm 21.50$} & \rw{$\mathbf{87.30 \pm 5.00}$} & \rw{$42.40 \pm 25.30$} & \rw{$80.40 \pm 5.70$} \\
\rw{Having Snack} & \rw{$0.00 \pm 0.00$} & \rw{$1.40 \pm 2.50$} & \rw{$\mathbf{12.20 \pm 11.30}$} & \rw{$6.40 \pm 11.10$} \\
\rw{Listening to Music} & \rw{$\mathbf{12.70 \pm 22.00}$} & \rw{$0.00 \pm 0.00$} & \rw{$0.00 \pm 0.00$} & \rw{$0.00 \pm 0.00$} \\
\rw{Napping} & \rw{$\mathbf{0.00 \pm 0.00}$} & \rw{$\mathbf{0.00 \pm 0.00}$} & \rw{$\mathbf{0.00 \pm 0.00}$} & \rw{$\mathbf{0.00 \pm 0.00}$} \\
\rw{Preparing Breakfast} & \rw{$23.20 \pm 8.20$} & \rw{$54.10 \pm 8.50$} & \rw{$57.40 \pm 2.50$} & \rw{$\mathbf{64.30 \pm 12.90}$} \\
\rw{Preparing Dinner} & \rw{$0.00 \pm 0.00$} & \rw{$0.00 \pm 0.00$} & \rw{$12.70 \pm 16.40$} & \rw{$\mathbf{51.80 \pm 7.10}$} \\
\rw{Preparing Lunch} & \rw{$24.70 \pm 11.80$} & \rw{$9.30 \pm 5.90$} & \rw{$\mathbf{57.60 \pm 9.40}$} & \rw{$47.60 \pm 13.60$} \\
\rw{Reading Book} & \rw{$0.00 \pm 0.00$} & \rw{$0.00 \pm 0.00$} & \rw{$0.00 \pm 0.00$} & \rw{$\mathbf{0.90 \pm 1.50}$} \\
\rw{Sleeping} & \rw{$24.20 \pm 25.80$} & \rw{$\mathbf{80.70 \pm 7.70}$} & \rw{$43.10 \pm 15.30$} & \rw{$77.30 \pm 7.90$} \\
\rw{Studying} & \rw{$20.80 \pm 14.50$} & \rw{$21.20 \pm 18.40$} & \rw{$\mathbf{25.50 \pm 21.30}$} & \rw{$21.70 \pm 18.80$} \\
\rw{Talking on the Phone} & \rw{$36.30 \pm 18.90$} & \rw{$\mathbf{50.40 \pm 3.70}$} & \rw{$37.40 \pm 20.90$} & \rw{$48.20 \pm 3.80$} \\
\rw{Toileting} & \rw{$41.70 \pm 7.90$} & \rw{$\mathbf{84.30 \pm 5.20}$} & \rw{$46.50 \pm 7.40$} & \rw{$59.90 \pm 16.60$} \\
\rw{Using Internet} & \rw{$30.50 \pm 1.90$} & \rw{$\mathbf{43.70 \pm 11.70}$} & \rw{$32.10 \pm 8.10$} & \rw{$39.20 \pm 14.40$} \\
\rw{Washing Dishes} & \rw{$8.80 \pm 5.70$} & \rw{$\mathbf{36.00 \pm 31.30}$} & \rw{$17.80 \pm 10.10$} & \rw{$33.50 \pm 30.10$} \\
\rw{Watching TV} & \rw{$64.70 \pm 6.90$} & \rw{$\mathbf{67.60 \pm 4.70}$} & \rw{$66.10 \pm 11.30$} & \rw{$65.50 \pm 3.90$} \\
\midrule
\rw{Accuracy} & \rw{$35.67 \pm 3.25$} & \rw{$\mathbf{57.66 \pm 6.77}$} & \rw{$41.43 \pm 7.30$} & \rw{$54.94 \pm 5.76$} \\
\rw{Weighted F1} & \rw{$33.18 \pm 5.78$} & \rw{$\mathbf{52.43 \pm 8.27}$} & \rw{$40.11 \pm 7.32$} & \rw{$51.56 \pm 5.94$} \\
\rw{Macro F1} & \rw{$19.24 \pm 2.77$} & \rw{$31.45 \pm 6.09$} & \rw{$27.88 \pm 4.23$} & \rw{$\mathbf{34.57 \pm 3.21}$} \\
\bottomrule
\end{tabular}%
}
\end{table*}

\subsubsection{\rw{Zero-Shot Multi-Resident Reasoning}}

\rw{We next retain the original multi-resident setting and examine whether the LLM can infer the activities of both residents from shared sensor observations. We modify the output format of the prompt to request a separate activity prediction for each resident. The input contains a structured summary of the shared ambient sensor events together with the contextual prior of each resident, allowing the model to reason about which interpretation is more plausible for each individual.}

\rw{Because the sensor stream does not identify which resident triggered each event, and our existing baselines produce only a single activity label rather than simultaneous resident-specific predictions, they cannot be directly compared under this setting. We therefore exclude upstream classifier predictions and provide only the sensor summary and the two resident-specific contextual descriptions to the LLM. The experiment is a zero-shot feasibility evaluation of contextual reasoning, rather than a direct evaluation of the complete \projectname{} pipeline.}

\begin{table}[t]
\centering
\small
\caption{\rw{Zero-shot resident-specific activity recognition results on the original multi-resident ARAS data. The LLM receives only the shared sensor summary and the contextual prior of each resident. Results are reported as mean$\pm$standard deviation in percentage points (\%).}}
\label{tab:aras_multi_resident_zeroshot}
\resizebox{0.7\columnwidth}{!}{%
\begin{tabular}{llccc}
\toprule
\rw{\textbf{Home}} & \rw{\textbf{Resident}} & \rw{\textbf{Accuracy}} & \rw{\textbf{Macro F1}} & \rw{\textbf{Weighted F1}} \\
\midrule
\rw{Home A} & \rw{R1} & \rw{$25.91 \pm 7.12$} & \rw{$13.39 \pm 2.76$} & \rw{$26.21 \pm 7.73$} \\
\rw{Home A} & \rw{R2} & \rw{$21.01 \pm 11.30$} & \rw{$9.42 \pm 3.99$} & \rw{$22.73 \pm 11.45$} \\
\rw{Home B} & \rw{R1} & \rw{$34.63 \pm 11.46$} & \rw{$8.29 \pm 3.17$} & \rw{$35.93 \pm 11.50$} \\
\rw{Home B} & \rw{R2} & \rw{$31.24 \pm 11.47$} & \rw{$5.92 \pm 2.03$} & \rw{$30.96 \pm 11.47$} \\
\bottomrule
\end{tabular}%
}
\end{table}

\rw{Overall, the zero-shot multi-resident results shown in Table~\ref{tab:aras_multi_resident_zeroshot} are poor across both homes and both residents, particularly in terms of macro F1. This outcome is expected because the ambient sensors observe the shared home environment but provide no information about which resident triggered each sensor event. When two residents perform activities concurrently or in close temporal proximity, their sensor events become interleaved, making it difficult for the model to associate the observed evidence with the correct resident.}

\rw{Although resident-specific contextual priors provide some information about the plausibility of different activities, they cannot reliably resolve this fundamental ambiguity in the sensor stream. The model must simultaneously infer both the activity and the resident responsible for the evidence without explicit identity information, leading to substantial cross-resident interference. These results highlight a limitation of directly extending \projectname{} to multi-resident environments and suggest that reliable multi-resident recognition requires resident-aware sensing or an explicit mechanism for assigning sensor events to individual residents.}
\section{Reproducibility Support}
\rw{This appendix consolidates the implementation details needed to reproduce our experiments. It reports the LLM and baseline configurations, the representation of user-specific contextual priors, and the prompt templates used in the cross-reference and multi-scale refinement stages of \projectname{}.}

\rw{\subsection{ LLM Configurations}}
\label{app:llm-configurations} 
\rw{To support reproducibility, we report the language models and inference parameters used in our experiments. 
We evaluated three models: GPT-5-mini, Qwen3-32B, and Llama-3.1-8B. GPT-5-mini was accessed through the OpenAI API, whereas Qwen3-32B and Llama-3.1-8B were deployed locally. 
For Qwen3-32B, we evaluated both reasoning-enabled and reasoning-disabled configurations. }

\rw{For GPT-5-mini, we used the fixed model snapshot \texttt{gpt-5-mini-2025-08-07}. 
The temperature was set to 1, which was the only supported value for the model configuration used in our experiments, and top-$p$ was set to 1. 
The maximum completion length was set to 4,096 tokens. 
The implementation initially passed \texttt{max\_tokens = 4096}. If this parameter was rejected by the GPT-5 endpoint, the same request was retried using \texttt{max\_completion\_tokens=4096}. Frequency and presence penalties were not explicitly specified, and therefore used the OpenAI API default values of 0. No random seed was specified. }

\rw{The two open-weight models were served locally through Ollama on a single NVIDIA RTX A6000 GPU. 
Qwen3-32B was deployed using the Ollama tag \texttt{qwen3:32b} and manifest digest 
\texttt{\seqsplit{030ee887880fc378860c2dd35101da424377520441ae4bfe7be6deff8ade7840}}.
The deployed model contained 32.8 billion parameters and used \texttt{Q4\_K\_M} quantization. For both its reasoning-enabled and reasoning-disabled configurations, the temperature was set to 0, top-$p$ to 0.95, top-$k$ to 20, the repetition penalty to 1.0, and the maximum generation length to 4,096 tokens. The presence penalty, frequency penalty, and random seed were not explicitly specified. The reasoning configuration was the only difference between the two Qwen3-32B settings. }

\rw{Llama-3.1-8B was deployed using the Ollama tag \texttt{llama3.1:8b} and manifest digest \texttt{\seqsplit{46e0c10c039e019119339687c3c1757cc81b9da49709a3b3924863ba87ca666e}}. The deployed model contained 8.0 billion parameters and used \texttt{Q4\_K\_M} quantization. 
The temperature was set to 0, top-$p$ to 0.9, top-$k$ to 40, the repetition penalty to 1.1, and the maximum generation length to 4,096 tokens. The presence penalty, frequency penalty, and random seed were not explicitly specified.}

\rw{All models were evaluated using fixed prompt templates and semantically equivalent task inputs. The prompts explicitly included the instruction \texttt{Be deterministic: the same input must yield the same output}. This instruction was intended to encourage consistent predictions. Because random seeds were not explicitly controlled and some inference parameters relied on API, model, or runtime defaults, exact token-level determinism was not assumed.}

\rw{\subsection{ Baseline Configurations}}

\label{app:baseline-configurations} 
\rw{We report the architecture and configurations of all neural baselines to support reproducibility. Unless otherwise stated, the same configuration is used across evaluation folds and datasets, while the output dimension is adjusted according to the activity label space of each dataset.}

\paragraph{\rw{Event-window baselines.}}
\rw{For the CASAS, Orange4Home, and UCI-ADL experiments, E-FCN, DeepCASAS, and TDOST operate on the event windows. E-FCN applies three one-dimensional convolutional layers with 128, 256, and 128 filters and kernel sizes of 8, 5, and 3, respectively. Each convolutional layer is followed by batch normalization and ReLU activation. Global average pooling and dropout with a rate of 0.5 are then applied before the final softmax classification layer.}
\rw{DeepCASAS-BiLSTM applies a single-layer bidirectional LSTM with a hidden size of 64 in each direction. The concatenated representation is passed to a linear classification layer.}
\rw{TDOST uses pre-trained DistilRoBERTa to generate embeddings from semantic event representation. The embeddings are processed by a two-layer bidirectional LSTM with a hidden size of 64 in each direction and dropout of 0.2. The resulting representation is passed to a linear classification layer.}

\paragraph{\rw{Duration-window baselines.}}
\rw{For MARBLE, all models use 6-second windows. DeepConvLSTM contains four one-dimensional convolutional layers, each with 64 filters and a kernel size of 3, followed by two LSTM layers with a hidden size of 128 and recurrent dropout of 0.2. A dropout layer with a rate of 0.5 is applied before classification.}
\rw{DualLSTM uses separate branches for wearable and environmental sensing. The wearable branch contains four two-dimensional convolutional layers with 64, 64, 128, and 128 output channels, respectively, using $2\times2$ kernels. The convolutional representation is processed by an LSTM with a hidden size of 256 and projected to a 64-dimensional feature vector. The environmental branch uses a two-dimensional convolutional layer with 8 output channels and a $2\times2$ kernel, followed by an LSTM with a hidden size of 128 and a 32-dimensional projection layer. The resulting 64- and 32-dimensional branch representations are concatenated and processed by a fusion head with dimensions $96\rightarrow64\rightarrow32\rightarrow C$, where $C$ denotes the number of activity classes.}

\subsection{Contextual Prior Format}
In this section, we describe the semi-structured format we used to construct user-specific contextual priors.
The contextual prior $C$ consists of four components. The first is a layout description, which characterizes the living environment in natural language. This description includes the house type (e.g., a one-bedroom-one-bathroom apartment or a studio), the number and types of rooms, and the activity-relevant furniture, appliances, or devices available in the home. In our implementation, the layout description is represented using a structured template:

\begin{quote}
\ttfamily
User is living in a <HOUSE\_TYPE>.\\
There is a <ROOM\_TYPE\_1> with <OBJECT\_LIST\_1>.\\
There is a <ROOM\_TYPE\_2> with <OBJECT\_LIST\_2>.
\end{quote}

The second component describes the typical times of common activities. Since some activities may contain more specific sub-activities (e.g., cooking may be categorized as preparing breakfast or preparing dinner), and their temporal boundaries are not always exact, we represent this information using a semi-structured format:
\begin{quote}
\ttfamily
<ACTIVITY>: usually occurs around <TIME\_1>. (may also occur around <TIME\_2>)\\
<ACTIVITY>: usually starts around <START\_TIME> and ends around <END\_TIME>.\\
<PARENT\_ACTIVITY> / <SUBACTIVITY>: usually occurs around <TIME>.
\end{quote}

The third component describes the typical locations of common activities. Since activities may also have multiple possible locations, or contain sub-activities associated with different locations, this component is represented using a semi-structured format:
\begin{quote}
\ttfamily
<ACTIVITY>: usually occurs in <LOCATION\_1>. (may also occur in <LOCATION\_2>)\\
<ACTIVITY>: usually occurs around <OBJECT> in <LOCATION>.\\
<PARENT\_ACTIVITY> / <SUBACTIVITY>: usually occurs in <LOCATION>.
\end{quote}

The fourth component captures user-specific habits, including common activity transitions, repeated patterns, and preferences. Since these priors are often more diverse and less regular than time or location patterns, we do not enforce a strict template for this component. Instead, we represent it using a loosely structured format, which allows multiple transition patterns, preferences, and exceptions to be expressed in flexible ways. Example entries include:
\begin{quote}
\ttfamily
<ACTIVITY\_1>: usually occurs after <ACTIVITY\_2>. \\
The user usually <PREFERENCE>. \\
The user occasionally <PREFERENCE\_OR\_EXCEPTION>.
\end{quote}

\subsection{Prompt Templates in ~\projectname{}}

In this section, we list the prompt template we used in ~\projectname{}.
The following prompt is used for the cross-reference stage. During actual execution, it will be combined with the corresponding input resource and raw data.
\begin{tcolorbox}[title=Prompt for Cross-Reference,breakable, colback=gray!5, colframe=black]
\begin{Verbatim}[breaklines=true, breakanywhere=true, fontsize=\small]
You are a deterministic multi-source sensor fusion engine for smart-home human activity recognition.
Your task is to infer the most likely activity label for each minute in a {{WINDOW_SIZE}}-minute window for home {{HOME_ID}}.
You must produce exactly {{WINDOW_SIZE}} minute-level predictions, one for each minute.
## INPUT
### Sensor Sources
The available sensor sources and their meanings are:
{correction_sources_text}
### Sensor Data
The sensor observations are provided as structured JSON:
{{SENSOR_BLOCK}}
### User Context
The user routine profile, habits, and contextual priors are:
{{CONTEXT_BLOCK}}
## INTERNAL REASONING PROCESS
For each minute, reason step by step using the following procedure and output labels from {POSSIBLE_ACTIVITY_LABELS}.
[IMPORTANT] Select the single best final label and keep the strongest alternative label based on the following criteria:
    - 1. Use location and time-of-day context to identify plausible activity labels.
    - 2. If there's direct interaction evidence (e.g., stove interaction->Cook, bathroom->Bed_to_Toilet), use it to determine the best label.
    - 3. If there's no direct evidence:
        - 3.1. Use the location to determine the possible label candidates.
        - 3.2. Use TDOST as a coarse hypothesis.
        - 3.3. Use the user routine/context as a prior.
    - 4. Only output "Sleep" if the wearable shows sleep. Do not output "Sleep" if the wearable is not sleeping.
    - 5. Do not output "Leave_Home" if the location is home.
    - 6. If no clear evidence, prefer a broader and more conservative label: "Other".
- - Write a short evidence-based reason mentioning the strongest direct evidence.
Rules:
- Output exactly {{WINDOW_SIZE}} items
- Strictly follow the priority rules and the reasoning process.
## OUTPUT FORMAT (STRICT JSON ONLY)
Output timestamp, labels, and reason for each minute, and do not output any other information.
Return only the final JSON object.
{
    "home_id": "...",
    "minute_predictions": [
        {
            "timestamp": "...",
            "labels": ["...", "..."],
            "reason": "...",
        },
    ]
}
\end{Verbatim}
\end{tcolorbox}

The following prompt is used for the multi-scale refinement stage. During actual execution, it will be combined with the fused results, inference history, and context data.
\begin{tcolorbox}[title=Prompt for Multi-Scale Refinement,breakable, colback=gray!5, colframe=black]
\begin{Verbatim}[breaklines=true, breakanywhere=true, fontsize=\small]
You are a deterministic temporal calibration engine for smart-home human activity recognition.
Your task is to revise the activity interpretation for the current refinement scope using minute-level predictions, prior activity history, and user context.
Important:
The current window is only the scope of evidence being refined, not a hard activity boundary.
A real activity may start before this window and continue through it, or begin inside this window and continue beyond it. Therefore, your goal is not to preserve window-local segmentation, but to produce the best revised segmentation for this scope. The output should be treated as the newest version of the activity results for this time range and may overwrite older fragmented results.
## INPUT DATA
### MINUTE-LEVEL PREDICTIONS
These are the minute-level predictions.
Each minute prediction contains:
- timestamp: the real clock time for that minute
- labels: a 2-item list where labels[0] is the primary label and labels[1] is the strongest alternative label
- reason: short explanation of why the label is chosen
The predictions are:
{{MINUTE_PREDICTION_BLOCK}}
### ACTIVITY_HISTORY
The recognized activity history immediately before this window is:
{{ACTIVITY_HISTORY_BLOCK}}
### USER_CONTEXT
The user routine profile, habits, and contextual priors are:
{{CONTEXT_BLOCK}}
### ALLOWED LABELS
Use only the following labels:
{POSSIBLE_ACTIVITY_LABELS}
## INTERNAL REFINEMENT PROCESS
### STEP 1: Local Temporal Smoothing
Refine the minute-level predictions within the current scope by enforcing local consistency.
- Temporal stabilization: prefer continuous activity segments over rapid label fluctuations, and smooth isolated conflicting minutes when surrounding evidence supports a stable activity.
- Duration awareness: treat unusually short isolated segments as likely noise unless they are clearly and consistently supported.
- Contextual consistency: when local evidence is ambiguous, prefer labels that are more consistent with nearby predictions and immediate context.
- Only preserve short high-confidence fragments when they are not strongly contradicted by the surrounding pattern.
### STEP 2: Cross-Window Continuity
Refine the boundary between ACTIVITY_HISTORY and the current scope by enforcing continuity across adjacent windows.
- Temporal continuity: do not treat the window boundary as a true activity boundary by default, and merge the same activity across the boundary when continuity is supported.
- Transition regularity: prefer transitions that are consistent with recent activity flow and reject abrupt boundary changes unless there is strong evidence for a true transition.
- Duration plausibility: consider whether the segment length is appropriate for the inferred activity type, and avoid assigning activities to durations that are highly implausible without strong supporting evidence.
- Activity relationship plausibility: consider whether neighboring segments form a behaviorally plausible sequence, and suppress brief inserted segments that strongly conflict with the surrounding activity unless they are clearly supported.
- Boundary correction: remove boundaries caused only by prior fragmentation or weak interruptions, and preserve a boundary only when the evidence clearly supports an activity change.
### STEP 3: Context and Routine Alignment
Refine the current interpretation using USER_CONTEXT as a soft long-term behavioral prior.
- Routine consistency: when evidence is weak or ambiguous, prefer activities that are more consistent with the user's routine and time-of-day pattern.
- Contextual plausibility: use time, location, and recent activity flow to resolve ambiguity and suppress context-conflicted activities that lack strong support.
- Activity-type plausibility: consider whether the inferred activity is behaviorally plausible given the surrounding activities, time-of-day, and user routine, and avoid interpretations that are locally possible but globally inconsistent without strong evidence.
- Context-conditioned specification: when supported by both local evidence and context, choose the most appropriate label within the allowed label set.
### STEP 4: Versioned Output
Return only the final revised segmentation for the current refinement scope.
- Latest version: treat the output as the newest version for the overlapping timestamps.
- Overwrite policy: use the revised result to replace older fragmented results rather than preserve previous local boundaries.
- Final segmentation only: do not output intermediate reasoning or intermediate segmentations.
## CRITICAL REQUIREMENTS
- Each activity must contain: home_id, start_timestamp, end_timestamp, version, label
- Combine consecutive minutes with the same final label into one segment
- Activities must be non-overlapping and cover the entire current refinement scope
- Use only these labels: {POSSIBLE_ACTIVITY_LABELS}
- Be deterministic: the same input must yield the same output
## OUTPUT FORMAT (STRICT JSON ONLY)
Return only a valid JSON object, no markdown, no code fences, no extra text.
{
  "revised_activities": [
    {
      "home_id": "...",
      "start_timestamp": "...",
      "end_timestamp": "...",
      "version": int,
      "label": "..."
    }
  ]
}
\end{Verbatim}
\end{tcolorbox}

    \vspace*{-0.5em}

\section{Detailed Results for Benchmark Evaluation}
\label{app:detailed_results}

\rw{\subsection{Additional Event-Window- and Activity-Segment-Based Evaluation}}
\label{sec:appendix_window_segment_evaluation}

\rw{In addition to the time-based and temporal evaluation presented in the main
paper, we report event-window-based and activity-segment-based
classification metrics. }

\rw{For the window-based evaluation, the minute-level predictions generated by
\projectname{} are projected onto the corresponding evaluation windows, and
one activity label is assigned to each window. Table~\ref{tab:combined_cross_domain}
reports accuracy, weighted F1, and macro F1 under the in-domain and cross-domain
settings. For the activity-segment-based evaluation, predictions within each
activity segment are aggregated into a single segment-based label.
Table~\ref{tab:combined_cross_domain} reports the corresponding results.}

\rw{Overall, \projectname{} demonstrates more balanced performance across activity
classes. It achieves the highest macro F1 in most of the evaluated settings at
both the window and activity-segment levels, with particularly clear
improvements in the Milan-to-Aruba, Kyoto7-to-Aruba, and Kyoto7-to-Milan
transfer settings. The segment-based results show a similar trend, indicating
that the improvements are not limited to minute-level predictions but
also extend to the recognition of complete activity episodes.}

\rw{It is important to note that window-based evaluation gives greater
weight to activities associated with dense sensor activations, because these
activities generate more evaluation windows. However, frequent sensor
triggering does not necessarily indicate that an activity occurs more often or
occupies a larger proportion of daily life. Consequently, the resulting
accuracy and weighted F1 may reflect the distribution of sensor events rather
than the actual temporal distribution of activities. For example, in the
Aruba-to-Kyoto7 setting, TDOST obtains substantially higher accuracy and
weighted F1 because its predictions are largely concentrated on the
sensor-dense and dominant \textit{Other} class, whereas \projectname{} achieves
a higher macro F1 by maintaining more balanced performance across activity
classes. Therefore, although window-based metrics remain useful for comparison
with conventional window-based HAR methods, they are not ideal for evaluating
continuous activity recognition, where performance should reflect activities
as they occur over time rather than how frequently they trigger sensors.
}

\begin{table*}[htbp]
\centering
\small
\caption{\rw{Event-window-based and activity-segment-based classification performance under in-domain and cross-domain settings. All metrics are reported as mean $\pm$ standard deviation across three evaluation folds. The best result is in bold.}}
\label{tab:combined_cross_domain}
\resizebox{0.9\textwidth}{!}{%
\begin{tabular}{llccc|ccc}
\toprule
& & \multicolumn{3}{c|}{\textbf{\rw{Window based}}}
& \multicolumn{3}{c}{\textbf{\rw{Segment based}}} \\
\cmidrule(lr){3-5}\cmidrule(lr){6-8}
\textbf{\rw{Setting}}
& \textbf{\rw{Model}}
& \textbf{\rw{Accuracy}}
& \textbf{\rw{Weighted F1}}
& \textbf{\rw{Macro F1}}
& \textbf{\rw{Accuracy}}
& \textbf{\rw{Weighted F1}}
& \textbf{\rw{Macro F1}} \\
\midrule


\multirow{4}{*}{\rw{CASAS Aruba}}
& \rw{E-FCN} & \rw{$74.72 \pm 1.95$} & \rw{$73.90 \pm 2.78$} & \rw{$48.80 \pm 2.66$} & \rw{$71.92 \pm 3.40$} & \rw{$68.84 \pm 4.33$} & \rw{$48.87 \pm 5.19$} \\
& \rw{DeepCASAS} & \rw{$\mathbf{77.50 \pm 2.52}$} & \rw{$\mathbf{77.00 \pm 2.88}$} & \rw{$\mathbf{54.07 \pm 2.36}$} & \rw{$73.50 \pm 4.02$} & \rw{$70.75 \pm 4.59$} & \rw{$\mathbf{53.14 \pm 5.78}$} \\
& \rw{TDOST} & \rw{$75.67 \pm 3.01$} & \rw{$74.27 \pm 4.11$} & \rw{$46.41 \pm 0.14$} & \rw{$\mathbf{73.68 \pm 4.78}$} & \rw{$70.29 \pm 5.51$} & \rw{$45.92 \pm 1.81$} \\
& \rw{\projectname{}} & \rw{$72.35 \pm 3.75$} & \rw{$73.01 \pm 3.41$} & \rw{$43.07 \pm 2.80$} & \rw{$73.66 \pm 3.13$} & \rw{$\mathbf{73.12 \pm 2.65}$} & \rw{$46.16 \pm 8.73$} \\
\midrule

\multirow{4}{*}{\rw{CASAS Milan}}
& \rw{E-FCN} & \rw{$\mathbf{37.05 \pm 4.83}$} & \rw{$\mathbf{36.91 \pm 3.63}$} & \rw{$13.22 \pm 0.61$} & \rw{$\mathbf{33.10 \pm 4.13}$} & \rw{$30.90 \pm 3.24$} & \rw{$13.05 \pm 0.88$} \\
& \rw{DeepCASAS} & \rw{$35.97 \pm 4.27$} & \rw{$36.37 \pm 3.49$} & \rw{$13.24 \pm 0.77$} & \rw{$31.50 \pm 2.96$} & \rw{$29.70 \pm 2.11$} & \rw{$11.52 \pm 0.89$} \\
& \rw{TDOST} & \rw{$36.08 \pm 4.39$} & \rw{$36.38 \pm 3.58$} & \rw{$12.54 \pm 0.59$} & \rw{$31.88 \pm 2.29$} & \rw{$30.20 \pm 1.61$} & \rw{$12.26 \pm 0.86$} \\
& \rw{\projectname{}} & \rw{$34.76 \pm 3.07$} & \rw{$33.78 \pm 3.91$} & \rw{$\mathbf{24.26 \pm 0.68}$} & \rw{$29.64 \pm 2.92$} & \rw{$\mathbf{33.18 \pm 4.51}$} & \rw{$\mathbf{24.82 \pm 1.94}$} \\
\midrule

\multirow{4}{*}{\rw{CASAS Kyoto7}}
& \rw{E-FCN} & \rw{$12.77 \pm 5.03$} & \rw{$15.72 \pm 4.85$} & \rw{$8.52 \pm 3.32$} & \rw{$18.42 \pm 3.38$} & \rw{$19.72 \pm 2.18$} & \rw{$8.24 \pm 1.50$} \\
& \rw{DeepCASAS} & \rw{$13.66 \pm 7.57$} & \rw{$16.24 \pm 8.03$} & \rw{$8.94 \pm 4.72$} & \rw{$19.04 \pm 2.79$} & \rw{$20.46 \pm 2.79$} & \rw{$8.53 \pm 1.55$} \\
& \rw{TDOST} & \rw{$9.27 \pm 2.92$} & \rw{$11.67 \pm 2.31$} & \rw{$6.10 \pm 1.62$} & \rw{$13.33 \pm 2.24$} & \rw{$15.11 \pm 1.63$} & \rw{$5.99 \pm 0.54$} \\
& \rw{\projectname{}} & \rw{$\mathbf{58.30 \pm 5.30}$} & \rw{$\mathbf{58.93 \pm 6.99}$} & \rw{$\mathbf{49.50 \pm 6.89}$} & \rw{$\mathbf{26.92 \pm 6.99}$} & \rw{$\mathbf{34.11 \pm 10.28}$} & \rw{$\mathbf{12.20 \pm 5.06}$} \\
\midrule

\multirow{4}{*}{\rw{Orange4Home}}
& \rw{E-FCN} & \rw{$90.37 \pm 0.54$} & \rw{$89.51 \pm 0.81$} & \rw{$61.94 \pm 2.22$} & \rw{$79.85 \pm 1.10$} & \rw{$77.44 \pm 1.72$} & \rw{$73.91 \pm 0.55$} \\
& \rw{DeepCASAS} & \rw{$91.24 \pm 0.84$} & \rw{$91.21 \pm 1.19$} & \rw{$67.33 \pm 2.49$} & \rw{$73.35 \pm 1.15$} & \rw{$71.50 \pm 2.51$} & \rw{$68.11 \pm 2.34$} \\
& \rw{TDOST} & \rw{$92.65 \pm 0.82$} & \rw{$92.49 \pm 1.06$} & \rw{$\mathbf{69.59 \pm 3.01}$} & \rw{$76.55 \pm 3.71$} & \rw{$73.76 \pm 5.10$} & \rw{$68.08 \pm 4.94$} \\
& \rw{\projectname{}} & \rw{$\mathbf{92.96 \pm 1.28}$} & \rw{$\mathbf{92.64 \pm 1.54}$} & \rw{$66.94 \pm 1.41$} & \rw{$\mathbf{83.16 \pm 2.18}$} & \rw{$\mathbf{81.41 \pm 2.78}$} & \rw{$\mathbf{78.09 \pm 3.16}$} \\
\midrule

\multirow{4}{*}{\rw{UCI-ADL Home A}}
& \rw{E-FCN} & \rw{$\mathbf{92.02 \pm 3.29}$} & \rw{$\mathbf{90.55 \pm 4.15}$} & \rw{$88.14 \pm 4.06$} & \rw{$91.88 \pm 3.80$} & \rw{$91.04 \pm 3.24$} & \rw{$86.79 \pm 1.03$} \\
& \rw{DeepCASAS} & \rw{$73.80 \pm 15.29$} & \rw{$73.28 \pm 14.99$} & \rw{$67.40 \pm 10.65$} & \rw{$77.68 \pm 12.62$} & \rw{$76.97 \pm 12.37$} & \rw{$69.09 \pm 10.39$} \\
& \rw{TDOST} & \rw{$82.26 \pm 14.88$} & \rw{$78.03 \pm 19.88$} & \rw{$73.05 \pm 22.00$} & \rw{$82.87 \pm 14.29$} & \rw{$78.48 \pm 19.14$} & \rw{$72.53 \pm 20.93$} \\
& \rw{\projectname{}} & \rw{$89.82 \pm 1.65$} & \rw{$89.22 \pm 1.64$} & \rw{$\mathbf{90.59 \pm 2.75}$} & \rw{$\mathbf{96.44 \pm 1.83}$} & \rw{$\mathbf{96.50 \pm 1.74}$} & \rw{$\mathbf{96.00 \pm 2.91}$} \\
\midrule

\multirow{4}{*}{\rw{UCI-ADL Home B}}
& \rw{E-FCN} & \rw{$84.99 \pm 0.60$} & \rw{$84.23 \pm 0.10$} & \rw{$75.92 \pm 0.25$} & \rw{$84.87 \pm 0.72$} & \rw{$84.10 \pm 0.36$} & \rw{$76.00 \pm 0.54$} \\
& \rw{DeepCASAS} & \rw{$81.54 \pm 1.48$} & \rw{$81.50 \pm 1.25$} & \rw{$75.83 \pm 1.31$} & \rw{$81.58 \pm 1.10$} & \rw{$81.55 \pm 0.88$} & \rw{$76.05 \pm 1.01$} \\
& \rw{TDOST} & \rw{$81.54 \pm 0.71$} & \rw{$79.91 \pm 0.66$} & \rw{$64.15 \pm 1.83$} & \rw{$81.20 \pm 0.79$} & \rw{$79.56 \pm 0.75$} & \rw{$63.99 \pm 1.85$} \\
& \rw{\projectname{}} & \rw{$\mathbf{86.20 \pm 2.04}$} & \rw{$\mathbf{85.43 \pm 1.42}$} & \rw{$\mathbf{76.23 \pm 0.83}$} & \rw{$\mathbf{86.29 \pm 2.09}$} & \rw{$\mathbf{85.53 \pm 1.45}$} & \rw{$\mathbf{76.62 \pm 0.83}$} \\
\midrule

\multicolumn{8}{c}{\textbf{\rw{Cross-domain evaluation}}} \\
\midrule

\multirow{4}{*}{\rw{\shortstack{Milan\\$\downarrow$\\Aruba}}}
& \rw{E-FCN} & \rw{$48.30 \pm 3.08$} & \rw{$40.42 \pm 2.31$} & \rw{$11.41 \pm 0.54$} & \rw{$45.03 \pm 1.09$} & \rw{$33.75 \pm 0.26$} & \rw{$13.45 \pm 1.73$} \\
& \rw{DeepCASAS} & \rw{$35.37 \pm 1.42$} & \rw{$32.93 \pm 1.87$} & \rw{$8.07 \pm 0.42$} & \rw{$35.22 \pm 0.75$} & \rw{$28.88 \pm 0.40$} & \rw{$8.49 \pm 0.37$} \\
& \rw{TDOST} & \rw{$58.07 \pm 4.01$} & \rw{$49.52 \pm 4.41$} & \rw{$19.13 \pm 0.88$} & \rw{$54.06 \pm 1.39$} & \rw{$43.17 \pm 1.06$} & \rw{$22.75 \pm 0.73$} \\
& \rw{\projectname{}} & \rw{$\mathbf{59.04 \pm 2.15}$} & \rw{$\mathbf{60.43 \pm 2.15}$} & \rw{$\mathbf{36.94 \pm 2.04}$} & \rw{$\mathbf{57.61 \pm 4.42}$} & \rw{$\mathbf{57.22 \pm 4.21}$} & \rw{$\mathbf{43.39 \pm 4.79}$} \\
\midrule

\multirow{4}{*}{\rw{\shortstack{Aruba\\$\downarrow$\\Milan}}}
& \rw{E-FCN} & \rw{$46.44 \pm 4.58$} & \rw{$\mathbf{40.12 \pm 4.49}$} & \rw{$11.86 \pm 0.70$} & \rw{$42.60 \pm 0.32$} & \rw{$32.02 \pm 1.05$} & \rw{$10.12 \pm 0.48$} \\
& \rw{DeepCASAS} & \rw{$47.65 \pm 2.77$} & \rw{$39.86 \pm 4.02$} & \rw{$11.14 \pm 0.44$} & \rw{$\mathbf{42.86 \pm 1.55}$} & \rw{$31.68 \pm 1.07$} & \rw{$9.61 \pm 0.34$} \\
& \rw{TDOST} & \rw{$\mathbf{48.68 \pm 3.39}$} & \rw{$39.87 \pm 4.09$} & \rw{$11.43 \pm 1.10$} & \rw{$41.51 \pm 3.27$} & \rw{$30.98 \pm 1.94$} & \rw{$9.57 \pm 0.23$} \\
& \rw{\projectname{}} & \rw{$39.43 \pm 4.52$} & \rw{$37.93 \pm 5.01$} & \rw{$\mathbf{28.15 \pm 1.40}$} & \rw{$32.37 \pm 2.08$} & \rw{$\mathbf{35.37 \pm 0.81}$} & \rw{$\mathbf{24.61 \pm 3.40}$} \\
\midrule

\multirow{4}{*}{\rw{\shortstack{Aruba\\$\downarrow$\\Kyoto7}}}
& \rw{E-FCN} & \rw{$19.15 \pm 2.40$} & \rw{$27.22 \pm 3.66$} & \rw{$9.04 \pm 1.21$} & \rw{$13.41 \pm 0.97$} & \rw{$16.25 \pm 1.87$} & \rw{$7.31 \pm 0.68$} \\
& \rw{DeepCASAS} & \rw{$17.04 \pm 3.52$} & \rw{$24.16 \pm 5.06$} & \rw{$7.84 \pm 1.00$} & \rw{$17.14 \pm 2.17$} & \rw{$19.69 \pm 2.69$} & \rw{$9.22 \pm 1.15$} \\
& \rw{TDOST} & \rw{$\mathbf{73.80 \pm 6.86}$} & \rw{$\mathbf{62.90 \pm 9.15}$} & \rw{$\mathbf{23.70 \pm 4.55}$} & \rw{$\mathbf{45.73 \pm 0.54}$} & \rw{$\mathbf{28.70 \pm 0.57}$} & \rw{$10.46 \pm 0.09$} \\
& \rw{\projectname{}} & \rw{$19.46 \pm 2.37$} & \rw{$16.31 \pm 2.43$} & \rw{$19.78 \pm 0.85$} & \rw{$29.44 \pm 11.89$} & \rw{$27.85 \pm 15.51$} & \rw{$\mathbf{12.36 \pm 7.39}$} \\
\midrule

\multirow{4}{*}{\rw{\shortstack{Milan\\$\downarrow$\\Kyoto7}}}
& \rw{E-FCN} & \rw{$\mathbf{36.99 \pm 4.99}$} & \rw{$\mathbf{44.88 \pm 5.87}$} & \rw{$13.47 \pm 1.67$} & \rw{$25.90 \pm 4.69$} & \rw{$\mathbf{24.28 \pm 3.43}$} & \rw{$8.83 \pm 1.15$} \\
& \rw{DeepCASAS} & \rw{$8.69 \pm 1.63$} & \rw{$15.17 \pm 2.80$} & \rw{$5.09 \pm 1.07$} & \rw{$6.06 \pm 1.97$} & \rw{$10.14 \pm 3.58$} & \rw{$3.81 \pm 1.39$} \\
& \rw{TDOST} & \rw{$14.30 \pm 5.58$} & \rw{$16.09 \pm 5.11$} & \rw{$9.46 \pm 5.09$} & \rw{$17.80 \pm 4.08$} & \rw{$12.64 \pm 5.76$} & \rw{$7.30 \pm 2.21$} \\
& \rw{\projectname{}} & \rw{$16.53 \pm 3.47$} & \rw{$14.60 \pm 2.57$} & \rw{$\mathbf{21.12 \pm 4.72}$} & \rw{$\mathbf{26.50 \pm 3.22}$} & \rw{$23.01 \pm 3.10$} & \rw{$\mathbf{9.33 \pm 1.64}$} \\
\midrule

\multirow{4}{*}{\rw{\shortstack{Kyoto7\\$\downarrow$\\Aruba}}}
& \rw{E-FCN} & \rw{$17.08 \pm 1.17$} & \rw{$17.02 \pm 0.93$} & \rw{$7.55 \pm 0.15$} & \rw{$18.25 \pm 1.00$} & \rw{$16.32 \pm 0.59$} & \rw{$10.35 \pm 0.43$} \\
& \rw{DeepCASAS} & \rw{$13.63 \pm 0.65$} & \rw{$13.03 \pm 0.96$} & \rw{$9.24 \pm 0.38$} & \rw{$14.69 \pm 0.54$} & \rw{$13.16 \pm 0.84$} & \rw{$9.55 \pm 0.72$} \\
& \rw{TDOST} & \rw{$14.37 \pm 0.54$} & \rw{$14.39 \pm 0.18$} & \rw{$8.95 \pm 0.71$} & \rw{$14.45 \pm 1.22$} & \rw{$11.82 \pm 1.57$} & \rw{$8.82 \pm 1.38$} \\
& \rw{\projectname{}} & \rw{$\mathbf{30.80 \pm 3.39}$} & \rw{$\mathbf{29.91 \pm 3.44}$} & \rw{$\mathbf{24.24 \pm 3.42}$} & \rw{$\mathbf{34.54 \pm 2.15}$} & \rw{$\mathbf{35.31 \pm 2.00}$} & \rw{$\mathbf{23.64 \pm 2.91}$} \\
\midrule

\multirow{4}{*}{\rw{\shortstack{Kyoto7\\$\downarrow$\\Milan}}}
& \rw{E-FCN} & \rw{$16.61 \pm 1.80$} & \rw{$15.36 \pm 1.38$} & \rw{$8.31 \pm 1.04$} & \rw{$17.55 \pm 1.84$} & \rw{$14.78 \pm 1.09$} & \rw{$11.54 \pm 0.68$} \\
& \rw{DeepCASAS} & \rw{$19.23 \pm 3.34$} & \rw{$17.43 \pm 3.21$} & \rw{$12.15 \pm 2.03$} & \rw{$19.39 \pm 1.59$} & \rw{$16.05 \pm 1.60$} & \rw{$12.73 \pm 1.25$} \\
& \rw{TDOST} & \rw{$19.56 \pm 2.59$} & \rw{$14.96 \pm 1.68$} & \rw{$9.81 \pm 0.35$} & \rw{$21.92 \pm 2.74$} & \rw{$14.28 \pm 2.38$} & \rw{$10.27 \pm 0.91$} \\
& \rw{\projectname{}} & \rw{$\mathbf{28.90 \pm 3.64}$} & \rw{$\mathbf{26.30 \pm 2.55}$} & \rw{$\mathbf{24.65 \pm 2.40}$} & \rw{$\mathbf{31.25 \pm 2.70}$} & \rw{$\mathbf{30.49 \pm 2.77}$} & \rw{$\mathbf{31.12 \pm 3.27}$} \\
\bottomrule
\end{tabular}%
}
\end{table*}

\subsection{Per-Class Results on Benchmarks}
To provide a more complete view of model behavior, we report per-class (label-level) F1 scores for all benchmarks. 

\begin{table*}[htbp]
\centering
\scriptsize
\caption{Per-class F1 scores (\%) of \projectname{} and baseline methods under in-domain and cross-domain settings on Aruba, Milan, and Kyoto7. Results are reported as mean$\pm$std over repeated runs. The symbol / denotes labels that are not available in the corresponding dataset.}
\label{tab:transfer_class_results_harmi}
\resizebox{\textwidth}{!}{
\begin{tabular}{llcccccccc}
\toprule
\textbf{Setting} & \textbf{Model} & \textbf{Other} & \textbf{Relax} & \textbf{Cook} & \textbf{Leave\_Home} & \textbf{Sleep} & \textbf{Eat} & \textbf{Work} & \textbf{Bed\_to\_Toilet} \\
\midrule

\multirow{4}{*}{\shortstack{Aruba\\$\downarrow$\\Aruba}}
& E-FCN     & 80.13$\pm$2.35 & 91.03$\pm$3.22 & 60.83$\pm$6.98 & 0.00$\pm$0.00 & 90.33$\pm$3.06 & 48.03$\pm$2.67 & 64.28$\pm$10.95 & 0.00$\pm$0.00 \\
& DeepCASAS & \textbf{84.97$\pm$1.15} & \textbf{92.42$\pm$3.56} & 67.46$\pm$7.59 & 0.00$\pm$0.00 & \textbf{94.87$\pm$1.19} & 57.45$\pm$6.12 & \textbf{72.94$\pm$10.74} & \textbf{5.27$\pm$7.45} \\
& TDOST     & 84.38$\pm$1.19 & 88.49$\pm$2.09 & 67.88$\pm$8.53 & 0.00$\pm$0.00 & 94.37$\pm$1.72 & \textbf{63.74$\pm$10.27} & 15.02$\pm$7.65 & 0.00$\pm$0.00 \\
& \projectname{}     & 82.73$\pm$2.72 & 90.37$\pm$1.87 & \textbf{72.33$\pm$6.48} & \textbf{1.55$\pm$2.19} & 88.87$\pm$2.86 & 52.17$\pm$25.24 & 28.16$\pm$10.50 & 0.00$\pm$0.00 \\

\midrule
\multirow{4}{*}{\shortstack{Milan\\$\downarrow$\\Milan}}
& E-FCN     & \textbf{26.30$\pm$11.96} & 15.83$\pm$1.80 & 8.40$\pm$0.93 & 3.31$\pm$4.62 & 7.83$\pm$5.60 & \textbf{0.42$\pm$0.60} & 4.20$\pm$1.17 & 0.00$\pm$0.00 \\
& DeepCASAS & 25.50$\pm$11.78 & 15.38$\pm$3.03 & 9.96$\pm$0.58 & 0.26$\pm$0.29 & 8.97$\pm$6.69 & 0.00$\pm$0.00 & 1.43$\pm$1.92 & 0.00$\pm$0.00 \\
& TDOST     & 24.13$\pm$10.36 & 13.78$\pm$1.18 & 9.60$\pm$0.87 & 0.06$\pm$0.06 & 8.65$\pm$7.39 & 0.00$\pm$0.00 & 2.18$\pm$1.36 & 0.00$\pm$0.00 \\
& \projectname{}     & 25.76$\pm$3.47 & \textbf{49.65$\pm$3.64} & \textbf{18.43$\pm$6.83} & \textbf{3.44$\pm$4.24} & \textbf{85.33$\pm$0.43} & 0.00$\pm$0.00 & \textbf{45.32$\pm$10.25} & \textbf{5.06$\pm$3.66} \\

\midrule
\multirow{4}{*}{\shortstack{Kyoto7\\$\downarrow$\\Kyoto7}}
& E-FCN     & 45.74$\pm$13.50 & 2.18$\pm$2.80 & 24.58$\pm$11.84 & / & 20.94$\pm$7.97 & / & 0.00$\pm$0.00 & 0.00$\pm$0.00 \\
& DeepCASAS & 16.62$\pm$19.22 & 1.26$\pm$1.52 & 25.51$\pm$11.75 & / & 17.80$\pm$9.90 & / & 14.58$\pm$6.42 & 4.04$\pm$5.29 \\
& TDOST     & 4.55$\pm$3.33 & 2.50$\pm$1.76 & 22.06$\pm$13.56 & / & \textbf{33.90$\pm$14.48} & / & 26.85$\pm$15.46 & 0.74$\pm$0.55 \\
& \projectname{}     & \textbf{53.80$\pm$0.39} & \textbf{11.68$\pm$3.89} & \textbf{52.22$\pm$9.85} & / & 18.32$\pm$5.48 & / & \textbf{66.70$\pm$9.04} & \textbf{11.89$\pm$1.83} \\

\midrule
\multirow{4}{*}{\shortstack{Milan\\$\downarrow$\\Aruba}}
& E-FCN     & 43.50$\pm$3.43 & 1.39$\pm$1.37 & 0.71$\pm$0.24 & 0.77$\pm$0.96 & 31.58$\pm$20.04 & 0.00$\pm$0.00 & 5.84$\pm$4.19 & 0.00$\pm$0.00 \\
& DeepCASAS & 33.54$\pm$3.63 & 0.00$\pm$0.00 & 1.70$\pm$1.11 & 0.01$\pm$0.01 & 28.99$\pm$14.74 & 0.00$\pm$0.00 & 1.47$\pm$1.51 & 0.00$\pm$0.00 \\
& TDOST     & 42.97$\pm$4.37 & 4.72$\pm$0.95 & 58.48$\pm$9.77 & 2.41$\pm$1.68 & 0.00$\pm$0.00 & 8.53$\pm$7.18 & 0.00$\pm$0.00 & 0.00$\pm$0.00 \\
& \projectname{}     & \textbf{50.92$\pm$5.68} & \textbf{57.69$\pm$5.44} & \textbf{62.06$\pm$10.22} & \textbf{5.49$\pm$5.12} & \textbf{67.79$\pm$3.69} & \textbf{68.08$\pm$22.59} & \textbf{66.03$\pm$27.99} & 0.00$\pm$0.00 \\

\midrule
\multirow{4}{*}{\shortstack{Aruba\\$\downarrow$\\Milan}}
& E-FCN     & 63.47$\pm$13.69 & 6.00$\pm$1.76 & 7.30$\pm$0.49 & 0.00$\pm$0.00 & 7.96$\pm$6.29 & 0.00$\pm$0.00 & 0.35$\pm$0.31 & 0.27$\pm$0.38 \\
& DeepCASAS & \textbf{63.70$\pm$13.25} & 6.95$\pm$0.64 & 7.02$\pm$1.16 & 0.00$\pm$0.00 & 6.46$\pm$7.14 & 0.00$\pm$0.00 & 0.00$\pm$0.00 & 0.00$\pm$0.00 \\
& TDOST     & 62.99$\pm$14.24 & 5.39$\pm$4.70 & 7.17$\pm$1.20 & 0.00$\pm$0.00 & 4.94$\pm$4.56 & 0.00$\pm$0.00 & 0.00$\pm$0.00 & 0.00$\pm$0.00 \\
& \projectname{}     & 38.11$\pm$4.66 & \textbf{48.94$\pm$6.46} & \textbf{8.66$\pm$3.49} & \textbf{3.13$\pm$4.43} & \textbf{85.52$\pm$0.99} & \textbf{12.90$\pm$18.25} & \textbf{52.81$\pm$23.00} & \textbf{4.39$\pm$3.27} \\

\midrule
\multirow{4}{*}{\shortstack{Aruba\\$\downarrow$\\Kyoto7}}
& E-FCN     & 35.63$\pm$11.99 & 26.97$\pm$32.55 & 2.66$\pm$3.70 & / & 14.55$\pm$5.76 & / & 0.00$\pm$0.00 & 0.00$\pm$0.00 \\
& DeepCASAS & 14.71$\pm$6.85 & 4.11$\pm$0.40 & \textbf{10.15$\pm$6.92} & / & \textbf{22.74$\pm$8.50} & / & \textbf{12.18$\pm$10.99} & 0.12$\pm$0.16 \\
& TDOST     & \textbf{53.67$\pm$11.34} & 0.00$\pm$0.00 & 0.08$\pm$0.06 & / & 0.21$\pm$0.29 & / & 0.00$\pm$0.00 & 0.00$\pm$0.00 \\
& \projectname{}     & 52.69$\pm$1.84 & \textbf{2.95$\pm$3.84} & 2.06$\pm$2.91 & / & 11.02$\pm$2.67 & / & 0.00$\pm$0.00 & \textbf{14.11$\pm$3.14} \\

\midrule
\multirow{4}{*}{\shortstack{Milan\\$\downarrow$\\Kyoto7}}
& E-FCN     & 42.77$\pm$17.70 & 6.09$\pm$5.27 & \textbf{17.65$\pm$6.04} & / & 0.00$\pm$0.00 & / & 0.00$\pm$0.00 & 0.00$\pm$0.00 \\
& DeepCASAS & 54.37$\pm$11.56 & 0.00$\pm$0.00 & 6.38$\pm$4.24 & / & 16.70$\pm$12.25 & / & \textbf{5.87$\pm$7.45} & 0.00$\pm$0.00 \\
& TDOST     & 35.72$\pm$22.43 & 7.39$\pm$3.49 & 0.00$\pm$0.00 & / & 0.00$\pm$0.00 & / & 0.04$\pm$0.05 & 0.00$\pm$0.00 \\
& \projectname{}     & \textbf{58.05$\pm$1.05} & \textbf{11.07$\pm$6.71} & 9.36$\pm$8.60 & / & \textbf{23.78$\pm$0.94} & / & 0.00$\pm$0.00 & \textbf{10.16$\pm$0.85} \\

\midrule
\multirow{4}{*}{\shortstack{Kyoto7\\$\downarrow$\\Aruba}}
& E-FCN     & 30.37$\pm$3.38 & 11.97$\pm$2.43 & \textbf{7.40$\pm$1.99} & / & \textbf{35.52$\pm$9.81} & / & 0.00$\pm$0.00 & 0.00$\pm$0.00 \\
& DeepCASAS & 26.22$\pm$1.57 & 15.36$\pm$2.68 & 6.54$\pm$0.90 & / & 1.57$\pm$1.45 & / & 4.98$\pm$4.54 & \textbf{1.17$\pm$0.87} \\
& TDOST     & 15.02$\pm$2.47 & 3.98$\pm$0.88 & 5.47$\pm$1.61 & / & 5.74$\pm$6.73 & / & 4.40$\pm$1.71 & 0.16$\pm$0.16 \\
& \projectname{}     & \textbf{65.53$\pm$0.57} & \textbf{42.21$\pm$5.40} & 3.43$\pm$4.85 & / & 29.02$\pm$1.22 & / & \textbf{5.57$\pm$7.87} & 1.10$\pm$1.56 \\

\midrule
\multirow{4}{*}{\shortstack{Kyoto7\\$\downarrow$\\Milan}}
& E-FCN     & 59.35$\pm$27.94 & 17.89$\pm$4.99 & 9.63$\pm$0.97 & / & \textbf{20.13$\pm$7.04} & / & 0.00$\pm$0.00 & 0.00$\pm$0.00 \\
& DeepCASAS & 6.49$\pm$2.46 & \textbf{26.59$\pm$5.12} & 9.44$\pm$1.95 & / & 3.07$\pm$4.28 & / & 4.53$\pm$0.76 & 2.20$\pm$3.11 \\
& TDOST     & 6.80$\pm$3.03 & 5.96$\pm$1.13 & 7.31$\pm$3.09 & / & 4.77$\pm$3.61 & / & \textbf{5.90$\pm$4.76} & 0.38$\pm$0.31 \\
& \projectname{}     & \textbf{61.54$\pm$2.49} & 24.80$\pm$3.15 & \textbf{24.35$\pm$6.20} & / & 11.51$\pm$2.23 & / & 5.02$\pm$2.94 & \textbf{6.63$\pm$3.64} \\

\bottomrule
\end{tabular}
}
\end{table*}

\begin{table}[t]
\centering
\small
\caption{\rw{Per-class F1 scores (\%) of E-FCN, DeepCASAS, TDOST, and \projectname{} on Orange4Home. All results are reported as mean$\pm$standard deviation in percentage points (\%). Bold values indicate the best result in each row.}}
\label{tab:orange_class_results}
\begin{tabular}{lcccc}
\toprule
\rw{\textbf{Class}} & \rw{\textbf{E-FCN}} & \rw{\textbf{DeepCASAS}} & \rw{\textbf{TDOST}} & \rw{\textbf{\projectname{}}} \\
\midrule
\rw{OTHER} & \rw{$0.00 \pm 0.00$} & \rw{$0.00 \pm 0.00$} & \rw{$0.00 \pm 0.00$} & \rw{$\mathbf{10.24 \pm 8.03}$} \\
\rw{ENTERING} & \rw{$80.36 \pm 7.45$} & \rw{$77.83 \pm 11.15$} & \rw{$71.32 \pm 13.97$} & \rw{$\mathbf{94.25 \pm 1.00}$} \\
\rw{GOING\_UP} & \rw{$\mathbf{57.07 \pm 7.03}$} & \rw{$54.93 \pm 12.65$} & \rw{$11.16 \pm 15.68$} & \rw{$53.08 \pm 7.59$} \\
\rw{SHOWERING} & \rw{$71.68 \pm 4.51$} & \rw{$84.91 \pm 3.49$} & \rw{$\mathbf{90.25 \pm 4.23}$} & \rw{$86.66 \pm 3.58$} \\
\rw{USING\_THE\_SINK} & \rw{$39.24 \pm 7.12$} & \rw{$54.89 \pm 5.39$} & \rw{$\mathbf{67.49 \pm 7.63}$} & \rw{$62.18 \pm 3.64$} \\
\rw{GOING\_DOWN} & \rw{$36.11 \pm 30.68$} & \rw{$\mathbf{75.91 \pm 3.04}$} & \rw{$65.01 \pm 17.80$} & \rw{$46.97 \pm 7.82$} \\
\rw{WATCHING\_TV} & \rw{$76.67 \pm 31.83$} & \rw{$\mathbf{99.55 \pm 0.17}$} & \rw{$78.63 \pm 27.45$} & \rw{$75.89 \pm 31.30$} \\
\rw{USING\_THE\_TOILET} & \rw{$33.18 \pm 5.79$} & \rw{$64.24 \pm 8.10$} & \rw{$38.75 \pm 26.64$} & \rw{$\mathbf{76.26 \pm 5.29}$} \\
\rw{COMPUTING} & \rw{$99.14 \pm 0.51$} & \rw{$\mathbf{99.71 \pm 0.12}$} & \rw{$99.60 \pm 0.07$} & \rw{$99.54 \pm 0.11$} \\
\rw{PREPARING} & \rw{$55.43 \pm 16.21$} & \rw{$49.68 \pm 9.82$} & \rw{$48.08 \pm 11.85$} & \rw{$\mathbf{72.81 \pm 14.16}$} \\
\rw{COOKING} & \rw{$84.63 \pm 1.59$} & \rw{$63.74 \pm 38.36$} & \rw{$\mathbf{96.50 \pm 1.68}$} & \rw{$93.24 \pm 1.54$} \\
\rw{EATING} & \rw{$79.61 \pm 8.63$} & \rw{$76.91 \pm 0.88$} & \rw{$\mathbf{90.62 \pm 0.58}$} & \rw{$90.60 \pm 2.93$} \\
\rw{WASHING\_THE\_DISHES} & \rw{$58.92 \pm 3.23$} & \rw{$49.21 \pm 8.11$} & \rw{$\mathbf{83.97 \pm 1.69}$} & \rw{$74.76 \pm 5.75$} \\
\rw{CLEANING} & \rw{$25.65 \pm 14.84$} & \rw{$\mathbf{37.29 \pm 12.70}$} & \rw{$23.12 \pm 8.00$} & \rw{$32.28 \pm 7.36$} \\
\rw{DRESSING} & \rw{$84.62 \pm 6.28$} & \rw{$82.60 \pm 6.10$} & \rw{$\mathbf{90.42 \pm 4.18}$} & \rw{$69.82 \pm 6.16$} \\
\rw{READING} & \rw{$63.01 \pm 5.24$} & \rw{$67.70 \pm 7.30$} & \rw{$70.17 \pm 8.31$} & \rw{$\mathbf{75.51 \pm 8.01}$} \\
\rw{NAPPING} & \rw{$35.62 \pm 23.97$} & \rw{$41.61 \pm 6.85$} & \rw{$39.89 \pm 5.76$} & \rw{$\mathbf{51.68 \pm 22.04}$} \\
\rw{LEAVING} & \rw{$97.61 \pm 3.36$} & \rw{$\mathbf{97.62 \pm 3.36}$} & \rw{$81.20 \pm 9.18$} & \rw{$97.61 \pm 3.36$} \\
\bottomrule
\end{tabular}
\end{table}

\begin{table*}[t]
\centering
\small
\caption{\rw{Per-class F1 scores (\%) of E-FCN, DeepCASAS, TDOST, and \projectname{} on UCI-ADL Homes A and B. All results are reported as mean$\pm$standard deviation in percentage points (\%). ``/'' indicates that the corresponding activity class is absent from the home. Bold values indicate the best result for each class in each home.}}
\label{tab:uci_adl_class_results}
\resizebox{\textwidth}{!}{%
\begin{tabular}{lcccc|cccc}
\toprule
& \multicolumn{4}{c|}{\rw{\textbf{Home A}}} & \multicolumn{4}{c}{\rw{\textbf{Home B}}} \\
\cmidrule(lr){2-5}\cmidrule(lr){6-9}
\rw{\textbf{Class}} & \rw{\textbf{E-FCN}} & \rw{\textbf{DeepCASAS}} & \rw{\textbf{TDOST}} & \rw{\textbf{\projectname{}}} & \rw{\textbf{E-FCN}} & \rw{\textbf{DeepCASAS}} & \rw{\textbf{TDOST}} & \rw{\textbf{\projectname{}}} \\
\midrule
\rw{BREAKFAST} & \rw{$\mathbf{100.00 \pm 0.00}$} & \rw{$69.90 \pm 42.50$} & \rw{$76.90 \pm 24.90$} & \rw{$\mathbf{100.00 \pm 0.00}$} & \rw{$59.30 \pm 24.40$} & \rw{$51.00 \pm 22.10$} & \rw{$\mathbf{61.00 \pm 8.40}$} & \rw{$58.40 \pm 23.60$} \\
\rw{DINNER} & \rw{/} & \rw{/} & \rw{/} & \rw{/} & \rw{$3.10 \pm 4.40$} & \rw{$\mathbf{32.80 \pm 30.60}$} & \rw{$0.50 \pm 0.80$} & \rw{$3.10 \pm 4.40$} \\
\rw{GROOMING} & \rw{$68.70 \pm 38.60$} & \rw{$57.70 \pm 41.60$} & \rw{$\mathbf{76.30 \pm 17.50}$} & \rw{$69.90 \pm 34.60$} & \rw{$73.40 \pm 16.80$} & \rw{$\mathbf{92.40 \pm 2.40}$} & \rw{$73.40 \pm 8.90$} & \rw{$79.30 \pm 14.60$} \\
\rw{LEAVING} & \rw{$\mathbf{100.00 \pm 0.00}$} & \rw{$\mathbf{100.00 \pm 0.00}$} & \rw{$\mathbf{100.00 \pm 0.00}$} & \rw{$\mathbf{100.00 \pm 0.00}$} & \rw{$\mathbf{99.00 \pm 1.40}$} & \rw{$96.70 \pm 4.60$} & \rw{$95.90 \pm 5.80$} & \rw{$98.90 \pm 1.40$} \\
\rw{LUNCH} & \rw{$\mathbf{97.80 \pm 3.10}$} & \rw{$61.80 \pm 28.40$} & \rw{$92.30 \pm 10.80$} & \rw{$\mathbf{97.80 \pm 3.10}$} & \rw{$52.80 \pm 35.40$} & \rw{$46.90 \pm 18.30$} & \rw{$\mathbf{57.50 \pm 41.50}$} & \rw{$52.50 \pm 34.00$} \\
\rw{SHOWERING} & \rw{$\mathbf{100.00 \pm 0.00}$} & \rw{$99.60 \pm 0.60$} & \rw{$60.80 \pm 43.60$} & \rw{$\mathbf{100.00 \pm 0.00}$} & \rw{$\mathbf{100.00 \pm 0.00}$} & \rw{$\mathbf{100.00 \pm 0.00}$} & \rw{$0.00 \pm 0.00$} & \rw{$96.10 \pm 5.50$} \\
\rw{SLEEPING} & \rw{$\mathbf{100.00 \pm 0.00}$} & \rw{$\mathbf{100.00 \pm 0.00}$} & \rw{$66.70 \pm 47.10$} & \rw{$\mathbf{100.00 \pm 0.00}$} & \rw{$99.30 \pm 0.50$} & \rw{$98.90 \pm 0.70$} & \rw{$99.10 \pm 0.40$} & \rw{$\mathbf{99.40 \pm 0.60}$} \\
\rw{SNACK} & \rw{$6.70 \pm 9.40$} & \rw{$0.00 \pm 0.00$} & \rw{$22.20 \pm 31.40$} & \rw{$\mathbf{60.00 \pm 30.30}$} & \rw{$69.80 \pm 1.60$} & \rw{$\mathbf{76.30 \pm 7.70}$} & \rw{$71.90 \pm 10.40$} & \rw{$68.60 \pm 2.30$} \\
\rw{SPARE\_TIME/TV} & \rw{$96.70 \pm 4.40$} & \rw{$94.40 \pm 7.00$} & \rw{$89.10 \pm 15.30$} & \rw{$\mathbf{98.00 \pm 2.60}$} & \rw{$\mathbf{97.30 \pm 2.00}$} & \rw{$95.60 \pm 2.10$} & \rw{$97.10 \pm 1.30$} & \rw{$\mathbf{97.30 \pm 1.70}$} \\
\rw{TOILETING} & \rw{$\mathbf{71.90 \pm 23.30}$} & \rw{$31.20 \pm 43.20$} & \rw{$40.50 \pm 31.50$} & \rw{$60.70 \pm 9.20$} & \rw{$70.90 \pm 4.40$} & \rw{$71.40 \pm 6.60$} & \rw{$72.90 \pm 7.30$} & \rw{$\mathbf{75.70 \pm 5.20}$} \\
\bottomrule
\end{tabular}%
}
\end{table*}

\subsection{~\projectname{} Confusion Matrix on Benchmarks}
\label{app:confusion_matrix}

To further analyze model behavior at the class level, we present the confusion matrices of \projectname{} on all benchmarks. 

\begin{figure*}[t]
\centering
\includegraphics[width=0.45\textwidth]{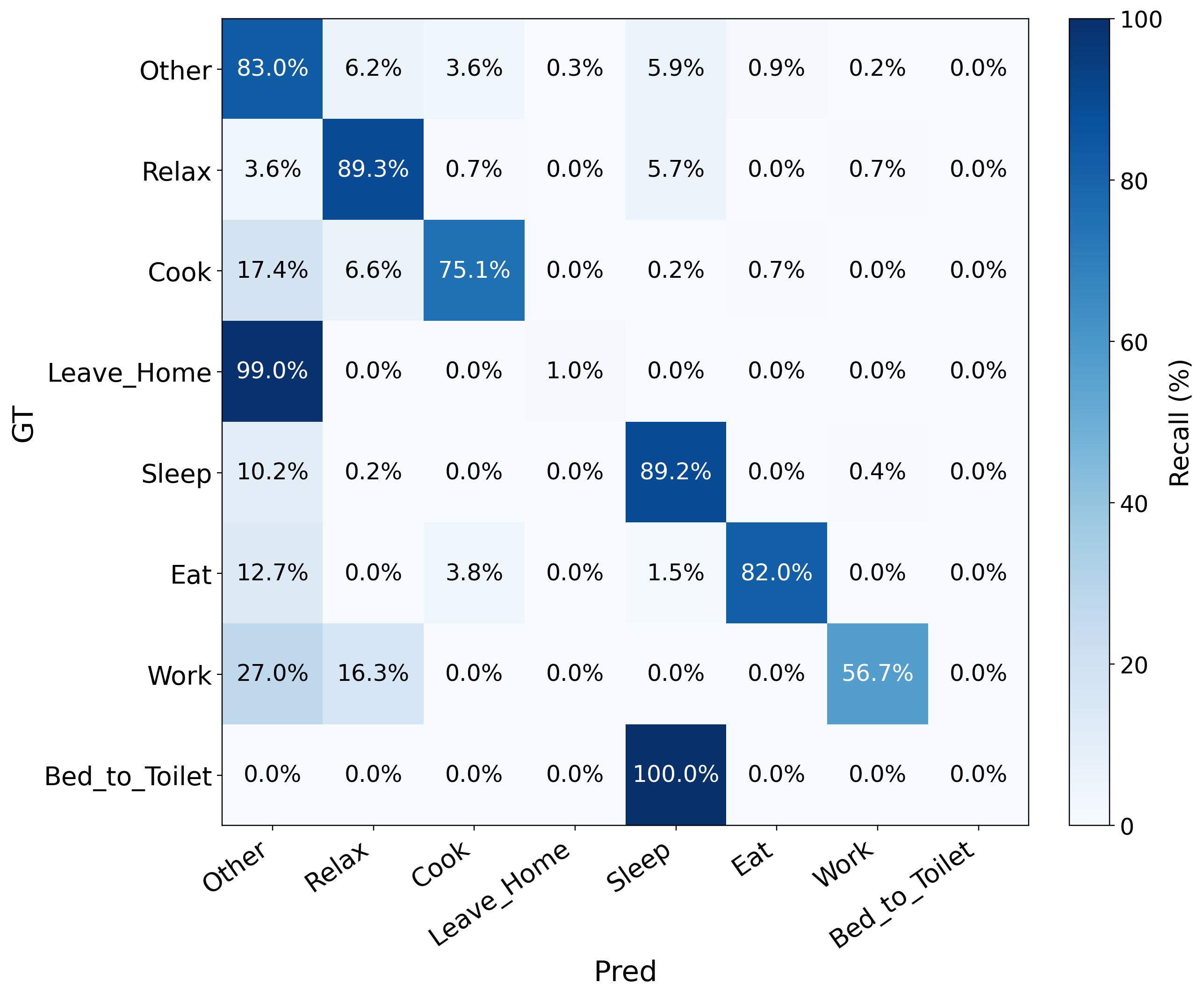}
\includegraphics[width=0.45\textwidth]{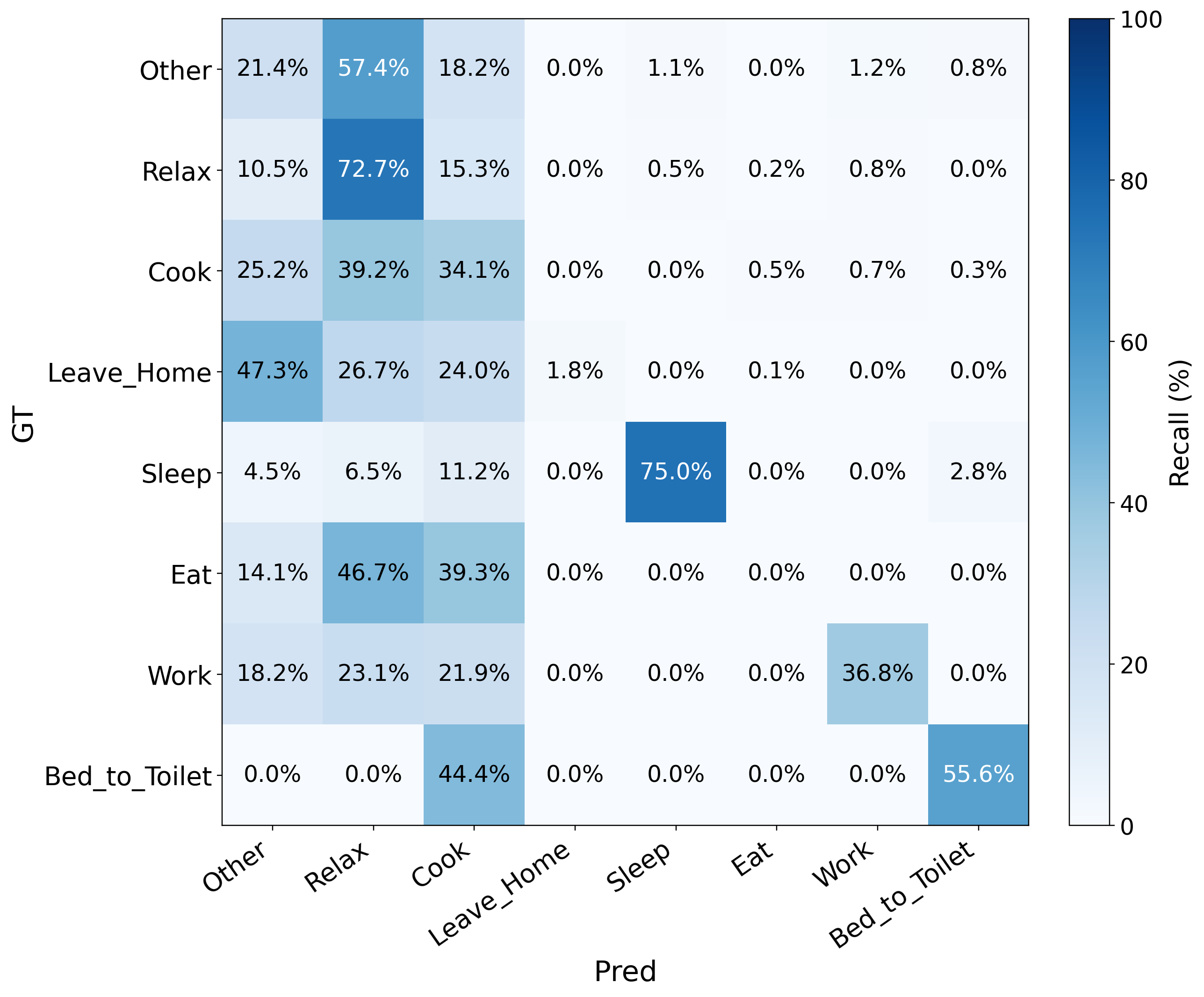}

\vspace{0.5em}
\includegraphics[width=0.45\textwidth]{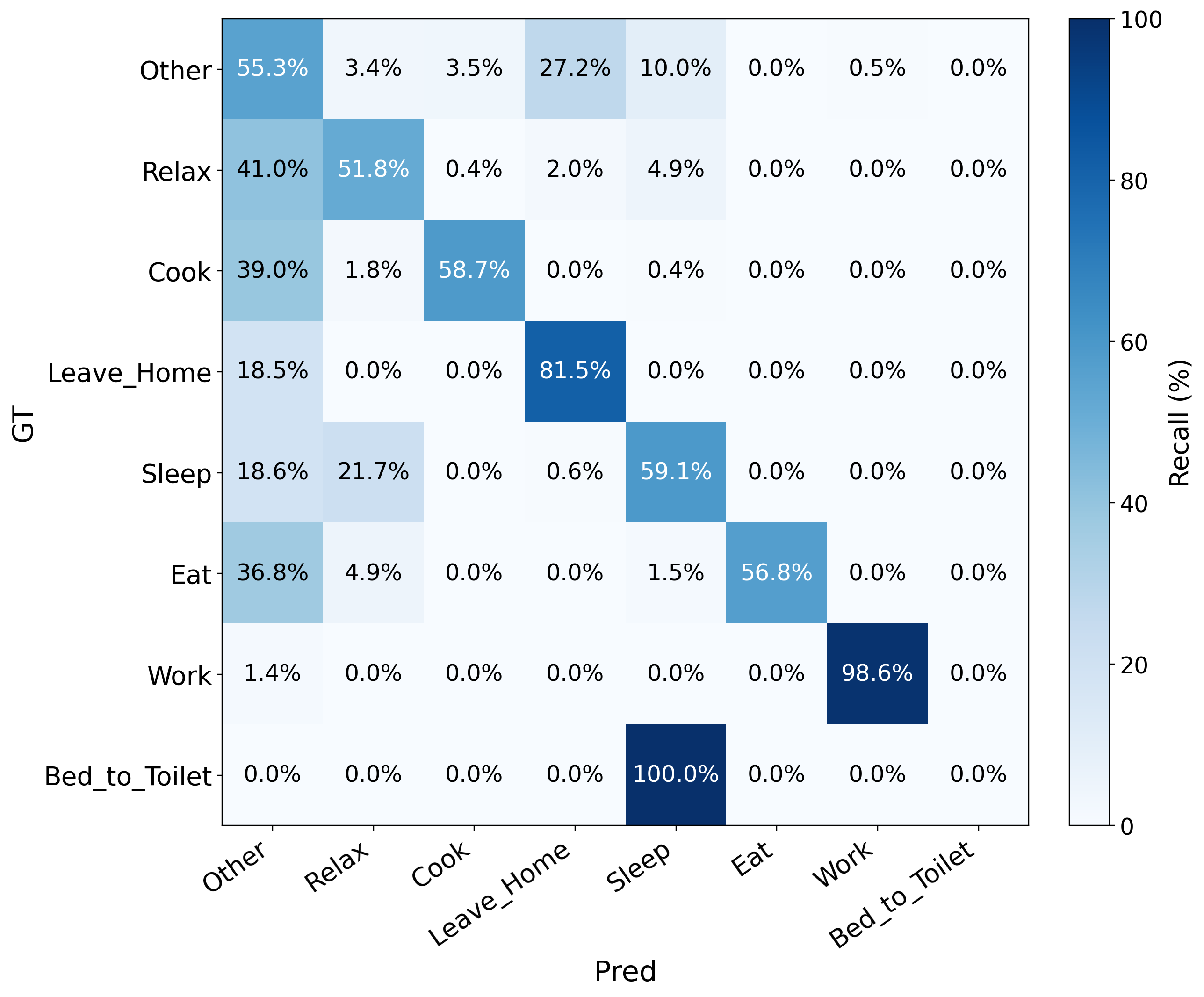}
\includegraphics[width=0.45\textwidth]{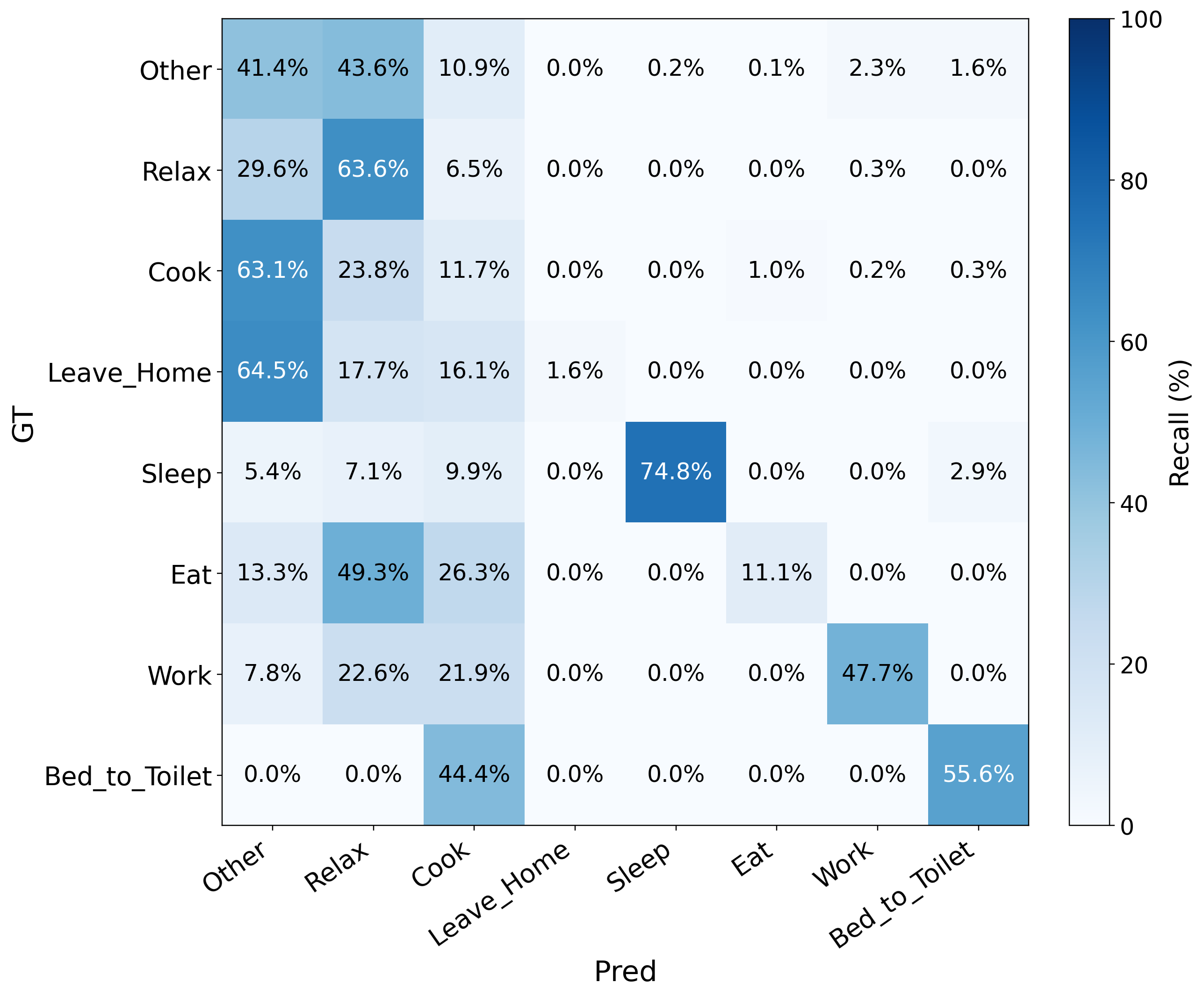}

\caption{
Row-normalized confusion matrices (\%) of \projectname{} for the Aruba--Milan evaluation settings:
Aruba$\rightarrow$Aruba (top-left), Milan$\rightarrow$Milan (top-right), Milan$\rightarrow$Aruba  (bottom-left), and Aruba$\rightarrow$Milan (bottom-right).
Each matrix is averaged over three runs after row-wise normalization.
}
\label{fig:confusion_am}
\end{figure*}

\begin{figure*}[t]
\centering

\includegraphics[width=0.45\linewidth]{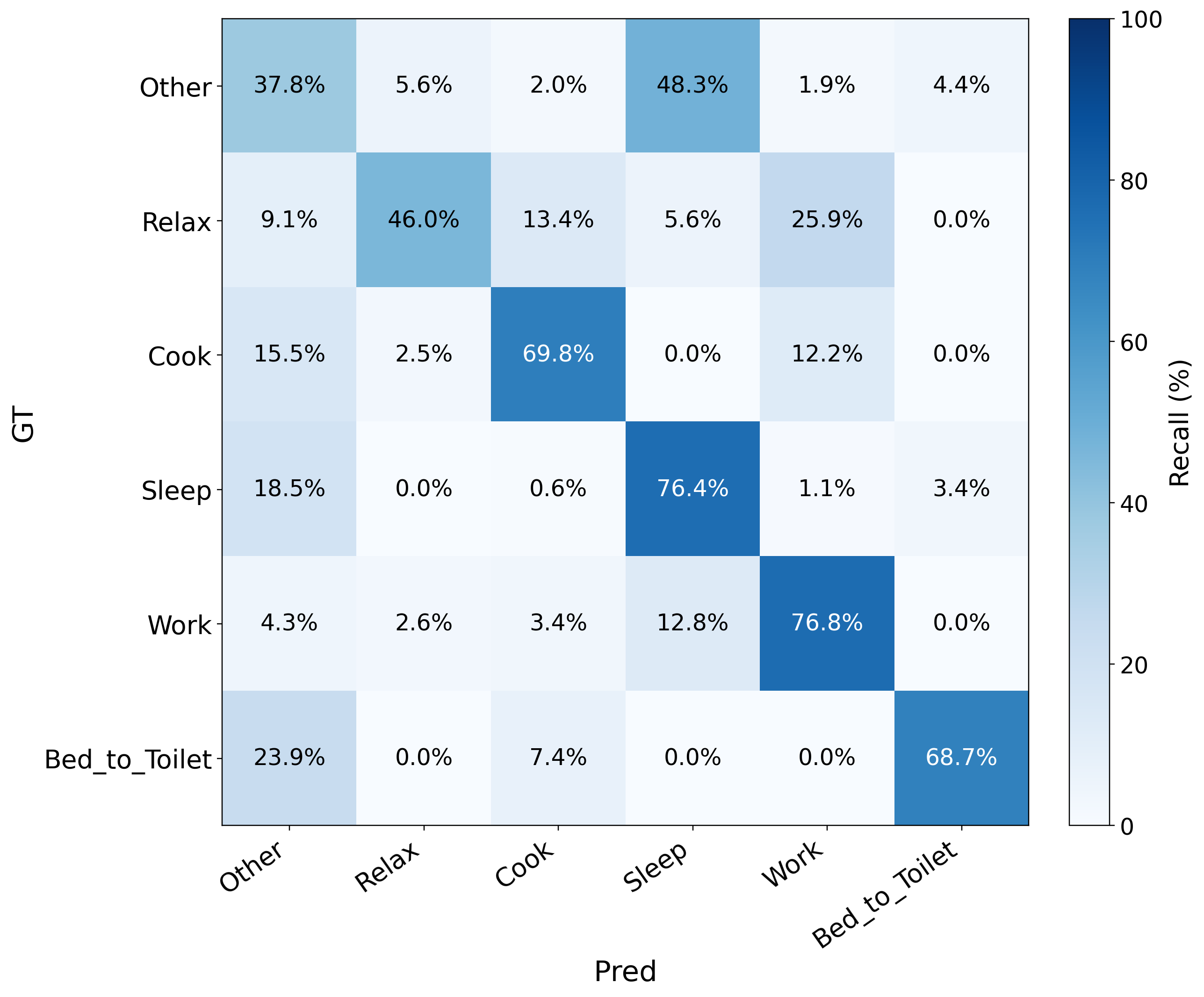}

\vspace{0.5em}

\includegraphics[width=0.45\linewidth]{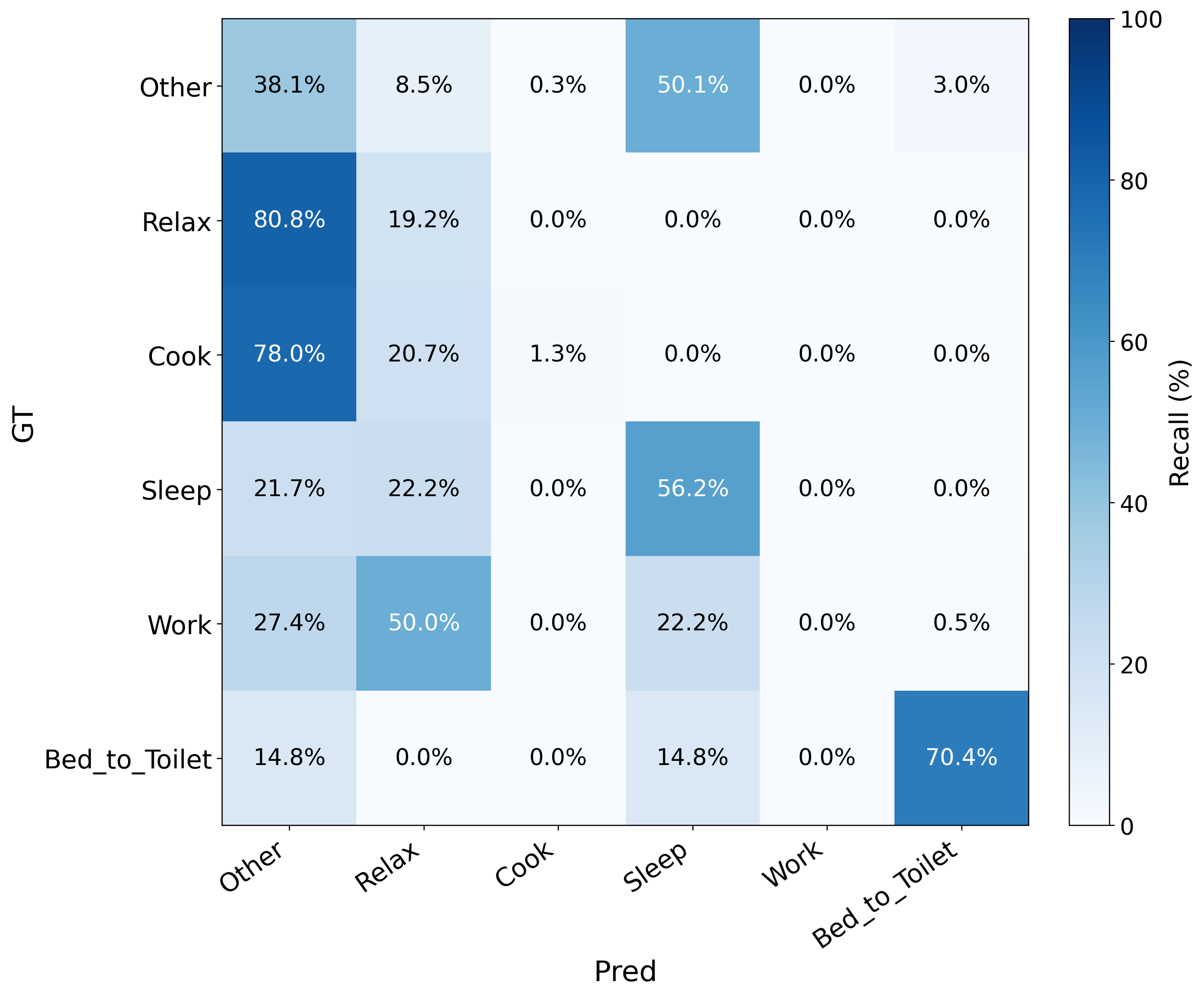}
\includegraphics[width=0.45\linewidth]{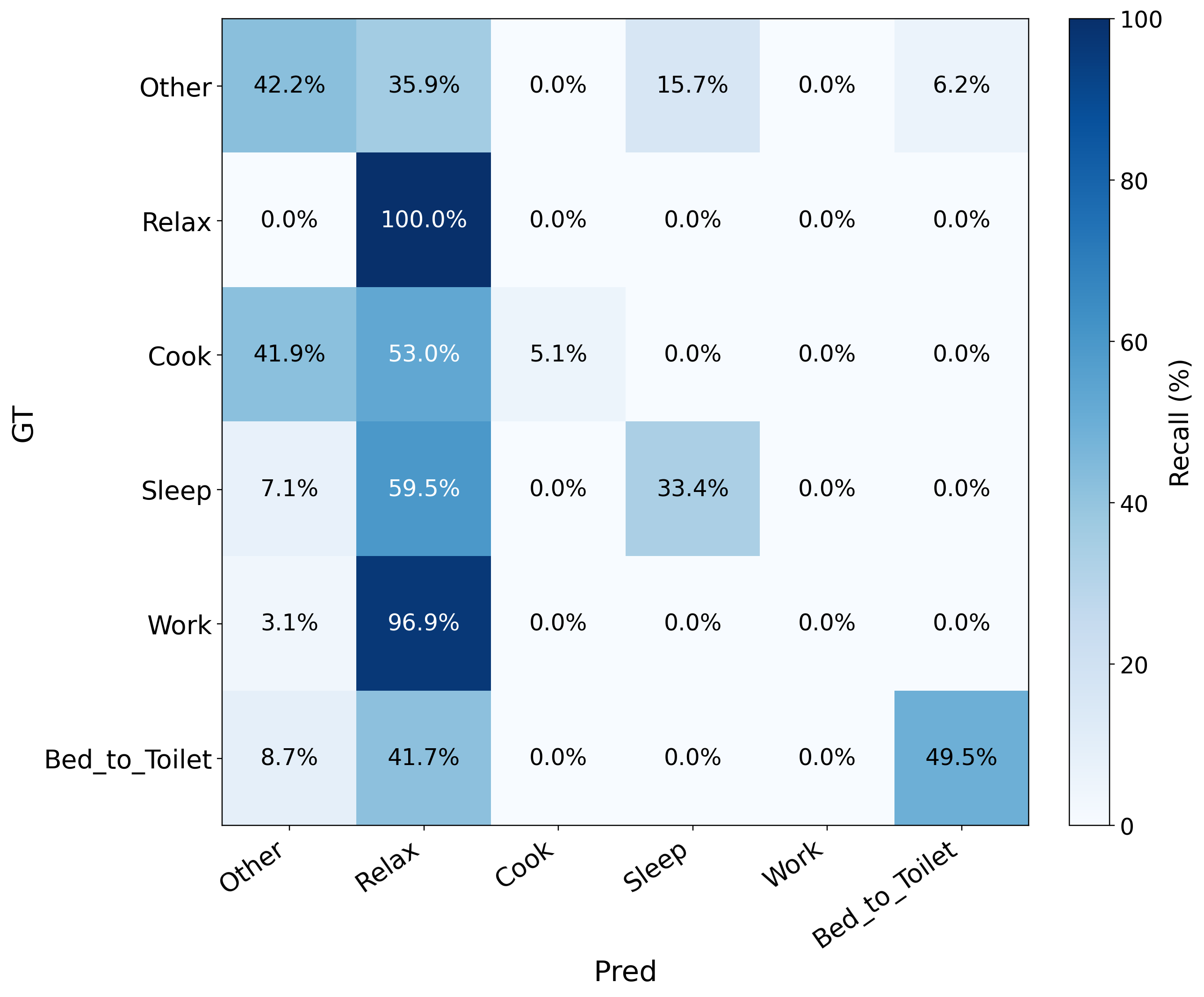}

\vspace{0.5em}

 \includegraphics[width=0.45\linewidth]{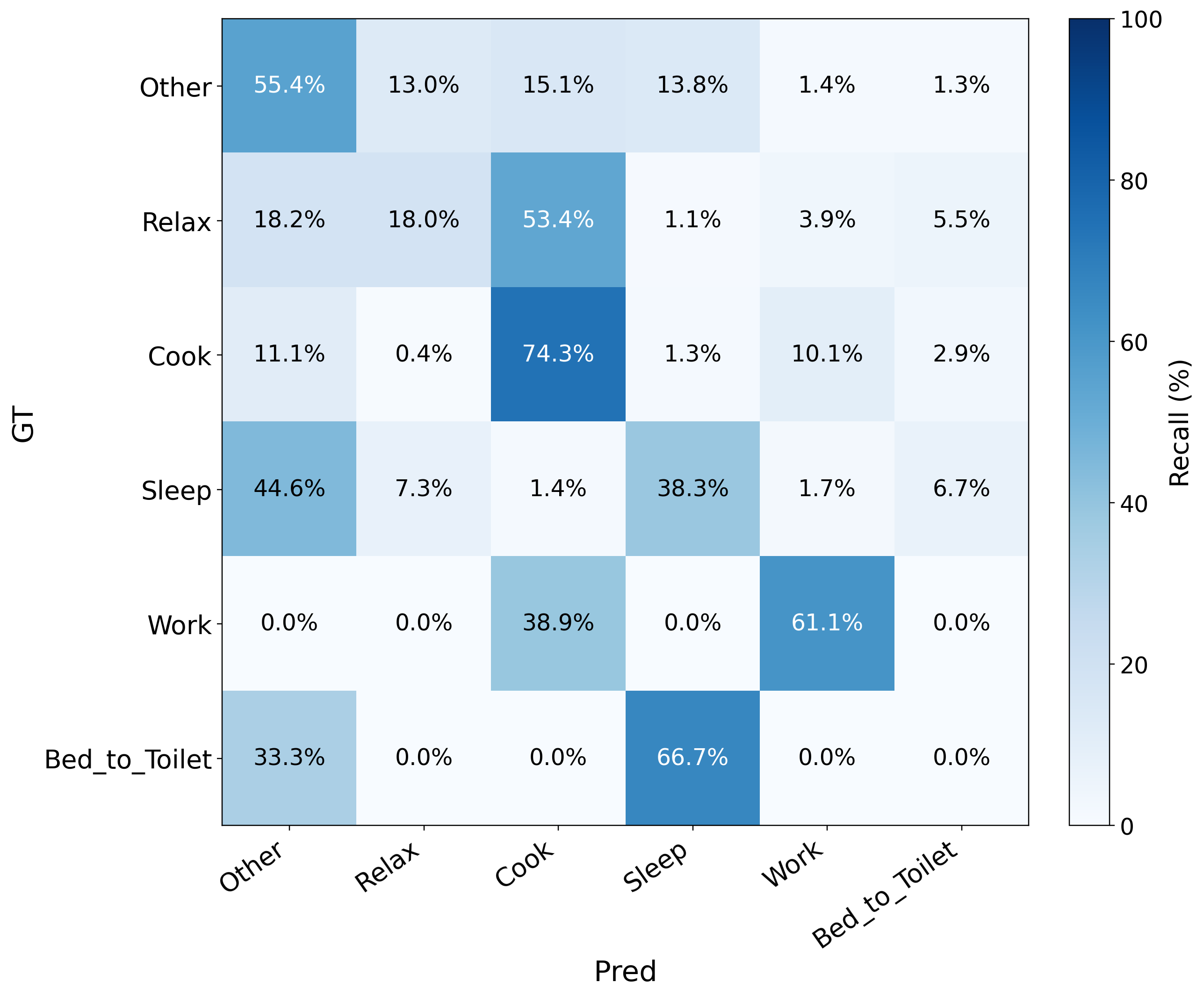}
 \includegraphics[width=0.45\linewidth]{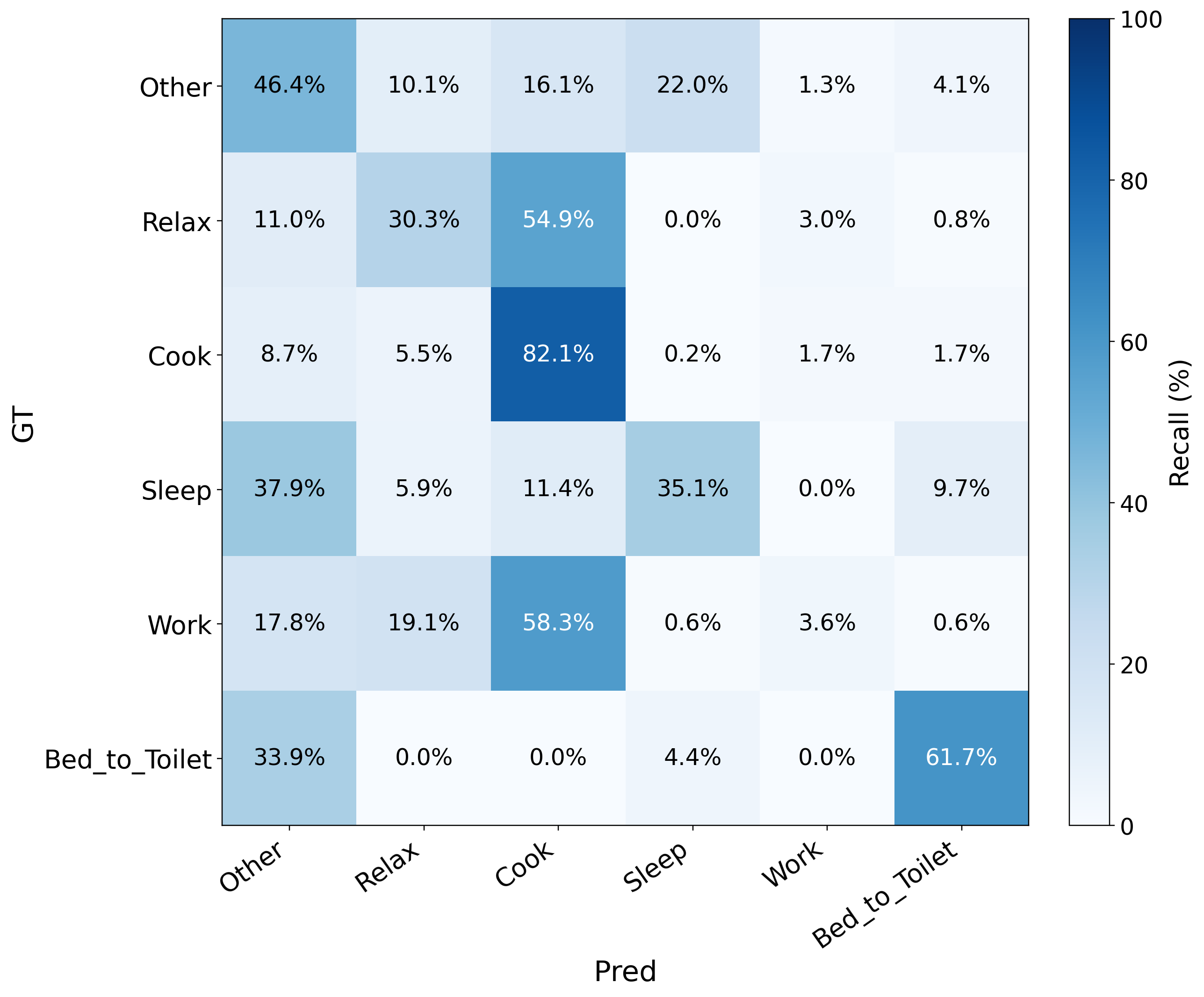}

\caption{
Row-normalized confusion matrices (\%) of \projectname{} for the evaluation settings involving Kyoto7:
Kyoto7$\rightarrow$Kyoto7 (top), Aruba$\rightarrow$Kyoto7 (mid-left), Milan$\rightarrow$Kyoto7 (mid-right), Kyoto7$\rightarrow$Aruba  (bottom-left), and Kyoto7$\rightarrow$Milan (bottom-right).
Each matrix is averaged over three runs after row-wise normalization.
}
\label{fig:confusion_amk}
\end{figure*}

\begin{figure*}[t]
\centering

\includegraphics[width=0.65\linewidth]{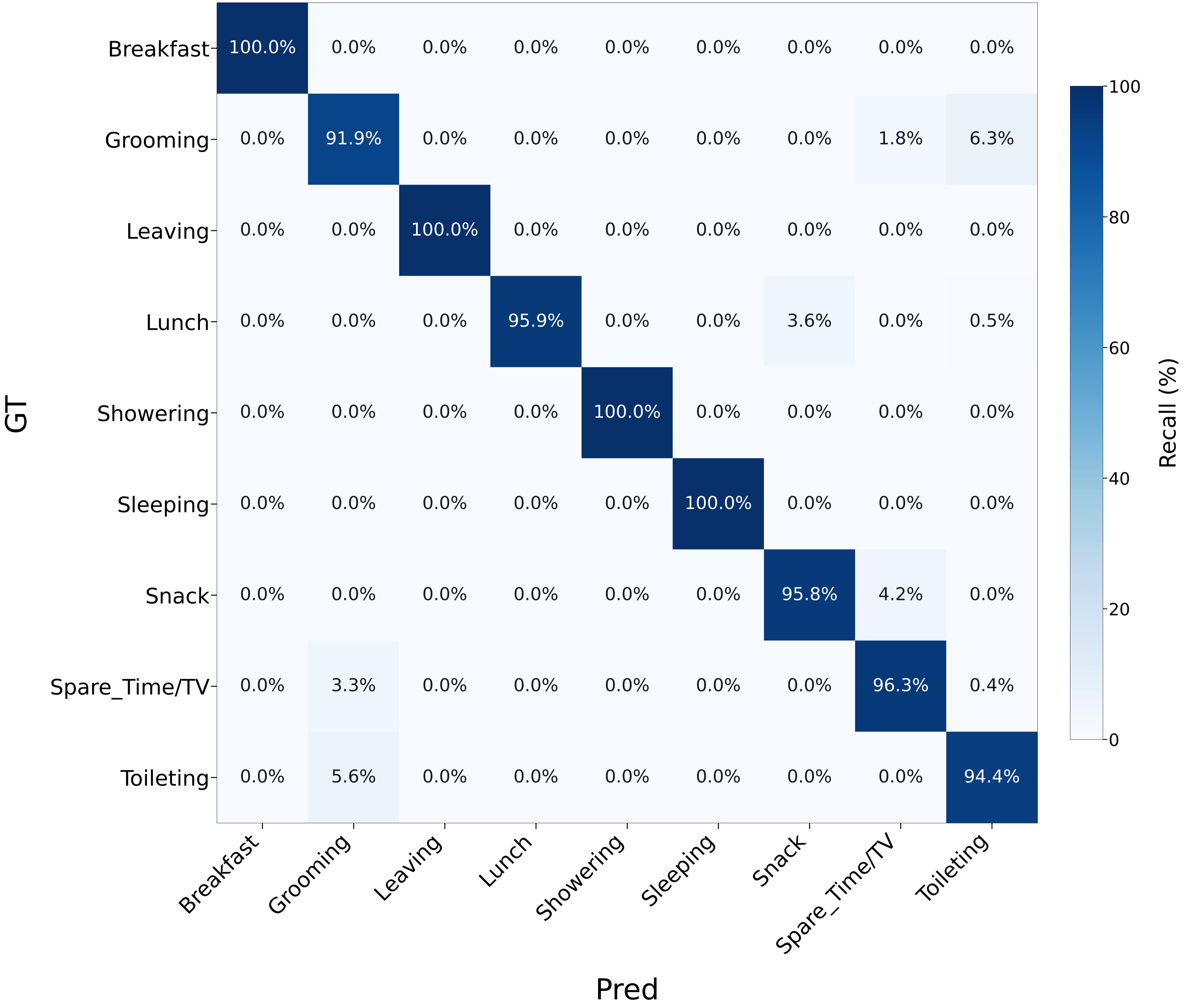}

\vspace{0.5em}

 \includegraphics[width=0.65\linewidth]{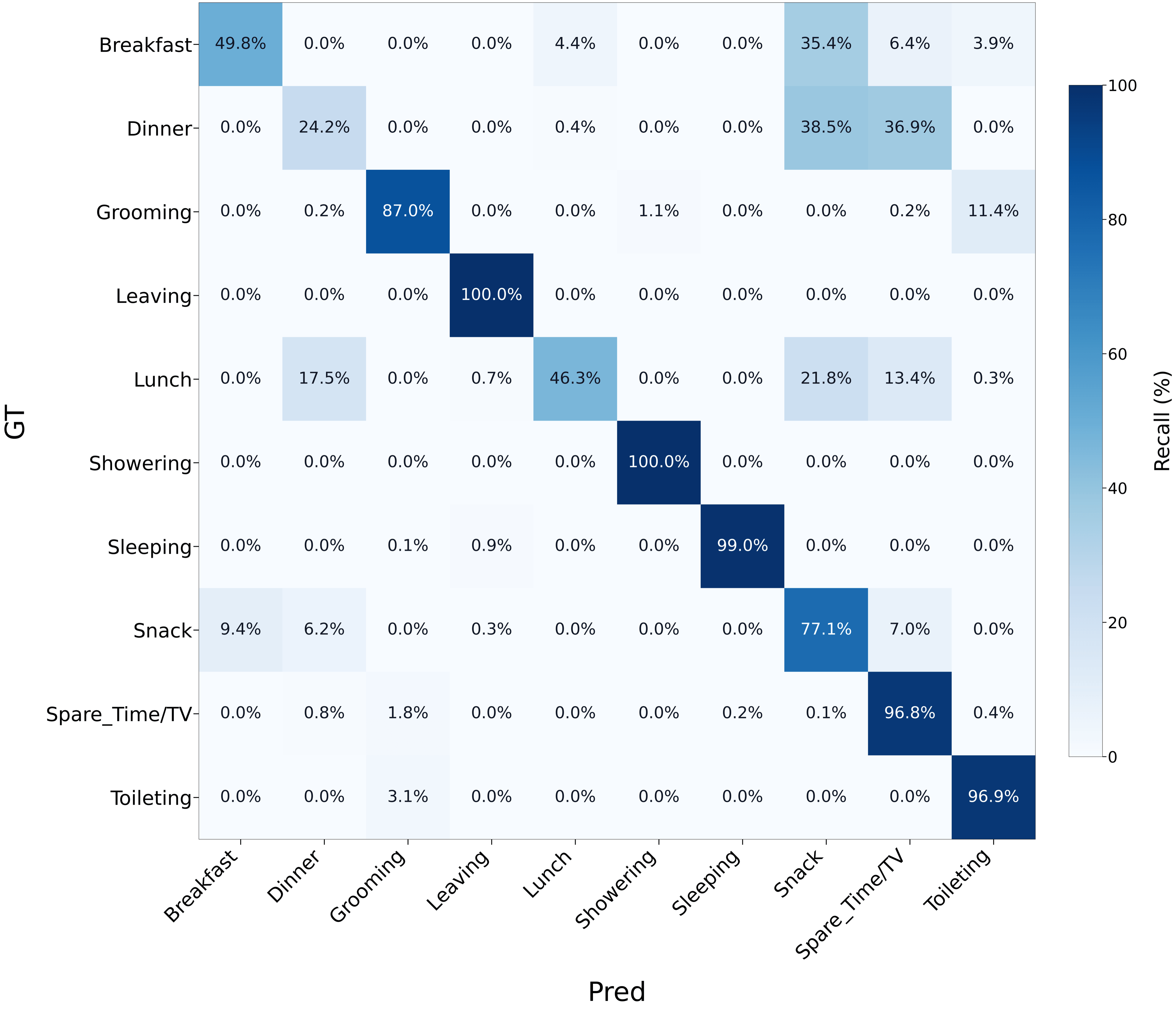}

\caption{\rw{Row-normalized confusion matrices (\%) of \projectname{} for UCI-ADL:
Home A (top) and Home B (bottom).
Each matrix is averaged over three runs after row-wise normalization.}}
\label{fig:confusion_adl}
\end{figure*}

\begin{figure*}[t]
\centering

 \includegraphics[width=\linewidth]{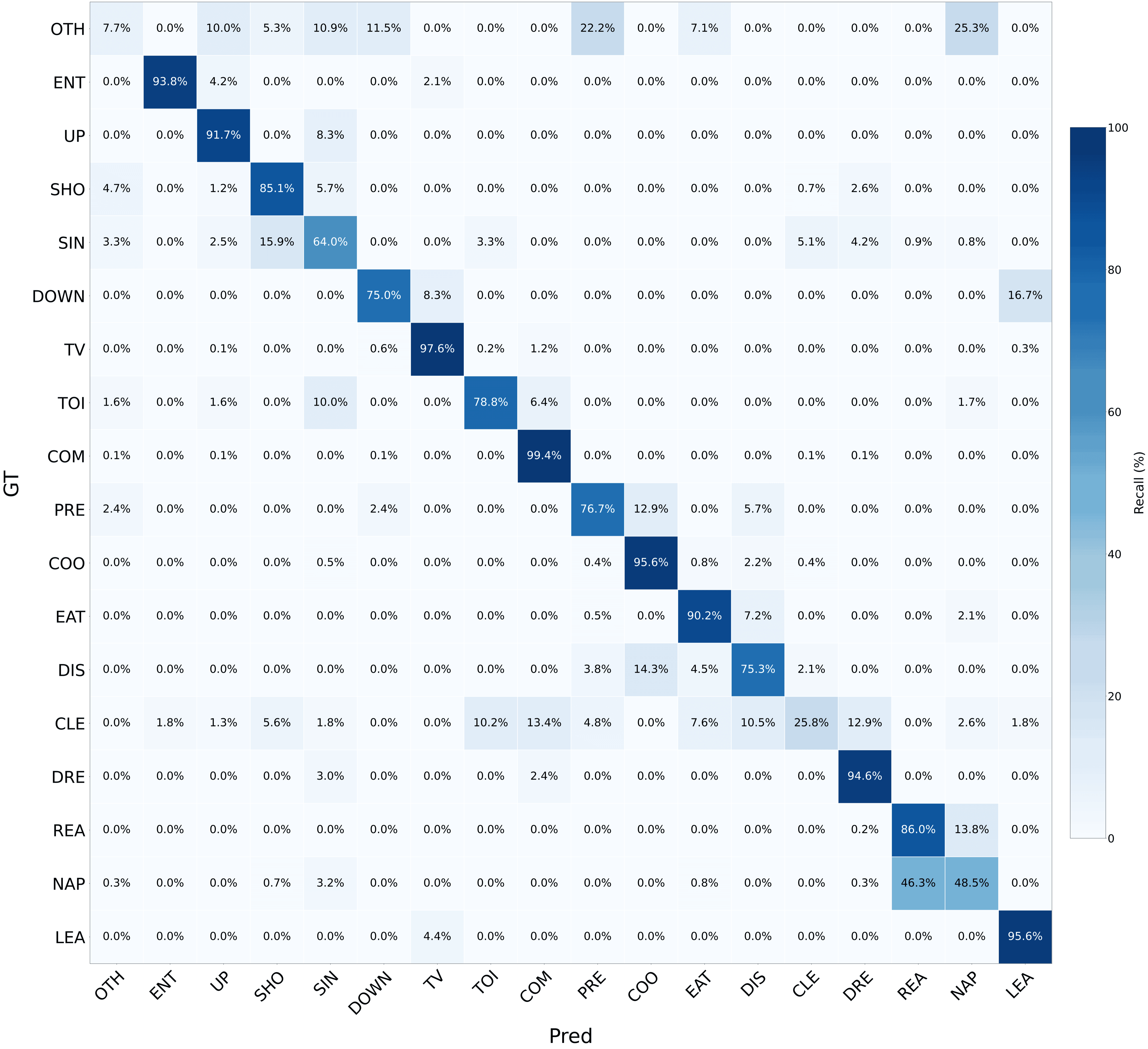}

\caption{\rw{Row-normalized confusion matrices (\%) of \projectname{} on Orange4Home. Each matrix is averaged over three runs after row-wise normalization. Class abbreviations are as follows: OTH--Other, ENT--Entering, UP--Going Up, SHO--Showering, SIN--Using the Sink, DOWN--Going Down, TV--Watching TV, TOI--Using the Toilet, COM--Computing, PRE--Preparing, COO--Cooking, EAT--Eating, DIS--Washing the Dishes, CLE--Cleaning, DRE--Dressing, REA--Reading, NAP--Napping, and LEA--Leaving.}}
\label{fig:confusion_adl}
\end{figure*}

\end{document}